\documentclass[aps,prb,twocolumn]{revtex4-1}
\usepackage{graphicx,amssymb,amsfonts,amsmath,dcolumn,bm}

\makeatletter
\newsavebox{\@brx}
\newcommand{\llangle}[1][]{\savebox{\@brx}{\(\m@th{#1\langle}\)}%
  \mathopen{\copy\@brx\mkern2mu\kern-0.9\wd\@brx\usebox{\@brx}}}
\newcommand{\rrangle}[1][]{\savebox{\@brx}{\(\m@th{#1\rangle}\)}%
  \mathclose{\copy\@brx\mkern2mu\kern-0.9\wd\@brx\usebox{\@brx}}}
\makeatother

\begin{document}

\title{Re-examining the Statistical Mechanics of an Interacting Bose Gas}

\author{A.M. Ettouhami} 


\email{ettouhami@gmail.com}

\affiliation{29 Lloyd Gibson Cres., Whitby, Ontario, L1R 2H6, Canada}

\date{\today}

\begin{abstract}
A physical system in the grand-canonical ensemble is in contact with a particle reservoir, hence the average number of particles in the system is not fixed, but depends on other thermodynamic variables, most notably the temperature $T$. Yet, in the field-theoretic formulation of the statistical mechanics of interacting bosons, the number of bosons that appears in the expressions of the grand-potential and of other thermodynamic quantities is taken to be a fixed quantity that is independent of temperature. In this paper, we re-examine the way in which Bogoliubov's theory of a dilute Bose gas at $T=0$ has been extended to describe the statistical mechanics of interacting bosons at finite temperatures, and show explicitly that the field-theoretic calculation of the grand partition function in this formulation amounts to a {\em canonical} trace over the eigenfunctions of the Bogoliubov Hamiltonian at {\em fixed} total number of bosons $N$, and that the additional trace over $N$ that is required in the grand-canonical formalism is never carried out. This implies that what usually passes as the grand-canonical treatment of the Bogoliubov Hamiltonian of interacting bosons is not quite grand-canonical, and is in fact a canonical treatment. We also show that the discontinuity in the condensate density predicted by previous formulations of this theory as the temperature $T$ goes past the critical transition temperature $T_c$ is a direct consequence of an inappropriate generalization of the Bogoliubov prescription to finite temperatures, and that this discontinuity disappears when this prescription is either used as a zero temperature approximation or avoided altogether. Armed with the above findings, we reformulate the statistical mechanics of interacting bosons in the  canonical ensemble using the variational number-conserving approach developed in a pevious publication [A. M. Ettouhami, {\em Prog. Theor. Phys.}, {\bf 127}, 453 (2012)], and derive the thermodynamics of the system. We then show how the canonical treatment can be used to setup a grand-canonical description of the statistical mechanics of a weakly interacting Bose gas where the average number of bosons in the system {\em does} vary with temperature. Consequences on the physics of interacting bosons are briefly discussed.

\end{abstract}




\maketitle

\section{Introduction} 

Bogoliubov's description of the quantum-mechanical ground state of dilute gases of interacting bosons \cite{Bogoliubov1947} 
forms the basis of our understanding of these systems.
\cite{Lee1957a,LHY1957,Lee1957b,Bruckner1957,Beliaev1958a,Beliaev1958b,Hugenholtz1959,Gavoret1964,Popov1965,Sawada1959,Hohenberg1965,Singh1967,Cheung1971,Szepfalusy1974,Wong1974,Bijlsma1997,Shi1998,Zagrebnov2001,Capogrosso2010} 
Yet, despite its widespread acceptance and aura of venerability, it has long been recognized that several aspects of Bogoliubov's theory
were not very well understood.\cite{Girardeau1959,Misawa1960,Gardiner1997,Girardeau1998,Castin1998,Leggett2001,Andersen2004,Yukalov2006,Suto2008,LeggettNJP,Kita2010,Bobrov2016}

In a recent paper, \cite{Ettouhami2012} we analyzed Bogoliubov's theory in quite some detail, 
and discussed several problematic aspects of this theory, which originate in the so-called Bogoliubov prescription on one hand,
and in the non-conservation of the number of bosons on the other. In particular, we have shown that Bogoliubov's theory
does not model a unique system of $N$ interacting bosons, but rather describes an independent collection of such systems, 
where in each one of these systems $N$ bosons are allowed to be in either one of the three single particle states with momenta ${\bf 0, \pm k}$.
Indeed, in Bogoliubov's theory, a truncated version $\hat{H}_B$ of the total Hamiltonian $\hat{H}$ of the system is considered, and this truncated Hamiltonian $\hat{H}_B$ is written as a sum $\hat{H}_B=\sum_{\bf k\neq 0}\hat{H}_{\bf k}$, where each $\hat{H}_{\bf k}$ describes a sytem of $N$ bosons that can only be in one of three momentum states: $-{\bf k}, {\bf 0}$, and ${\bf k}$. Then, the Hamiltonians $\hat{H}_{\bf k}$ are diagonalized independently from one another, leading to a theory that describes, as we mentioned above, not a unique system of $N$ interacting bosons, where the same particle in the condensed state with ${\bf k}=0$ can be excited to any one of the other single-particle states with ${\bf k}\neq 0$, but a juxtaposition of independent systems, each with its own reservoir of $N$ bosons, and where a boson in the condensate with ${\bf k}={\bf 0}$ can only be excited to one of the single particle states $\pm{\bf k}$. Reformulating the theory so as to describe a unique system of $N$ bosons such that a boson in the condensate with ${\bf k}={\bf 0}$ can now be excited to any one of single-particle states with momenta $\{\pm{\bf k}_1,\ldots,\pm{\bf k}_\infty\}$, and taking care to preserve the conservation of particle number, we have explicitly shown\cite{Ettouhami2012} that the depletion of bosons is significantly reduced with respect to the Bogoliubov case, and that the elementary excitations of the truncated Hamiltonian $\hat{H}_B$ become gapped, in contradiction with the results of the standard Bogoliubov theory.  

As has been noted elsewhere, \cite{Suto2008} a gapped excitation spectrum does not necessarily indicate that the theory is incorrect or that there is a problem with the underlying variational procedure. Actually, most number-conserving theories tend to predict a gapped excitation spectrum. \cite{Suto2008} Of course, experimental evidence dictates that in a complete theory the elementary excitations of a uniform, translationally invariant system of interacting bosons should not have a gap. This, however, should not induce us to turn a blind eye to all the conceptual difficulties of Bogoliubov's theory and accept it as ``the" correct formulation merely because it predicts a gapless excitation spectrum. To do so would violate the integrity of the whole process of scientific inquiry, in which the end does not justify the means, and every approximate theory needs to be conceptually sound before its results can be accepted. So, rather than continuing to use an imperfect formulation which predicts a gapless spectrum using highly questionable intermediate steps, a proper approach would consist in using a theory with sound foundations and no internal inconsistencies even if such a theory predicts elementary excitations that have a finite gap, and work toward improving this theory to eventually reach a refined version that predicts no gap. After all, progress in research is most of the time incremental, and there is a better chance of a conceptually sound gapless theory to emerge from a correctly formulated gapped approach than from an incorrectly formulated gapless one.

In this article, we want to resume the project we started in Ref. \onlinecite{Ettouhami2012} and re-examine the way in which Bogoliubov's theory of a dilute Bose gas at $T=0$ has been extended to describe the statistical mechanics of interacting bosons at finite temperatures. Here again, we find upon close scrutiny that several aspects of the conventional formulation of the statistical mechanics of Bose systems are quite questionable and do not rest on sound foundations. The main manifestation of this is our finding that what usually passes as the grand-canonical treatment of the Bogoliubov Hamiltonian of interacting bosons is not a grand-canonical treatment at all, and is in fact a canonical one. A clue that betrays the canonical nature of the standard Bogoliubov-Beliaev-Popov (BBP) theory at finite temperature is the presence of the density of condensed bosons $n_0(T)$ 
in the expressions of the thermodynamic quantities derived within this approach when in fact in a grand-canonical formulation the number of condensed bosons $N_0$ should be traced over and should not even appear in these expressions. Another clue is the fact that the density of bosons $n_B$ in these formulations is taken to be a constant that does not vary with temperature, while in a grand-canonical description the average density of bosons is not fixed, but depends on other thermodynamic variables, most notably the temperature.
Below, we will show that the main results of the BBP approach can be derived within a purely canonical formulation, and we show how one can extend this theory to the grand-canonical case where the system is in contact with a particle reservoir and the average density of bosons $n_B$ is not a constant, but depends on the chemical potential $\mu$, the volume $V$ and the temperature $T$,  $n_B = n_B(\mu, V, T)$, and explicitly derive an expression of this dependence.

The rest of this paper is organized as follows. In Sec. \ref{Sec:Review} we review the standard formulation of the statistical mechanics of a dilute Bose gas. This formulation is based on the use of the so-called Bogoliubov prescription, where the imaginary-time boson field $\Psi({\bf r},\tau)$ is approximated as the sum of a condensed part $\sqrt{n_0}$, treated as a constant in $({\bf r},\tau)$-space, and a fluctuating field $\psi({\bf r},\tau)$ representing depleted bosons:
\begin{equation}
\Psi({\bf r},\tau) \simeq \sqrt{n_0} + \psi({\bf r},\tau). 
\label{Eq:BogPres}
\end{equation}
In Ref. \onlinecite{Ettouhami2012}, we saw that the zero temperature version of this prescription was the root cause of many shortcomings and inconsistencies of Bogoliubov's theory. Here, we will show that the finite temperature version of Bogoliubov's prescription creates one more major inconsistency by artificially leading to the appearance of a jump discontinuity in
the density of condensed bosons $n_0(T)$ at the critical transition temperature $T_c$. A discussion of how this happens and how this problem can be avoided will be presented in Sec. \ref{Sec:Discontinuity}, where we 
argue that the quantity $n_0$ inside the square root on the {\em rhs} of Eq. (\ref{Eq:BogPres}) should not depend on temperature, because it is the result of the ad-hoc prescription (\ref{Eq:BogPres}) and not the result of a thermal averaging process. We furthermore show that interpreting $n_0$ on the {\em rhs} of Eq. (\ref{Eq:BogPres}) as the density of condensed bosons at zero temperature
leads to a critical transition temperature $T_c$ which is higher than the transition temperature $T_{c0}$ of an ideal Bose gas, with the difference $\Delta T_c = T_c - T_{c0}$ varying with the parameter $n_Ba^3$, where $n_B$ is the density of bosons and $a$ the $s$-wave scattering length, according to a relation of the form:
\begin{equation}
\frac{\Delta T_c }{T_{c0}} \propto (n_Ba^3)^\eta, 
\end{equation}
with an exponent:
\begin{equation}
\eta = {\frac{1}{6}}.
\end{equation}
We then show in Sec. \ref{Sec:Canonical} that the formulation of the BBP theory of Sec. \ref{Sec:Review} is in fact canonical in nature, and not grand-canonical as is commonly thought. We do so  by going back to first principles, and deriving fundamental results of this theory by taking the trace over the orthonormal basis of eigenfunctions of the system at a fixed boson number $N$.  When the theory is reformulated in this fashion, where we avoid the use of the sophisticated but not too transparent apparatus of field-theory, we are led to realize that the results of the standard theory can be recovered by taking a canonical trace, and hence previous formulations of the statistical mechanics of interacting bosons are all essentially canonical.
In Sec. \ref{Sec:GSEnergyFock}, we generalize the variational theory of Ref. \onlinecite{Ettouhami2012} to include Fock interactions between depleted bosons, and we find the zero temperature ground state
and elementary excitations of the system when these Fock interactions between depleted bosons are taken into account. 
Armed with the knowledge of the energy levels of the system, in Sec. \ref{Sec:StatMechCanonicalFormulation} we take canonical traces over these levels to derive the thermodynamics of the interacting Bose gas in the canonical ensemble. In Sec. \ref{Sec:GrandCanonicalFormulation} we show how the statistical mechanics of interacting bosons can be formulated properly within the grand-canonical ensemble by doing the extra step of taking the trace over the toal number of bosons $N$, and in Sec. \ref{Sec:Conclusions} we present our conclusions.

\section{Bogoliubov prescription for an interacting Bose gas} 
\label{Sec:Review}

Let us start our study by reviewing the standard formulation of the statistical mechanics of a condensed Bose gas. The traditional formulation uses imaginary-time equations of motion for Heisenberg field operators to derive a set of Dyson equations for the normal and anomalous Green's functions. \cite{Beliaev1958a,FetterWalecka,Shi1998} Modern formulations use the path-integral formalism to derive the same results, and that is the formalism we shall use in the rest of this Section.

Our starting point will be the imaginary-time grand-canonical partition function of the system, which is given by:
\begin{align}
Z_G = \int [d\Psi^*({\bf r},\tau)][d\Psi({\bf r},\tau)]\, e^{-S[\Psi^*,\Psi]/\hbar}, 
\label{Eq:DefZg}
\end{align}
where $S$ is the imaginary-time action:
\begin{align}
&S = \int_0^{\beta\hbar}d\tau\Big\{\int d{\bf r}\,\Psi^*({\bf r},\tau)\Big(\hbar\partial_\tau
-\frac{\hbar^2\nabla^2}{2m} -\mu \Big)\Psi({\bf r},\tau)
\notag\\
&+\frac{1}{2}\int d{\bf r}d{\bf r}'\,\Psi^*({\bf r},\tau)\Psi^*({\bf r}',\tau)V({\bf r}-{\bf r}')
\Psi({\bf r}',\tau)\Psi({\bf r},\tau)\Big\}.
\end{align}
In the above equation, $\hbar$ is Planck's constant $h$ divided by $2\pi$, $m$ is the mass of the bosons and $\mu$ is the chemical potential. On the other hand, $\beta=1/(k_BT)$, where $k_B$ is Boltzmann's constant and $T$ is the temperature. In this Section, we shall assume for simplicity that the interaction potential $V(\bf r)$ between bosons is repulsive, and that it can be approximated by a short-range, point-contact interaction of the form $V({\bf r}) = g\delta({\bf r})$, where the interaction strength $g$ can expressed in terms of the $s$-wave scattering length $a$ through the relation: 
\begin{equation}
g = \frac{4\pi\hbar^2 a}{m}.
\label{Eq:scatteringLength}
\end{equation}
Below the critical temperature $T_c$ the zero momentum state is macroscopically populated, and the boson field $\Psi({\bf r},\tau)$ is rewritten as the sum: 
\begin{equation}
\Psi({\bf r},\tau) \simeq \sqrt{n_0} + \psi({\bf r},\tau),
\label{Eq:BogPrescription}
\end{equation}
where $n_0$ is the density of the condensate, and where the field $\psi({\bf r},\tau)$ describes excited bosons. If we denote by $\{\varphi_{\bf k}({\bf r})\}$ the set of eigenfunctions of the non-interacting Hamiltonian $\hat{H}_0=-\hbar^2\nabla^2/2m$ such that 
$\varphi_{\bf k}({\bf r})=e^{i{\bf k}\cdot{\bf r}}/\sqrt{V}$ and $\hat{H}_0\varphi_{\bf k}({\bf r})=(\hbar^2k^2/2m)\varphi_{\bf k}({\bf r})$, then $\psi({\bf r},\tau)$ can be written as a Fourier expansion of the form 
\begin{equation}
\psi({\bf r},\tau) =\sum_{\bf k\neq 0}\psi({\bf k},\tau)\varphi_{\bf k}({\bf r}).
\end{equation}
The function $\psi({\bf k},\tau)$ can in turn be expanded in a Matsubara Fourier series, 
\begin{equation}
\psi({\bf k},\tau) = \frac{1}{\sqrt{\beta\hbar}}\sum_{n=-\infty}^\infty \psi({\bf k},\omega_n)e^{-i\omega_n\tau}
\end{equation}
with $\omega_n=2\pi n/\beta\hbar$, so that $\psi({\bf r},\tau)$ becomes:
\begin{equation}
\psi({\bf r},\tau) =\frac{1}{\sqrt{\beta\hbar V}}\sum_{\bf k\neq 0}\sum_{n=-\infty}^\infty
\psi({\bf k},\omega_n)e^{i({\bf k}\cdot{\bf r} - \omega_n\tau)}.
\label{Eq:psiMatsubara}
\end{equation}
Using the prescription (\ref{Eq:BogPrescription}) and the 
decomposition (\ref{Eq:psiMatsubara}) in the expression of the action, and noticing that $\int d{\bf r}\psi({\bf r},\tau) = 0$, we find that $S$ can be written as a sum of a part $S_0$ that is quadratic in $\psi$ and a part $S_1$ that contains cubic and quartic contributions. These are given by:
\begin{subequations}
\begin{align}
S_0 & = \beta\hbar V(-\mu n_0 + \frac{1}{2} g n_0^2) 
\nonumber\\
&+ \sum_{\bf k\neq 0}\sum_{n=-\infty}^\infty\Big\{
\psi^*({\bf k},\omega_n)\big(-i\hbar\omega_n + \varepsilon_{\bf k} - \mu + 2gn_0\big)\psi({\bf k},\omega_n)
\notag\\
& + \frac{gn_0}{2}\big[
\psi({\bf k},\omega_n)\psi(-{\bf k},-\omega_n) +  
\psi^*({\bf k},\omega_n)\psi^*(-{\bf k},-\omega_n)
\big]\Big\},
\label{Eq:S0Fourier}
\\
S_1 & = \frac{g}{2}\int_0^{\beta\hbar}d\tau \int d{\bf r}\Big\{
2\sqrt{n_0}\big[\psi({\bf r},\tau) + \psi^*({\bf r},\tau)\big]|\psi({\bf r},\tau)|^2 
\nonumber\\
&+ |\psi({\bf r},\tau)|^4
\Big\}.
\end{align}
\end{subequations}
At this point, it is convenient to write the quadratic part $S_0$ using matrix notation in the form:
\begin{align}
&S_0 = \beta\hbar V(-\mu n_0 + \frac{1}{2} g n_0^2) 
\notag\\
&+ \sum_{\bf k\neq 0}\sum_{n=-\infty}^\infty \left\{
\Big[\psi({\bf k},\omega_n) \, \psi^*({\bf k},\omega_n)\Big]
G^{-1}({\bf k},\omega_n)
\left[
\begin{array}{c}
\psi({\bf k},\omega_n) \\
\psi^*({\bf k},\omega_n) \\
\end{array}
\right]\right\},
\label{Eq:S0MatrixForm}
\end{align}
where the inverse Green's function matrix $G^{-1}({\bf k},\omega_n)$ is given by:
\begin{align}
G^{-1}({\bf k},\omega_n) = \left(
\begin{array}{cc}
(G^{-1})_{11}  & (G^{-1})_{12} \\
(G^{-1})_{21} & (G^{-1})_{22}\\
\end{array}
\right),
\label{Eq:Ginv_matrix}
\end{align}
with:
\begin{subequations}
\begin{align}
(G^{-1})_{11}  & = -i\hbar\omega_n + \varepsilon_{\bf k} - \mu + 2gn_0,
\\
(G^{-1})_{12} & = (G^{-1})_{21}  = gn_0,
\\
(G^{-1})_{22} & = i\hbar\omega_n + \varepsilon_{\bf k} - \mu + 2gn_0.
\end{align}
\label{Eq:G^-1}
\end{subequations}
Inverting $G^{-1}$ gives the following expression of the Green's function matrix $G({\bf k},\omega_n)$:
\begin{align}
G({\bf k},\omega_n) = \frac{1}{\hbar^2\omega_n^2 + E_{\bf k}^2}\left(
\begin{array}{cc}
(G^{-1})_{22}  & -(G^{-1})_{12} \\
-(G^{-1})_{21} & (G^{-1})_{11}\\
\end{array}
\right),
\label{Eq:res_G}
\end{align}
where we defined the enrgy spectrum $E_{\bf k}$ such that:
\begin{equation}
E_{\bf k} = \sqrt{\big(\varepsilon_{\bf k} + 2gn_0 - \mu \big)^2 - (gn_0)^2}.
\label{Eq:Ekraw}
\end{equation}
Requiring this excitation spectrum to vanish as $k\to 0$ imposes the constraint:
\begin{equation}
\mu = gn_0,
\label{Eq:resultmu}
\end{equation}
upon which the expression of $E_{\bf k}$ becomes:
\begin{equation}
E_{\bf k} = \sqrt{\varepsilon_{\bf k}\big(\varepsilon_{\bf k} + 2gn_0 \big)}.
\label{Eq:Ek}
\end{equation}
Using the expression of the quadratic action $S_0$ from Eq. (\ref{Eq:S0MatrixForm}) and the Green's function matrix $G({\bf k},\omega_n)$ from Eq. (\ref{Eq:res_G}), 
one can derive the following results for the so called normal and anomalous thermal averages, $\langle\psi({\bf k}_1,\omega_n)\psi^*({\bf k}_2,\omega_m)\rangle$
and $\langle\psi({\bf k}_1,\omega_n)\psi({\bf k}_2,\omega_m)\rangle$, respectively:
\begin{subequations}
\begin{align}
\langle\psi({\bf k}_1,\omega_n)\psi^*({\bf k}_2,\omega_m)\rangle = \hbar G_{11}({\bf k},\omega_n)\delta_{\bf k_1, k_2}\delta_{n,m},
\label{Eq:avgNormal}
\\
\langle\psi({\bf k}_1,\omega_n)\psi({\bf k}_2,\omega_m)\rangle = \hbar G_{12}({\bf k},\omega_n)\delta_{\bf k_1, k_2}\delta_{n,m}.
\label{Eq:avgAnomalous}
\end{align}
\label{Eq:avgPsiPsi}
\end{subequations}
It then follows that the average number of bosons $N_{\bf k} = \langle \hat{N}_{\bf k}\rangle$ with momentum ${\bf k}$ is given by:
\begin{subequations}
\begin{align}
N_{\bf k} & = \frac{1}{\beta\hbar}\lim_{\eta \to 0^+}\sum_{n=-\infty}^\infty \hbar G_{11}({\bf k},\omega_n)\, e^{i\omega_n\eta},
\label{Eq:N_k:a}
\\
& =\frac{\varepsilon_{\bf k} + gn_0}{2E_{\bf k}}\coth\Big(\frac{\beta E_{\bf k}}{2}\Big) - \frac{1}{2}.
\label{Eq:N_k:b}
\end{align}
\end{subequations}
The density of depleted bosons $n_d$ is given by:
\begin{equation}
n_d = \frac{1}{V}\sum_{\bf k\neq 0} N_{\bf k},
\end{equation}
and transforming the sum into an integral, we obtain:
\begin{equation}
n_d(T) = \int\frac{d\bf k}{(2\pi)^3}\left[
\frac{\varepsilon_{\bf k} + gn_0}{2E_{\bf k}}\coth\Big(\frac{\beta E_{\bf k}}{2}\Big) - \frac{1}{2}
\right].
\label{Eq:def:n1}
\end{equation}
From the above expession of the density of depleted bosons, one can obtain the density of condensed bosons 
\begin{equation}
n_0(T) = n_B - n_d(T).
\label{Eq:def:n0}
\end{equation} 
With the knowledge of $n_0(T)$, the normal and anomalous Green's functions $G_{11}$ and $G_{12}$ are now fully specified and can be used to derive the thermodynamic properties of the Bose gas.

Now, in previous literature the thermodynamics of condensed Bose systems has been studied using the method summarized above mainly at temperatures well below the transition temperature $T_c$. This is because this formulation predicts a jump discontinuity in the condensate density $n_0(T)$ at $T_c$. In the following Section, we want to examine the origin of this discontinuity, which we will find can be traced back to the Bogoliubov prescription used in this formulation, and more precisely to the improper way this prescription is used at finite temperatures. This will be done next.

\section{Origin of the jump discontinuity in the density of condensed bosons at $T=T_c$} 
\label{Sec:Discontinuity}

As mentioned above, one of the salient features of the standard formulation of the statistical mechanics of interacting bosons is the fact that the density of condensed bosons $n_0(T)$ has a discontinuity\cite{Shi1998} 
at the critical transition temperature $T_c$. We now show that this discontinuity is a direct consequence of the inappropriate use of the Bogoliubov prescription, Eq. (\ref{Eq:BogPrescription}), at finite temperatures. Indeed, when we write the decomposition $\Psi({\bf r},\tau) \simeq \sqrt{n_0} + \psi({\bf r},\tau)$ in the expression of the action $S[\Psi,\Psi^*]$, $\sqrt{n_0}$ is merely the $\Psi({\bf k}=0,\omega_n=0)$ component of $\Psi({\bf r},\tau)$. As such, it is {\em not} the result of a thermal averaging process, and hence it does not and cannot represent the number of condensed bosons at any finite temperature $T$. In fact, the quantity $n_0$ that appears in the decomposition (\ref{Eq:BogPrescription}) can, at the most, be interpreted as the density of condensed bosons at $T=0$. Interpreting $n_0$ as $n_0(T)$ in Eq. (\ref{Eq:BogPrescription}) is not only unjustified, it also has an unphysical and rather annoying consequence in that it causes the density of codensed bosons to have a jump discontinuity at the critical transition temperature $T_c$. In the following, we shall discuss how this phenomenon takes places, and how correctly interpreting $n_0$ in Eq. (\ref{Eq:BogPrescription}) as $n_0(T=0)$ eliminates this jump discontinuity and leads to a continuous variation of $n_0(T)$ across the Bose condensation critical temperature $T_c$.

Let us for definiteness rewrite here the density of depleted bosons at temperature $T$ obtained in the previous Section, Eq. (\ref{Eq:def:n1}) -- (note that we use the shorthand notation $\int_{\bf k}\equiv \int d{\bf k}/(2\pi)^3$):
\begin{equation}
n_d(T) = \int_{\bf k}\left[
\frac{\varepsilon_{\bf k} + gn_0}{2E_{\bf k}}\coth\Big(\frac{\beta E_{\bf k}}{2}\Big) - \frac{1}{2}
\right].
\nonumber
\end{equation}
Close to the transition temperature $T_c$, we shall follow Ref. \onlinecite{Shi1998} and calculate the difference:
\begin{align}
n_d - n_{cr} = \int_{\bf k}\left[
\frac{\varepsilon_{\bf k} + gn_0}{2E_{\bf k}}\coth\Big(\frac{\beta E_{\bf k}}{2}\Big) 
 - \frac{1}{2}\coth\Big(\frac{\beta\varepsilon_{\bf k}}{2}\Big)
\right].
\label{Eq:n1-ncr}
\end{align}
Here $n_{cr}(T)$ is the density of excited particles of an ideal Bose gas at a given temperature $T$, and is given by:
\begin{equation}
n_{cr}(T) = \int_{\bf k}\,\frac{1}{e^{\beta\varepsilon_{\bf k}} - 1} = \zeta(3/2)\left(\frac{mk_BT}{2\pi\hbar^2}\right)^{\frac{3}{2}},
\label{Eq:ncr}
\end{equation}
where $\zeta(x)$ is Riemann's Zeta function, and $\zeta(3/2)\simeq 2.612$.
To calculate the difference in Eq. (\ref{Eq:n1-ncr}), we note that
the main contribution to the integral comes from the region of ${\bf k}$-space where the energy spectrum $E_{\bf k}$ differs significantly from the free-particle energy $\varepsilon_{\bf k}$, which in this case is the region $\varepsilon_{\bf k} \leq gn_0$. Assuming that $gn_0 \ll k_BT_c$, we can approximate $\coth(\beta E_{\bf k}/2)$ by $(2/\beta E_{\bf k})$ and $\coth(\beta\varepsilon_{\bf k}/2)$ by $(2/\beta\varepsilon_{\bf k})$ to obtain:\cite{Shi1998}
\begin{equation}
n_d(T) - n_{cr}(T) = -\frac{k_BT}{8\pi}\left(\frac{2m}{\hbar^2}\right)^{\frac{3}{2}} \sqrt{2gn_0}.
\label{Eq:n1-ncr-2}
\end{equation}
We now need to use the relationship between the total number of bosons $n_B$, the number of condensed bosons $n_0(T)$ and the number of depleted bosons $n_d(T)$,
\begin{equation}
n_B = n_0(T) + n_d(T),
\label{Eq:nBn0n1}
\end{equation}
to eliminate $n_d(T)$ from Eq. (\ref{Eq:n1-ncr-2}) and rewrite it solely in terms of $n_B$ and $n_0(T)$, with the result:
\begin{equation}
n_B = n_0(T) + n_{cr}(T) - \frac{k_BT}{8\pi}\left(\frac{2m}{\hbar^2}\right)^{3/2}\sqrt{2gn_0}.
\label{Eq:res_n(T)}
\end{equation}
In the standard formulation of the statistical mechanics of an interacting Bose gas, the quantity $n_0$ inside the square root, which is the one coming directly from 
the Bogoliubov prescription of Eq. (\ref{Eq:BogPrescription}), is taken to be a function of temperature. In this case, Eq. (\ref{Eq:res_n(T)}) can be seen as a quadratic equation for $\sqrt{n_0(T)}$ which has two solutions:
\begin{equation}
\sqrt{n_0^{(\pm)}(T)} = \frac{1}{2}\left[\sqrt{n_g(T)} \pm \sqrt{n_g(T) + 4(n_B-n_{cr})} \right].
\end{equation}
Here, we followed Ref. \onlinecite{Shi1998} and called $\sqrt{n_g(T)}$ the coefficient of $\sqrt{n_0}$ in Eq. (\ref{Eq:res_n(T)}), namely:
\begin{equation}
\sqrt{n_g(T)} =  \frac{k_BT}{8\pi}\left(\frac{2m}{\hbar^2}\right)^{3/2}\sqrt{2g}.
\end{equation}
Below the transition temperature, $n_{cr} < n_B$, and so the solution $n_0^{(-)}(T)$ in Eq. (\ref{Eq:res_n(T)}) is negative and hence unphysical.
We then can immediately see that the physical solution
\begin{equation}
n_0(T) = \frac{1}{4}\left[\sqrt{n_g(T)} + \sqrt{n_g(T) + 4(n_B-n_{cr})} \right]^2,
\end{equation}
being the square of a strictly positive number, does not vanish for any value of the temperature $T$.
At $T=T_c$, $n_{cr}(T_c) = n_B$ and hence $n_0(T)$ is discontinuous at the transition:\cite{Shi1998}
\begin{subequations}
\begin{align}
n_0(T) & = n_g(T_c), \quad & T \to T_c^-,
\\
n_0(T) & = 0, \quad & T\to T_c^+.
\end{align}
\end{subequations}
At this point, we will note that a similar jump discontinuity of $n_0(T)$ at $T=T_c$ is also found to take place in the formulation of Hartree-Fock theory \cite{Zhang2006}
 that uses Bogoliubov's prescription with a temperature-dependent $n_0$ (more on this below).

Having reviewed how the standard derivation gives rise to a jump discontinutiy in the density of condensed bosons at $T=T_c$,
we now go back to Eq. (\ref{Eq:res_n(T)}) and observe that the $n_0$ inside the square root of this last equation is the same $n_0$ that appears in the expression
of the Green's function $G_{11}({\bf k},\omega_n)$, and, as such, should be independent of temperature (the other $n_0(T)$ on the {\em rhs} of Eq. (\ref{Eq:res_n(T)}) comes from using the relation
between $n_B$, $n_0$ and $n_1$, Eq. (\ref{Eq:nBn0n1}), and has to have a temperature dependence). In other words, 
we will rewrite the Bogoliubov prescription of Eq. (\ref{Eq:BogPrescription}) in the form:
\begin{equation}
\Psi({\bf r},\tau) \simeq \Psi_0 + \psi({\bf r},\tau),
\label{Eq:ModBogPrescription}
\end{equation}
where $\Psi_0$ does {\em not} depend on temperature because it is the result of the ad-hoc prescription (\ref{Eq:ModBogPrescription}) and  {\em not} the result of a thermal averaging process.
Then the action of Eq. (\ref{Eq:S0MatrixForm}) and hence the Green's function matrix of Eq. (\ref{Eq:res_G}) will all be written in terms of the temperature-independent quantity $\Psi_0$ (instead of $\sqrt{n_0}$ which one can easily be misled to interpret as a quantity that varies with temperature), and so would the {\em rhs} of Eq. (\ref{Eq:def:n1}) giving the density of depleted bosons at temperature $T$. 
In other words, Eq. (\ref{Eq:def:n1}) should now read:
\begin{equation}
n_d(T) = \int_{\bf k}\left[
\frac{\varepsilon_{\bf k} + g|\Psi_0|^2}{2E_{\bf k}}\coth\Big(\frac{\beta E_{\bf k}}{2}\Big) - \frac{1}{2}
\right],
\label{Eq:def:n1Psi0}
\end{equation}
with the energy spectrum
\begin{equation}
E_{\bf k} = \sqrt{\varepsilon_{\bf k}(\varepsilon_{\bf k} + 2g|\Psi_0|^2)},
\label{Eq:ModBogSpectrum}
\end{equation}
where $|\Psi_0|^2$ on the {\em rhs} of these last two equations does not depend on temperature.
Comparing the excitation spectrum (\ref{Eq:ModBogSpectrum}) with the standard $T=0$ expression $E_{\bf k}=\sqrt{\varepsilon_{\bf k}(\varepsilon_{\bf k} + 2gn_0(T=0))}$, we see that we can identify $|\Psi_0|^2$ to be the density  of condensed bosons at zero temperature:\cite{Remark}
\begin{equation}
|\Psi_0|^2 = n_0(T=0).
\end{equation}
It then follows that a more appropriate way to write Eq. (\ref{Eq:res_n(T)}) is as follows:
\begin{equation}
n_B = n_0(T) + n_{cr} - \frac{k_BT}{8\pi}\left(\frac{2m}{\hbar^2}\right)^{3/2}\sqrt{2gn_0(0)},
\label{Eq:res_n(T)corr}
\end{equation}
where we now write the density of condensed bosons inside the square root as $n_0(0)$ to emphasize that this is the quantity that is coming directly from the decomposition (\ref{Eq:ModBogPrescription}) in the imaginary-time action, and hence is the $T=0$ value of the density of condensed bosons (below in Sec. \ref{Sec:GSEnergyFock} we will show that this quantity should in fact be $n_B$). Solving for $n_0(T)$, we obtain:
\begin{align}
& n_0(T)  = n_B - n_{cr}(T) + \frac{k_BT}{8\pi}\left(\frac{2m}{\hbar^2}\right)^{\frac{3}{2}}\sqrt{2gn_0(0)},
\notag\\
& = n_B - \zeta\left(\frac{3}{2}\right)\left(\frac{mk_BT}{2\pi\hbar^2}\right)^{\frac{3}{2}} + \frac{k_BT}{8\pi}\left(\frac{2m}{\hbar^2}\right)^{\frac{3}{2}} \sqrt{2gn_B},
\end{align}
where in going from the first line to the second one we used the expression (\ref{Eq:ncr}) of $n_{cr}(T)$, and we approximated $n_0(0)$ inside the square root with the total density of bosons $n_B$.
Dividing both sides of the above equation by $n_B$, we can write:
\begin{equation}
\frac{n_0(T)}{n_B} = 1 -\frac{ \zeta\left(\frac{3}{2}\right)}{n_B}\left(\frac{mk_BT}{2\pi\hbar^2}\right)^{\frac{3}{2}} + \frac{k_BT}{8\pi}\left(\frac{2m}{\hbar^2}\right)^{\frac{3}{2}} \sqrt{\frac{2g}{n_B}}.
\label{Eq:n0/nB}
\end{equation}
At $T=T_c$, we expect $n_0(T_c)=0$. Assuming this to be true, we can write:
\begin{equation}
1 -\frac{ \zeta\left(\frac{3}{2}\right)}{n_B}\left(\frac{mk_BT_c}{2\pi\hbar^2}\right)^{\frac{3}{2}} + \frac{k_BT_c}{8\pi}\left(\frac{2m}{\hbar^2}\right)^{\frac{3}{2}} \sqrt{\frac{2g}{n_B}} = 0.
\label{Eq:Tc-gn}
\end{equation}
The critical temperature for a free Bose gas $T_{c0}$ can be obtained by letting $g=0$ in the above equation, and satisfies the following relation:
\begin{equation}
\frac{ \zeta\left(\frac{3}{2}\right)}{n_B}\left(\frac{mk_BT_{c0}}{2\pi\hbar^2}\right)^{\frac{3}{2}} = 1,
\label{Eq:Tc0}
\end{equation}
which gives the well-known result:
\begin{equation}
T_{c0} = \frac{2\pi\hbar^2}{mk_B}\Big[\frac{n_B}{\zeta(3/2)}\Big]^{2/3}.
\label{Eq:def:Tc0}
\end{equation}
Going back to the condensate fraction of Eq. (\ref{Eq:n0/nB}), if we scale the temperature $T$ by $T_{c0}$ we obtain after a few manipulations:
\begin{align}
\frac{n_0(T)}{n_B} = 1 - \left(\frac{T}{T_{c0}}\right)^\frac{3}{2} + \frac{\sqrt{4\pi}(n_Ba^3)^\frac{1}{6}}{\big[\zeta(3/2)\big]^\frac{2}{3}}\left(\frac{T}{T_{c0}}\right).
\end{align}
At $T=T_c$, $n_0(T) = 0$, and so:
\begin{align}
1 - \left(\frac{T_c}{T_{c0}}\right)^\frac{3}{2} + \frac{\sqrt{4\pi}(n_Ba^3)^\frac{1}{6}}{\big[\zeta(3/2)\big]^\frac{2}{3}}\left(\frac{T_c}{T_{c0}}\right) = 0.
\label{Eq:Tc-gn2}
\end{align}
Now, let $\Delta T_c = T_c - T_{c0}$, so that $T_c = T_{c0}(1+\Delta T_c/T_{c0})$. Then, if we assume that the difference between critical temperatures is small, {\em i.e.} $\Delta T_c \ll T_{c0}$, then we can write:
\begin{equation}
 \left(\frac{T_c}{T_{c0}}\right)^\frac{3}{2} \simeq 1 + \frac{3}{2}\frac{\Delta T_c}{T_{c0}},
\end{equation}
and Eq. (\ref{Eq:Tc-gn2}) has the approximate solution:
\begin{align}
\frac{\Delta T_c}{T_{c0}} = \frac{ c (n_Ba^3)^\frac{1}{6} }{1 - c (n_Ba^3)^\frac{1}{6}},
\label{Eq:interm3}
\end{align}
where we call $c$ the numerical constant:
\begin{equation}
c =\frac{4}{3} \frac{\sqrt{\pi}}{\big[\zeta(3/2)\big]^\frac{2}{3}}\simeq 0.219874.
\end{equation}

We now want to check this result numerically. Going back to Eq. (\ref{Eq:def:n1}), we see that we can write for the condensate fraction $n_0(T)/n_B = 1 - n_d(T)/n_B$ the following expression:
\begin{widetext}
\begin{align}
\frac{n_0(T)}{n_B} = 1 - \frac{2^\frac{5}{2}}{\sqrt{\pi}}
(n_Ba^3)^\frac{1}{2}
\int_0^\infty d\tilde{k}\,\tilde{k}^2\left[
\frac{\tilde{k}^2 +1}{\sqrt{\tilde{k}^2(\tilde{k}^2+2)}}
\coth\Big(\frac{[\zeta(3/2)]^{2/3}(n_Ba^3)^{1/3}}{T/T_{c0}}\sqrt{\tilde{k}^2(\tilde{k}^2+2)}\Big) - 1
\right],
\label{Eq:condfracBog}
\end{align}
\end{widetext}
where we introduced the dimensionless wave-vector $\tilde{\bf k}$ such that:
\begin{equation}
\tilde{\bf k} = \frac{\bf k}{k_0},
\label{Eq:def:tildek}
\end{equation}
with 
\begin{equation}
k_0=\frac{\sqrt{2mgn_B}}{\hbar} = (8\pi n_B a)^{\frac{1}{2}}.
\label{Eq:def:k0}
\end{equation}

\begin{figure}[tb]
\includegraphics[width=8.09cm, height=5.5cm]{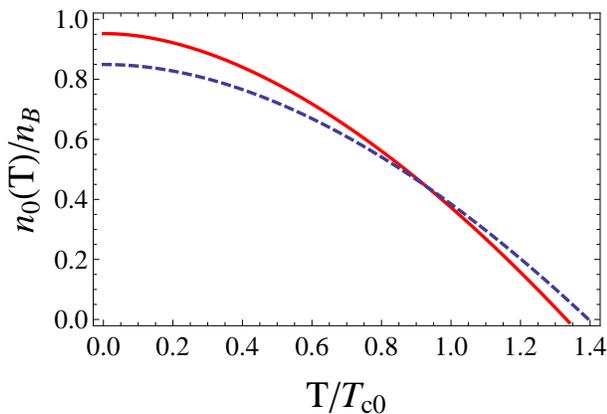}
\caption[]{(Color online) Plot of the condensate fraction $n_0(T)/n_B$ as a function of the reduced temperature $T/T_{c0}$ in the naive Bogoliubov theory for $n_Ba^3 = 0.001$ (solid line) and $n_Ba^3=0.01$ (dashed line).
}\label{Fig:plotn0Bog}
\end{figure}

In Fig. \ref{Fig:plotn0Bog}, we plot the condensate fraction as obtained from the above equation for two different values of the parameter $n_Ba^3$. We find that for $n_Ba^3=0.001$, the condensate fraction vanishes at $T_c/T_{c0}=1.33$, and for $n_Ba^3=0.01$, the condensate fraction vanishes at $T_c/T_{c0}=1.39$. 

\begin{figure}[tb]
\includegraphics[width=8.09cm, height=5.5cm]{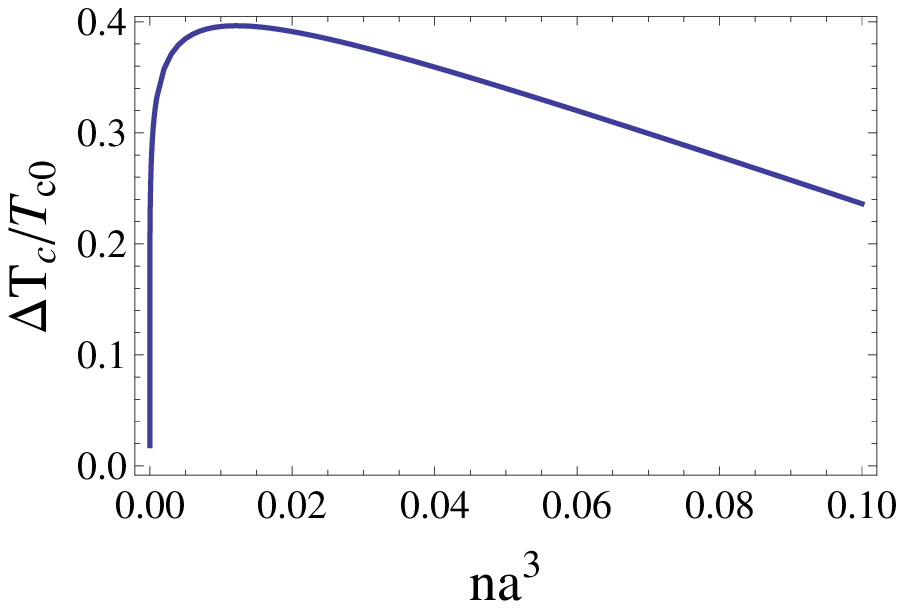}
\includegraphics[width=8.09cm, height=5.5cm]{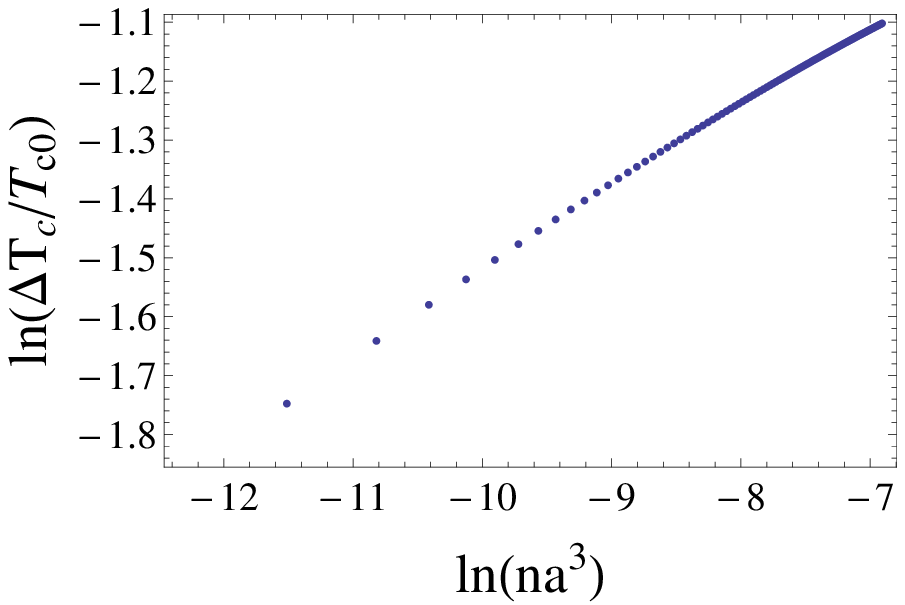}
\caption[]{Upper panel: Plot of $\Delta T_c/T_{c0}$ vs. $n_Ba^3$ as obtained from the numerical solution of the equation $n_0(T)/n_B=0$ using Eq. (\ref{Eq:condfracBog}). Lower panel: Plot of $\ln(\Delta T_c/T_{c0})$ vs. $\ln(n_Ba^3)$
as obtained using the data in the upper panel for values of $na^3\leq 0.001$. The plot shows a linear dependence with a best fit slope $\approx 0.158$, in pretty good agreement with the slope of $1/6=0.166$ predicted by Eq. (\ref{Eq:interm3}).}\label{Fig:Tcna3}
\end{figure}

The expression above for the condensate fraction, Eq.  (\ref{Eq:condfracBog}), can be used to study the dependence of $T_c$ on $n_Ba^3$ numerically by looking for the temperature $T$ that solves the equation $n_0(T)/n_B=0$.
The upper panel of Fig. \ref{Fig:Tcna3} shows a plot of $\Delta T_c/T_{c0}$ vs. $n_Ba^3$ found using this method, where the infinite slope near the origin indicates a power-law relationship with an exponent that is less than unity. In the lower panel, we plot $\ln(\Delta T_c/T_{c0})$ vs. $\ln(n_Ba^3)$, which shows that there is a linear relationship between these two logarithms with a slope of $0.158\simeq 1/6$, in perfect agreement with the analytical result of Eq. (\ref{Eq:interm3}).

It is worth taking a pause at this juncture to note the vast body of literature which treated the problem of the shift in the transition temperature due to repulsive interactions between bosons. Beginning with the early work of Lee and Yang, \cite{Lee1957b} various methods of calculations over the years have produced widely dissimilar results, 
\cite{Glassgold1960,Luban1962,FetterWalecka,Kanno1969a,Kanno1969b,Toyoda1982,Stoof1992,Bijlsma1996,Huang1999,Mueller2001,Gruter1997,Holzmann1999a,Holzmann1999b,Baym1999,Baym2000,Wilkens2000,Arnold2000,deSouza2001,Kashurnikov2001,Arnold2001,Holzmann2001,Arnold2002,Baym2001,deSouza2002,Kneur2002,Braaten2002,Kastening2003}
with positive and negative shifts in $T_c$ that were predicted to be proportional to $a$, 
$\sqrt{a}$, $a\ln(a)$, etc. It is not our intention here to give support to one of these results or another, our main focus in this paper not being to give an accurate or definitive calculation of the shift in transition temperature, but merely to correct a misconception about the standard Beliaev BBP predicting a jump discontinuity of $n_0(T)$ at the transition. It is interesting to note that the positive shift we obtain
\begin{equation}
\frac{\Delta T_c}{T_{c0}} \propto (n_B a^3)^\eta
\end{equation}
with the exponent
\begin{equation}
\eta = \frac{1}{6}
\end{equation}
has been predicted previously, \cite{Lee1957b,Glassgold1960,Huang1999} although the author is not aware of this result having been obtained within an approach similar to the one used in the current study.

Now that we got rid of the jump discontinuity in the condensate fraction at the critical temperature $T_c$ predicted by the standard, field-theoretic formulation of the statistical mechanics of interacting bosons, 
one may ask how we should interpret the quantity $n_B$ that appears in Eq. (\ref{Eq:condfracBog}). Indeed, in the grand-canonical ensemble, where the system is in contact with a particle reservoir, the density of bosons $n_B$ should depend on temperature. Hence, the question arises as to whether we should interpret $n_B$ in Eq. (\ref{Eq:condfracBog}) to be the total density of bosons {\it (i)} at zero temperature, {\it (ii)} at the critical temperature $T_c$, or {\it (iii)} at some intermediate temperature between $T=0$ and $T=T_c$. The fact is, the plots in Figs. \ref{Fig:plotn0Bog} and \ref{Fig:Tcna3} were drawn with the assumption that $n_B$ does not depend on temperature, and even if we are willing to assume a temperature dependence for $n_B$, there is simply not enough information in the BBP formulation to find what this temperature dependence exactly is. 


In the following, we want to examine the physical content of the BBP formulation by trying to derive some of its major results using first-principles, {\em i.e.} using traces over eigenstates of the Bogoliubov Hamiltonian. In doing so, we will establish that what passes as a grand-canonical treatment of the interacting Bose gas is in fact a canonical treatment, not a grand-canonical one as is commonly thought, and hence that we do not have to worry about the temperature dependence of $n_B$ in Eq. (\ref{Eq:condfracBog}) because this quantity, in the canonical ensemble, is fixed and does not in any way vary with temperature.


\section{Statistical mechanics of interacting bosons: canonical formulation}
\label{Sec:Canonical}

\subsection{Back to basics: thermal averages as traces over eigenvectors of the Bogoliubov Hamiltonian} 
\label{Sub:Canonical}

In this Section, we want to derive fundamental building blocks of the BBP theory, namely the expressions of the normal and anomalous thermal averages $\langle a_{\bf k}^\dagger a_{\bf k}\rangle$ 
and $\langle a_{\bf k} a_{-\bf k}\rangle$ respectively, using a first-principles type of approach.
Our starting point will be the canonical partition function of the Bogoliubov model, which for a system of $N$ bosons is given by (note that we are purposely adding a subscript $N$ to the symbol $\mbox{Tr}$ to indicate that we are taking a canonical trace with a fixed number of bosons $N$):
\begin{align}
Z_C(N,V,T) = \mbox{Tr}_N\Big( e^{-\beta \hat{H}_B}\Big).
\label{Eq:Z:H_B}
\end{align}
In the above equation, $\hat{H}_B$ is the truncated part of the total Hamiltonian $\hat{H}$ that can be diagonalized using 
Bogoliubov's canonical transformations, and is given in Fourier space by the following expression:
\begin{align}
\hat{H}_B & = \sum_{\bf k\neq 0}\Big[
\varepsilon_{\bf k} a_{\bf k}^\dagger a_{\bf k} 
+ \frac{v({\bf k})}{2V}\big ( a_0^\dagger a_0 a_{\bf k}^\dagger a_{\bf k} 
+ a_0^\dagger a_0 a_{-\bf k}^\dagger a_{-\bf k}
\nonumber\\
&+ a_{\bf k}^\dagger a_{-\bf k}^\dagger a_0 a_0 + a_{\bf k} a_{-\bf k} a_0^\dagger a_0^\dagger\big)\Big].
\label{Eq:H_B:standard}
\end{align}
After the above Hamiltonian is diagonalized, it can be written in the form:
\begin{align}
\hat{H}_B = E_0 + \sum_{\bf k\neq 0}E_{\bf k}
\alpha_{\bf k}^\dagger\alpha_{\bf k},
\label{Eq:H_B}
\end{align}
where $E_0$ is the ground state energy of the system, and $E_{\bf k}$ is the energy of an excitation of wavevector ${\bf k}$. In the standard Bogoliubov formulation, these two quantities are given by:
\begin{subequations}
\begin{align}
E_0 & = - \frac{1}{2}\sum_{\bf k\neq 0} \Big[ \varepsilon_{\bf k} + n_B v({\bf k}) - E_{\bf k}\Big] < 0 ,
\\
E_{\bf k} & = \sqrt{\varepsilon_{\bf k}\big(\varepsilon_{\bf k} + 2 n_B v({\bf k})\big)}.
\end{align}
\end{subequations}
On the other hand, $\alpha_{\bf k}^\dagger$ and $\alpha_{\bf k}$ are the creation and annihilation operators of an elementary excitation of momentum ${\bf k}$:
\begin{subequations}
\begin{align}
\alpha_{\bf k}^\dagger = u_{\bf k}a_{\bf k}^\dagger + v_{\bf k} a_{-\bf k},
\\
\alpha_{\bf k} = u_{\bf k}a_{\bf k} + v_{\bf k} a_{-\bf k}^\dagger,
\end{align}
\label{Eq:alphas}
\end{subequations}
where the so-called coherence factors $u_{\bf k}$ and $v_{\bf k}$ are given by:
\begin{subequations}
\begin{align}
u_{\bf k}^2 &= \frac{1}{2}\Big(\frac{\varepsilon_{\bf k} + n_Bv({\bf k})}{E_{\bf k}}+1\Big), \label{Eq:def:u}
\\
v_{\bf k}^2 &= \frac{1}{2}\Big(\frac{\varepsilon_{\bf k} + n_Bv({\bf k})}{E_{\bf k}} - 1\Big). \label{Eq:def:v}
\end{align}
\label{Eq:coh-factors}
\end{subequations}

We shall denote by $|\Psi_B(N)\rangle$ the ground state of Bogoliubov's Hamiltonian $\hat{H}_B$, such that $\alpha_{\bf k}|\Psi_B(N)\rangle = 0$ for all wavevectors ${\bf k}\neq 0$. It then follows that: 
\begin{equation}
\hat{H}_B |\Psi_B(N)\rangle = E_0|\Psi_B(N)\rangle,
\end{equation}
confirming our statement above that $E_0$ in Eq. (\ref{Eq:H_B}) is the ground state energy of $\hat{H}_B$.
On the other hand, we shall denote by $|\Psi_{\{m_i\}}(N)\rangle = |m_1,m_2,\ldots, m_M\rangle$ the eigenfunction of $\hat{H}_B$ corresponding to $m_i$ excitations of wavevector ${\bf k}_i$, $i = 1,\ldots, M$ ($M$ here is the number of momentum modes kept in the Hilbert space, and will eventually be sent to infinity).
These eigenfunctions can be written in the form:
\begin{align}
|\Psi_{\{m_i\}}(N)\rangle = \frac{\big(\alpha_{{\bf k}_1}^\dagger\big)^{m_1}}{\sqrt{m_1!}}\times\cdots\times
\frac{\big(\alpha_{{\bf k}_M}^\dagger\big)^{m_M}}{\sqrt{m_M!}}|\Psi_B(N)\rangle,
\label{Eq:def:Psi_m}
\end{align}
and one can easily verify by direct calculation that $|\Psi_{\{m_i\}}(N)\rangle$ satisfies the following equation:
\begin{align}
\hat{H}_B|\Psi_{\{m_i\}}(N)\rangle &= \Big(E_0 + \sum_{i\neq 0} m_i E_{{\bf k}_i}\Big) |\Psi_{\{m_i\}}(N)\rangle.
\label{Eq:HBPsim}
\end{align}
We now go back to Eq. (\ref{Eq:Z:H_B}) and rewrite the canonical partition function $Z_C(N, V, T)$ in the form:
\begin{align}
Z_C = \sum_{\{ m_i \}}
\langle m_1,m_2,\ldots, m_M| e^{-\beta\hat{H}_B}|m_1,m_2,\ldots, m_M\rangle.
\end{align}
In the above expression, the summation extends over all values of $m_i$ that are smaller than the dimension of the matrix representation of the Hamiltonian $\hat{H}_B$. 
In the thermodynamic limit $N\to \infty$ and $M\gg 1$, the summations over the $m_i$'s can be extended all the way to infinity, with the result:
\begin{subequations}
\begin{align}
{Z_C} &=e^{-\beta E_0} \sum_{m_1=0}^\infty\cdots  \sum_{m_M=0}^\infty\prod_{i=1}^M
\Big(e^{-\beta E_{{\bf k}_i}}\Big)^{m_i},
\\
& = e^{-\beta E_0} 
\prod_{i=1}^M\left( \sum_{m_i=0}^\infty \Big( e^{-\beta E_{{\bf k}_i}} \Big)^{m_i}\right),
\\
& = e^{-\beta E_0} 
\prod_{i=1}^M\frac{1}{1 - e^{-\beta E_{{\bf k}_i} }},
\label{Eq:resZN}
\end{align}
\end{subequations}
where, in going from the second to the third line use has been made of the geometric summation formula:
\begin{equation}
\sum_{m_i=0}^\infty \Big(e^{-\beta E_{{\bf k}_i}}\Big)^{m_i} = \frac{1}{1 - e^{-\beta E_{{\bf k}_i}}}.
\label{Eq:geom:sum}
\end{equation}
Now, the thermal average of the operator $\hat{N}_{{\bf k}_i}=a_{{\bf k}_i}^\dagger a_{{\bf k}_i}$ in the canonical ensemble
is given by:
\begin{equation}
\big\langle a_{{\bf k}_i}^\dagger a_{{\bf k}_i}\big\rangle = \frac{1}{{Z_C}}\mbox{Tr}_N\Big(
a_{{\bf k}_i}^\dagger a_{{\bf k}_i} e^{-\beta\hat{H}}
\Big).
\label{Eq:def:canonicalAvg}
\end{equation}
Although it is possible to conduct the calculation of the above thermal average
in the number-conserving formalism of Ref. \onlinecite{Ettouhami2012}, we here for simplicity shall use the number non-conserving formalism in which the relation between the operators $a_{{\bf k}_i}$ and $\{\alpha_{{\bf k}_i}, \alpha_{{\bf k}_i}^\dagger\}$ is linear, and can be obtained by inverting Eqs. (\ref{Eq:alphas}), with the result:
\begin{subequations}
\begin{align}
a_{{\bf k}_i} & = u_{{\bf k}_i}\alpha_{{\bf k}_i} - v_{{\bf k}_i}\alpha_{{-\bf k}_i}^\dagger,
\\
a_{{\bf k}_i}^\dagger & = u_{{\bf k}_i}\alpha_{{\bf k}_i}^\dagger - v_{{\bf k}_i}\alpha_{{-\bf k}_i}.
\end{align}
\label{Eq:a:vs:alpha}
\end{subequations}
The operator $\hat{N}_{\bf k}=a_{\bf k}^\dagger a_{\bf k}$ is therefore given by:
\begin{align}
\hat{N}_{\bf k} = u_{\bf k}^2 \alpha_{\bf k}^\dagger \alpha_{\bf k} + v_{\bf k}^2 \alpha_{-\bf k}\alpha_{-\bf k}^\dagger
- u_{\bf k}v_{\bf k}(\alpha_{\bf k}^\dagger \alpha_{-\bf k}^\dagger + \alpha_{-\bf k}\alpha_{\bf k}).
\label{Eq:Nkalphak}
\end{align}
Let us find the trace of ${\hat N}_{\bf k}\exp(-\beta \hat{H}_B)$  in the base formed by the $|\Psi_{\{ m_i \}}\rangle$. In order to do that, we shall first use Eq. (\ref{Eq:HBPsim})
to calculate the action of this operator on the ket $|\Psi_{\{ m_i \}}\rangle$, with the result:
\begin{align}
\hat{N}_{{\bf k}_j}e^{-\beta\hat{H}_B}|\Psi_{\{ m_i \}}\rangle & =  e^{ -\beta E_0}
\prod_{i=1}^M\Big(e^{-\beta E_{{\bf k}_i}}\Big)^{m_i}
N_{{\bf k}_j} |\Psi_{\{ m_i \}}\rangle.
\end{align}
Hence:
\begin{align}
\langle \Psi_{\{ m_i \}}|&\hat{N}_{{\bf k}_j}e^{-\beta\hat{H}_B}|\Psi_{\{ m_i \}}\rangle  =  e^{-\beta E_0}
\prod_{i=1}^M\Big(e^{-\beta E_{{\bf k}_i}}\Big)^{m_i}
\notag\\
&\times\langle \Psi_{\{ m_i \}}| \hat{N}_{{\bf k}_j} |\Psi_{\{ m_i \}}\rangle.
\label{Eq:NkExpavg}
\end{align}
Using the expression of $\hat{N}_{{\bf k}_j}$ in terms of the $\alpha_{\bf k}$'s, Eq. (\ref{Eq:Nkalphak}), in the above expression, we obtain:
\begin{align}
\langle \Psi_{\{ m_i \}}|\hat{N}_{{\bf k}_j}|\Psi_{\{ m_i \}}\rangle  & =
\langle \Psi_{\{ m_i \}}| (u_{{\bf k}_j}^2\alpha_{{\bf k}_j}^\dagger \alpha_{{\bf k}_j}
\notag\\
&+ v_{{\bf k}_j}^2 \alpha_{-{\bf k}_j}\alpha_{-{\bf k}_j}^\dagger) |\Psi_{\{ m_i \}}\rangle. 
\label{Eq:avgNk_0}
\end{align}
We now use the fact that $\alpha_{{\bf k}_j}^\dagger\alpha_{{\bf k}_j}$ simply counts the number of excitations of wavevector ${\bf k}_j$ in the excited state $|\Psi_{\{ m_i \}}(N)\rangle$, which is nothing but $m_j$, to write:
\begin{equation}
\alpha_{{\bf k}_j}^\dagger \alpha_{{\bf k}_j}|\Psi_{\{ m_i \}}(N)\rangle = m_j |\Psi_{\{ m_i \}}(N)\rangle.
\label{Eq:alphasPsim1}
\end{equation} 
Also, from the canonical commutation relations of the $\alpha_{\bf k}$'s, we see that $\alpha_{-{\bf k}_j}\alpha_{-{\bf k}_j}^\dagger = 1 + \alpha_{-{\bf k}_j}^\dagger\alpha_{-{\bf k}_j}$, and hence:
\begin{equation}
\alpha_{-{\bf k}_j} \alpha_{-{\bf k}_j}^\dagger|\Psi_{\{ m_i \}}(N)\rangle = ( 1+m_{-j}) |\Psi_{\{ m_i \}}(N)\rangle,
\label{Eq:alphasPsim2}
\end{equation} 
where we denote by $m_{-j}$ the number of excitations of wavevector $-{\bf k}_j$. Now, if we use Eqs. (\ref{Eq:alphasPsim1}) and (\ref{Eq:alphasPsim2}) in Eq. (\ref{Eq:avgNk_0}), this last equation becomes:
\begin{equation}
\langle \Psi_{\{ m_i \}}| \hat{N}_{{\bf k}_j}|\Psi_{\{ m_i \}}\rangle  = m_j u_{{\bf k}_j}^2 + (1+m_{-j})v_{{\bf k}_j}^2.
\end{equation}
Replacing this result back into Eq. (\ref{Eq:NkExpavg}) and taking the trace, we obtain: 
\begin{widetext}
\begin{align}
\langle \Psi_{\{ m_i \}}| \hat{N}_{{\bf k}_j}e^{-\beta\hat{H}_B}|\Psi_{\{ m_i \}}\rangle  & =  e^{-\beta E_0}\Big\{
u_{{\bf k}_j}^2\sum_{m_j=0}^\infty m_j \Big(e^{-\beta E_{{\bf k}_j}}\Big)^{m_j}
\times
\prod_{i =1(\neq j)}^M\sum_{m_i=0}^\infty \Big(e^{-\beta E_{{\bf k}_i}}\Big)^{m_i}
\notag\\
&+v_{{\bf k}_j}^2\sum_{m_{-j}=0}^\infty(1+m_{-j})\Big( e^{-\beta E_{-{\bf k}_j}} \Big)^{m_{-j}}
\times\prod_{i=1(\neq -j)}^M\sum_{m_i=0}^\infty\Big(e^{-\beta E_{{\bf k}_i}}\Big)^{m_i}
\Big\}.
\label{Eq:NkExpTrace}
\end{align}
\end{widetext}
Dividing by ${Z_C}$, as in Eq. (\ref{Eq:def:canonicalAvg}), we obtain that the thermal average of $\hat{N}_{\bf k}$ in the canonical ensemble is given by:
\begin{align}
\langle \hat{N}_{{\bf k}_j}\rangle & = u_{{\bf k}_j}^2\frac{\sum_{m_j=0}^\infty m_j\big(e^{-\beta E_{{\bf k}_j}}\big)^{m_j}}{\sum_{m_j=0}^\infty \big(e^{-\beta E_{{\bf k}_j}}\big)^{m_j}}
\notag\\
& + v_{{\bf k}_j}^2\frac{\sum_{m_{-j}=0}^\infty (1+m_{-j})\big(e^{-\beta E_{-{\bf k}_j}}\big)^{m_{-j}}}{\sum_{m_{-j}=0}^\infty \big(e^{-\beta E_{-{\bf k}_j}}\big)^{m_{-j}}}.
\end{align}
Recalling that $E_{-{\bf k}_j}=E_{{\bf k}_j}$, this last equation becomes:
\begin{align}
\langle \hat{N}_{{\bf k}_j}\rangle = v_{{\bf k}_j}^2 + (u_{{\bf k}_j}^2 +  v_{{\bf k}_j}^2)
\frac{\sum_{m_j=0}^\infty m_j\big(e^{-\beta E_{{\bf k}_j}}\big)^{m_j}}{\sum_{m_j=0}^\infty \big(e^{-\beta E_{{\bf k}_j}}\big)^{m_j}}.
\end{align}
Using Eq. (\ref{Eq:geom:sum}) and the fact that
\begin{equation}
\sum_{n=0}^\infty n x^n = \frac{x}{(1-x)^2},
\label{Eq:geom:2}
\end{equation}
we finally obtain:
\begin{equation}
\langle \hat{N}_{{\bf k}_j}\rangle = v_{{\bf k}_j}^2 + \frac{u_{{\bf k}_j}^2 + v_{{\bf k}_j}^2}
{e^{\beta E_{{\bf k}_j}}-1}.
\label{Eq:avgNk}
\end{equation}
Now, using the expression of the coherence factors $u_{{\bf k}_i}^2$ and $v_{{\bf k}_i}^2$ from Eq. (\ref{Eq:coh-factors}), we can rewrite $\langle \hat{N}_{{\bf k}_j}\rangle$
in the form:
\begin{align}
\langle \hat{N}_{{\bf k}_j}\rangle = \frac{\varepsilon_{{\bf k}_j} + n_Bv({\bf k}_j)}{2E_{{\bf k}_j}} \coth\Big(\frac{\beta E_{{\bf k}_j}}{2}\Big)- \frac{1}{2}.
\label{Eq:avgNkn_B}
\end{align}
This is the exact same expression that was obtained previously from the field-theortic formulation of the statistical mechanics of an interacting Bose gas, see Eq. (\ref{Eq:N_k:b}). The fact that we obtained this expression within a strictly {\em canonical} formalism, where we traced over the basis of eigenstates $|\Psi_{\{m_i\}}(N)\rangle$ corresponding to a fixed number of bosons $N$, is quite remarkable. In fact, and as we already mentioned in the beginning of this section, a quite conspicuous clue that the usual treatment is canonical  is the presence of the density of bosons $n_B$ on the {\em rhs} of Eq. (\ref{Eq:avgNkn_B}), for in a correctly formulated grand-canonical treatment the number of bosons $N$ would be traced over, and the {\em rhs} of this last equation would no longer contain the density of bosons $n_B=N/V$. (Strictly speaking, in the standard Bogoliubov derivation $n_0$ appears on the {\em rhs} of Eq. (\ref{Eq:avgNkn_B}) rather than $n_B$, and the same argument can be made that in a grand-canonical formulation $n_0$ should be traced over and hence should not appear on the {\em rhs} of this last equation.)


We now want to calculate the anomalous thermal average $\langle a_{{\bf k}_j} a_{-{\bf k}_j}\rangle$. Following the analysis we made in Ref. \onlinecite{Ettouhami2012} of the meaning of this anomalous average at zero temperature, where we showed that it can be interpreted as the expectation value $\langle\Psi(N-2)|a_{{\bf k}_j}a_{-{\bf k}_j}|\Psi(N)\rangle$, $|\Psi(N)\rangle$ and $|\Psi(N-2)\rangle$ being the Bogoliubov ground states of a system of $N$ and $(N-2)$ bosons respectively, we here will define the canonical trace in the following way:
\begin{subequations}
\begin{align}
& \langle a_{{\bf k}_j} a_{-{\bf k}_j}\rangle  = \frac{1}{Z_C}\,\mbox{Tr}\left( a_{{\bf k}_j} a_{-{\bf k}_j} e^{-\beta\hat{H}_B}\right),
\\
& = \frac{1}{Z_C}\sum_{\{m_i\}} \langle\Psi_{\{m_i\}}(N-2) | a_{{\bf k}_j}a_{-{\bf k}_j} e^{-\beta\hat{H}_B}|\Psi_{\{m_i\}}(N)\rangle.
\end{align}
\end{subequations}
Using Eq. (\ref{Eq:HBPsim}), we can write:
\begin{align}
\langle a_{{\bf k}_j} a_{-{\bf k}_j}\rangle & =  e^{-\beta E_0}\frac{1}{Z_C}\,\sum_{m_1=0}^\infty\cdots\sum_{m_M=0}^\infty \,\prod_{i=1}^M \left( e^{-\beta E_{{\bf k}_i}} \right)^{m_i}
\nonumber\\
&\times \langle\Psi_{\{m_i\}}(N-2) | a_{{\bf k}_j}a_{-{\bf k}_j}|\Psi_{\{m_i\}}(N)\rangle.
\label{Eq:can:akak}
\end{align}
We now use Eq. (\ref{Eq:a:vs:alpha}) to express $a_{\bf k}a_{-\bf k}$ in terms of the $\alpha_{\bf k}$'s, with the result:
\begin{align}
a_{{\bf k}_j} a_{-{\bf k}_j} & = u_{{\bf k}_j}^2\alpha_{{\bf k}_j}\alpha_{-{\bf k}_j} + v_{{\bf k}_j}^2\alpha_{{\bf k}_j}^\dagger \alpha_{-{\bf k}_j}^\dagger 
- u_{{\bf k}_j} v_{{\bf k}_j}\alpha_{{\bf k}_j}\alpha_{{\bf k}_j}^\dagger
\nonumber\\
& - u_{{\bf k}_j} v_{{\bf k}_j}\alpha_{-{\bf k}_j}^\dagger\alpha_{-{\bf k}_j}.
\end{align}
Taking the quantum expectation value, we obtain:
\begin{subequations}
\begin{align}
\langle\Psi_{\{m_i\}} | a_{{\bf k}_j}a_{-{\bf k}_j}|\Psi_{\{m_i\}}\rangle & = - u_{{\bf k}_j}v_{{\bf k}_j}\langle\Psi_{\{m_i\}} | (1+ \alpha_{{\bf k}_j}^\dagger \alpha_{{\bf k}_j}
\nonumber\\
&+ \alpha_{-{\bf k}_j}^\dagger \alpha_{-{\bf k}_j} )|\Psi_{\{m_i\}}\rangle,
\\
& = - u_{{\bf k}_j} v_{{\bf k}_j} (1 + m_j + m_{-j}).
\end{align}
\end{subequations}
Replacing this result in Eq. (\ref{Eq:can:akak}), and using expression (\ref{Eq:resZN}) of the canonical partition function $Z_C$, we can write:
\begin{align}
\langle a_{{\bf k}_j} a_{-{\bf k}_j}\rangle & = -u_{{\bf k}_j} v_{{\bf k}_j}\Bigg[
1 + \frac{ \sum_{m_j=0}^\infty m_j   \Big(e^{-\beta E_{{\bf k}_j} }\Big)^{m_j}}{\sum_{m_j=0}^\infty  \Big(e^{-\beta E_{{\bf k}_j} }\Big)^{m_j}}
\nonumber\\
& + \frac{ \sum_{m_{-j}=0}^\infty m_{-j} \Big(e^{-\beta E_{{-\bf k}_j} }\Big)^{m_{-j}}}{\sum_{m_{-j}=0}^\infty  \Big(e^{-\beta E_{{-\bf k}_j} }\Big)^{m_{-j}}}
\Bigg].
\end{align}
Performing the summations with the help of Eqs. (\ref{Eq:geom:sum}) and (\ref{Eq:geom:2}), we finally obtain:
\begin{subequations}
\begin{align}
\langle a_{{\bf k}_j} a_{-{\bf k}_j}\rangle & = - u_{{\bf k}_j} v_{{\bf k}_j}\Big[
1 + \frac{2}{e^{\beta E_{{\bf k}_j} }-1}
\Big], 
\\
& = - \frac{n_B v({\bf k}_j)}{2E_{{\bf k}_j}}\Big[
1 + \frac{2}{e^{\beta E_{{\bf k}_j} }-1}
\Big].
\end{align}
\end{subequations}
This is again the exact same expression as the one obtained from the standard Bogoliubov theory,\cite{Shi1998,FetterWalecka} except that the total density of bosons $n_B$ appears on the {\em rhs} instead of $n_0$. We emphasize that this is so because we used $n_B$ instead of $n_0$ in the expressions of the coherence factors $u_{\bf k}$ and $v_{\bf k}$, Eq. (\ref{Eq:coh-factors}).

The fact that the thermal averages $\langle a_{\bf k}^\dagger a_{\bf k}\rangle$ and $\langle a_{\bf k}a_{-\bf k}\rangle$ that are derived in the BBP approach can be derived using canonical traces is quite remarkable. It tells us that thermal averages of physical observables (one or two-body operators that can be expressed as combinations of $a_{\bf k}^\dagger a_{\bf k}$ and $a_{\bf k}a_{-\bf k}$ and their hermitian conjugates) can be obtained in the canonical ensemble, and hence that the whole BBP formulation is canonical in nature, and not grand canonical. In the following subsection, we want to analyze the various tracing schemes that one can use to obtain partition functions for interacting bosons, and try to understand how in the BBP, starting from a supposedly grand-canonical partition function we end up with results that can be derived in the canonical ensemble.


\subsection{Analyzing the various possible tracing schemes}

We now are in a good position to discuss what exactly is not quite right about the standard field-theoretic formulations of the statistical mechanics of interacting bosons. These formulations are so elegant and so mathematically sophisticated that it is hard to imagine that they actually do anything other than what they purport to be doing. Unfortunately, mathematical sophistication is no guarantee of correctness, and below we will show that in the standard formulation of BBP's theory, the number of condensed bosons $N_0$ is never traced over, and this incomplete tracing operation effectively reduces what is supposed to be a grand-canonical trace to a canonical one instead. 

Let us for example have a detailed look at the manipulations that led us to Eq. (\ref{Eq:res_G}). When we write the action $S_0$ as a quadratic form in the fields $\psi$ and $\psi^*$, as we did in Eq. (\ref{Eq:S0MatrixForm}), and take the inverse of the matrix $G^{-1}$ to find the Green's function matrix $G$, these two quick steps are thought to be equivalent to the following ones in first-quantized methods:
\begin{enumerate}
\item
The eigenvalues and eigenfunctions of the Bogoliubov Hamiltonian are found for a system of $N$ bosons;

\item
A canonical trace of the Boltzmann weight $e^{-\beta(\hat{H} - \mu \hat{N})}$ is taken over these eigenfunctions at constant number of bosons $N$;

\item
A second trace is taken over the total number of bosons $N$ from $N=0$ to $N = \infty$ to find the grand-canonical partition function.
\end{enumerate}

The fact that the above three steps do actually take place behind the scenes is never doubted and is generally taken for granted.
It turns out, unfortunately, that taking the inverse of the matrix $G^{-1}$ to go from Eq. (\ref{Eq:G^-1}) to Eq. (\ref{Eq:res_G}) succeeds in completing steps 1 and 2 of the above program, but fails to complete step 3. In other words, finding the Green's function matrix $G$ is equivalent to taking the trace over the eigenfunctions $|\Psi_{\{m_i\}}(N)\rangle$ of the Bogoliubov Hamiltonian $\hat{H}_B$ {\em at fixed boson number $N$}, which is a canonical trace, but is not equivalent to taking the extra step of doing a trace over $N$ that would make the approach grand-canonical.

We now want to discuss the various kinds of trace that one may use to evaluate partition functions in the canonical and grand-canonical ensembles.

\subsubsection{Canonical ensemble: tracing over the basis of single-particle states}

Let us first consider the relatively simple situation of a gas of $N$ bosons in the canonical ensemble.
The canonical partition function ${Z_C} = \mbox{Tr}_N\left(e^{-\beta\hat{H}}\right)$ can be written as the following trace
over boson single-particle states:
\begin{align}
{Z_C} & = \sum_{N_0=0}^N \cdots \sum_{N_M=0}^N \langle N_0,\ldots,N_M| e^{-\beta\hat{H}}|N_0,\ldots,N_M\rangle
\notag\\
&\times \delta_{\sum_{i=0}^M N_i,N},
\label{Eq:ZNTrace}
\end{align}
where the ket $|N_0,\ldots,N_M\rangle$ represents the state having $N_0$ bosons with momentum ${\bf k}_0={\bf 0}$, $N_1$ bosons with momentum ${\bf k}_1$, and so on, {\em i.e.}:
\begin{equation}
|N_0,\ldots,N_M\rangle = \prod_{i=0}^M\frac{\big(a_{{\bf k}_i}^\dagger\big)^{N_i}}{\sqrt{N_i!}}|0\rangle.
\end{equation} 
Note how, in Eq. (\ref{Eq:ZNTrace}), we chose to take the trace of the Hamiltonian $\hat{H}_B$ over the complete orthonormal basis composed of the eigenstates of the non-interacting Hamiltonian $\hat{H}_0 = \sum_{\bf k} \varepsilon_{\bf k}a_{\bf k}^\dagger a_{\bf k}$ at fixed number of bosons $N$, this last constraint being enforced by the Kronecker delta on the {\em rhs} of this last equation, which
ensures that the total number of particles in all the basis kets $|N_0,N_1,\ldots,N_M\rangle$ is equal to $N$.

\subsubsection{Canonical ensemble: tracing over the basis of elementary excitations}

Now, after we diagonalize the Hamiltonian $\hat{H}_B$ using a number-conserving formalism such as the one developed in Ref. \onlinecite{Ettouhami2012}, we end up with another basis of eigenfunctions describing elementary excitations of the system of $N$ bosons, with one such eigenfunction $|\Psi_{\{m_i\}}(N)\rangle = |m_1,m_2,\ldots, m_M\rangle$ describing a state having $m_i$ excitations in the state of momentum ${\bf k}_i$, with $i=1,\ldots,M$, see Eq. (\ref{Eq:def:Psi_m}).
The trace being independent of the orthonormal basis chosen, we observe that there is a second, equivalent way to calculate the partition function $Z_C$ which consists in taking a trace over the basis of eigenfunctions 
$\{|\Psi_{\{m_i\}}(N)\rangle\}$ at {\em fixed} total number of bosons $N$:
\begin{equation}
{Z_C} = \sum_{m_1=0}^{D(N,M)}\cdots \sum_{m_M=0}^{D(N,M)} \langle m_1,\ldots,m_M| e^{-\beta\hat{H}}|m_1,\ldots,m_M\rangle.
\label{Eq:ZNTrace2}
\end{equation}
In the above equation, the $m_i$'s take integer values $m_i=0,1,2,\ldots,D(N,M)$, corresponding to the number of excitations of wavevector ${\bf k}_i$. The upper limit of the summation over any single quantum number $m_i$ is the dimension $D(N,M)$ of the matrix representation of the Hamiltonian $\hat{H}_B$ and depends on the number of bosons and the number of momentum modes considered.
Assuming that our system has $N$ bosons and $M$ momentum modes, 
any given state of the Hilbert space can be written in the form $|N_0, N_1,N_2,\ldots, N_M\rangle$, and hence
the dimension of the Hilbert space is the number of ways we can solve the equation:
\begin{equation}
N_0 + N_1 + N_2 + \cdots + N_{M} = N.
\label{Eq:sumN}
\end{equation}
This is a well-known combinatorial problem, which can be mapped onto the problem of finding the number $D(N,M)$ of arrangements of $N$ balls and $M-1$ dividers (all indistinguishable), and is given by:
\begin{equation}
D(N,M) ={N + M - 1 \choose M - 1} = \frac{(N + M - 1)!}{(M-1)! N!}.
\end{equation}
It can be shown that $D(N,M)$ becomes extremely large for values of $N$ and $M$ such that $N + M\gg N, M$, and hence for all practical purposes we can extend the summations in Eq. (\ref{Eq:ZNTrace2}) all the way to infinity:
\begin{equation}
{Z_C} = \sum_{m_1=0}^\infty \cdots \sum_{m_M=0}^\infty \langle m_1,\ldots,m_M| e^{-\beta\hat{H}}|m_1,\ldots,m_M\rangle.
\label{Eq:ZNTrace3}
\end{equation}
Note the fundamental difference between Eq. (\ref{Eq:ZNTrace}) and Eq. (\ref{Eq:ZNTrace3}). For while in the former there is a need for a Kronecker delta on the {\em rhs} to ensure that the constraint (\ref{Eq:sumN}) is satisfied, 
no such constraint is imposed on the $m_i$'s. This has the important consequence that while it is notoriously difficult to calculate the canonical trace directly from Eq. (\ref{Eq:ZNTrace}), precisely because of the constraint
$\sum_i n_i=N$ on the {\em rhs} of this equation, in the basis of Bogoliubov eigenfunctions there is no such constraint and the sums on the {\em rhs} of Eq. (\ref{Eq:ZNTrace2})  can be decoupled into a product of geometric series and hence can be easily evaluated.

\subsubsection{Grand-canonical ensemble: which trace is used in the BBP formalism ?}

We now want to discuss the grand-canonical ensemble, where the grand-canonical partition function can be obtained from the canonical one using the following relation:
\begin{equation}
Z_G(\mu, V, T) = \sum_{N=0}^\infty {Z_C}(N,V,T)\, e^{\beta\mu N}.
\label{Eq:ZgTrace}
\end{equation}
Inserting the expression of ${Z_C}$ from Eq. (\ref{Eq:ZNTrace}) into Eq. (\ref{Eq:ZgTrace}), one can show that the summation over all values of $N$ from $N=0$ to $\infty$ is equivalent to extending the summations over the $N_i$'s from $0$ to $\infty$ in Eq. (\ref{Eq:ZNTrace}), with the result:\cite{Remark-ideal}
\begin{align}
Z_G & = \sum_{N_0=0}^\infty \cdots \sum_{N_M=0}^\infty \langle N_0,\ldots,N_M| e^{-\beta(\hat{H}-\mu\hat{N})}|N_0,\ldots,N_M\rangle.
\label{Eq:ZgTrace2}
\end{align}
Alternatively, if we start from the expression (\ref{Eq:ZNTrace2}) of the canonical partition function, where the trace is evaluated in the basis of the $|\Psi_{\{m_i\}}(N)\rangle$'s, the grand partition function takes the form:
\begin{align}
Z_G & = \sum_{N=0}^\infty \sum_{m_1=0}^\infty \cdots \sum_{m_M=0}^\infty \langle m_1,\ldots,m_M| 
\nonumber\\
&\times e^{-\beta(\hat{H}-\mu\hat{N})}|m_1,\ldots,m_M\rangle.
\label{Eq:ZgTrace2b}
\end{align}
The two tracing schemes in Eqs. (\ref{Eq:ZgTrace2}) and (\ref{Eq:ZgTrace2b}) are mathematicall equivalent and have to produce the same result for the grand partition function $Z_G$. Notice from Eq. (\ref{Eq:ZgTrace2}) that {\em all} single particle occupation numbers $N_0, \ldots,N_M$ are traced over and should not appear in the expression of $Z_G$. And, since the total number of bosons $N$ can be decomposed as the sum $N= N_0 + \cdots + N_M$, $N$ itself should also not appear in the expression of $Z_G$. Similarly, since Eq. (\ref{Eq:ZgTrace2b}) should give the exact same result at Eq. (\ref{Eq:ZgTrace2}), we conclude that the {\em rhs} of Eq. (\ref{Eq:ZgTrace2b}) should not depend on any one of the single particle occupation numbers $N_0, \ldots,N_M$, and should also not depend on the total number of bosons $N$.

We now can make an attempt at explaining the reason why the result we obtained in Sec. \ref{Sec:Review} for the thermal average $\big\langle a_{\bf k}^\dagger a_{\bf k}\big\rangle$, which was supposedly derived in the grand-canonical ensemble, coincides with the result of Eq. (\ref{Eq:avgNkn_B}), which was derived in the canonical ensemble (provided $n_0$ is replaced with $n_B$). 
This reason has to do with the Bogoliubov prescription, which replaces \cite{Ettouhami2012} the density of bosons $n_B$ in the expression of the ground state and excitation energies with $n_0$ (we shall indeed see in Sec. \ref{Sec:GSEnergyFock} below that when the Bogoliubov prescription is avoided, the condensate density $n_0$ does not appear in the expressions of the ground state and excitation energies of the system, and that only $n_B$ appears in these expressions), and with the fact that the occupation number of the condensate $N_0$ is kept as an immutable constant and never traced over. In other words, the BBP formulation is equivalent to the canonical trace in Eq. (\ref{Eq:ZNTrace3}), as the final trace over $N$ in Eq. (\ref{Eq:ZgTrace2b}) is never performed in this approach. The fact that the trace is taken in an incomplete way leads to a canonical result instead of a grand-canonical one for the average $N_{\bf k}=\big\langle a_{\bf k}^\dagger a_{\bf k}\big\rangle$ of the number of bosons in the single-particle state of momentum ${\bf k}$.

The realization that the statistical treatments of the interacting Bose gas are canonical in nature and not grand-canonical as is commonly thought is the main result of this paper. In what follows, we want to extend the work we did in Ref. \onlinecite{Ettouhami2012} and generalize the variational number-conserving theory we developed in that reference to finite temperatures. However, before we can do that, we need to find a way to take Fock interactions between depleted bosons into account. This will be the subject of the following Section.

\section{Ground state energy and excitation spectrum of interacting bosons: taking Fock interactions between depleted bosons into account}
\label{Sec:GSEnergyFock}

In Sec. \ref{Sec:StatMechCanonicalFormulation} below, we will want 
to examine the canonical formulation of the statistical mechanics of interacting bosons by extending the variational approach developed in Ref. \onlinecite{Ettouhami2012} to finite temperatures.
The motivation behind our desire to use the variational approach resides in the fact that it will allow us to avoid the use of Bogoliubov's prescription, and hence will allow us to seamlessly avoid 
issues such as the jump discontinuity in the condensed density $n_0(T)$ at the critical temperature $T_c$, and unambiguously demonstrate that $n_0$ which appears in the expression of the density of 
depleted bosons should in fact be replaced by $n_B$. 

Because we will be interested in the thermodynamics of the Bose gas in the whole range of temperatures between $T=0$ and the critical transition temperature $T_c$, we will need to take 
into account the Fock interactions between depleted bosons, which are neglected in the $T=0$ diagonalization of Bogoliubov's Hamiltonian.
This can be done in two different ways. One way consists in treating these Fock interactions as a perturbation to the quadratic action $S_0$, and performing a perturbative expansion of the partition function in terms of this perturbation. 
Another, more compact way to treat the Fock terms, which avoids the difficulties of perturbation theory and mirrors more closely the inner workings of an exact solution,
consists in trying to generalize the variational approach of Ref. \onlinecite{Ettouhami2012} to find the ground state and excitation energies in presence of
the Fock interactions between depleted bosons, and is the way we are going to follow in this Section.
To this end, let us consider the following Hamiltonian:
\begin{equation}
\hat{H} = \hat{H}_B + \hat{H}_F'.
\label{Eq:fullH}
\end{equation}
Here $\hat{H}_B$ is Bogoliubov's Hamiltonian of Eq. (\ref{Eq:H_B:standard}), 
and $\hat{H}_F'$ is the part of the total Hamiltonian that represents the Fock interaction between depleted bosons, 
\begin{equation}
\hat{H}_F' = \frac{1}{2V}\sum_{\bf k\neq 0}\sum_{\bf k' \neq 0, k} v({\bf k}-{\bf k}') a_{\bf k}^\dagger a_{\bf k} a_{\bf k'}^\dagger a_{\bf k'}.
\end{equation}
Henceforth, we shall write the Hamiltonian $\hat{H}$ in the form:
\begin{equation}
\hat{H} = \sum_{\bf k\neq 0}\hat{H}_{\bf k},
\label{Eq:fullH2}
\end{equation}
where the single-mode Hamiltonian $\hat{H}_{\bf k}$ is given by:
\begin{align}
\hat{H}_{\bf k} & = \varepsilon_{\bf k} a_{\bf k}^\dagger a_{\bf k} 
+ \frac{v({\bf k})}{2V}( a_0^\dagger a_0 a_{\bf k}^\dagger a_{\bf k} 
+ a_0^\dagger a_0 a_{-\bf k}^\dagger a_{-\bf k}
\nonumber\\
&+ a_{\bf k}^\dagger a_{-\bf k}^\dagger a_0 a_0 + a_{\bf k} a_{-\bf k} a_0^\dagger a_0^\dagger)
\nonumber\\
&+  \frac{1}{2V}\sum_{\bf k' \neq 0, k} v({\bf k}-{\bf k}') a_{\bf k}^\dagger a_{\bf k} a_{\bf k'}^\dagger a_{\bf k'}.
\label{Eq:H_k}
\end{align}
We shall first find the variational, number-conserving ground state of the total Hamiltonian $\hat{H}$ in Subsection \ref{Sub:ConservingCanonical}
before deriving the excitation spectrum of this Hamiltonian in Subsection \ref{Sub:ExcEn}.

\subsection{Variational ground state energy}
\label{Sub:ConservingCanonical}

As we mentioned above, our aim in this Subsection is to generalize the variational approach of Ref. \onlinecite{Ettouhami2012} to the Hamiltonian $\hat{H}$ of Eqs. (\ref{Eq:fullH2})-(\ref{Eq:H_k}),
which includes the Fock interaction between depleted bosons. To this end, we shall use for $|\Psi(N)\rangle$ an expression of the
form (we again here denote by $M$ the total number of momentum modes kept in the calculation, which will eventually be sent to infinity):
\begin{align}
|\Psi(N)\rangle & = \sum_{n_1=0}^{n_{1max}} \ldots\sum_{n_M=0}^{n_{Mmax}}
C_{n_1}C_{n_2}\ldots C_{n_M}
\nonumber\\
&\times |N-2\sum_{i=1}^M n_i; n_1,n_1;\ldots;n_M,n_M\rangle,
\label{Eq:fullPsi(N)}
\end{align}
where the normalized basis wavefunctions are given by:
\begin{align}
|N-2\sum_{i=1}^M n_i;& n_1,n_1;\ldots;n_M,n_M\rangle  = 
\frac{\big(a_0^\dagger\big)^{N-2\sum_{i=1}^M n_i}}{\sqrt{[N-2\sum_{i=1}^M n_i]!}}
\nonumber\\
&\times\prod_{i=1}^M \frac{\big(a_{{\bf k}_i}^\dagger)^{n_i}}{\sqrt{n_i!}}
\frac{\big(a_{{-\bf k}_i}^\dagger)^{n_i}}{\sqrt{n_i!}}|0\rangle.
\end{align}
As we explained in detail in Ref. \onlinecite{Ettouhami2012}, the ground state wavefunction in Eq. (\ref{Eq:fullPsi(N)}) is {\em not} a simple product of 
ground state wavefunctions of the single-mode Hamiltonians $\hat{H}_{\bf k}$, as was the case in previous approaches to this problem\cite{Lee1957a,HuangBook,Leggett2001}
where the kets $|N-2\sum_{i=1}^M n_i; n_1,n_1;\ldots;n_M,n_M\rangle$ on the {\em rhs} of Eq. (\ref{Eq:fullPsi(N)})
were replaced with $|N_0; n_1,n_1;\ldots;n_M,n_M\rangle$, {\em i.e.} the number of bosons in the single-particle state of momentum ${\bf k}=0$ was kept as
an {\em immutable constant} $N_0$ {\em regardless} of the number of bosons $n_1, n_2,\ldots, n_M$ in the other single-particle states ${\bf k}_i\neq 0$. 
Here by contrast, the presence of the quantity $\big[N-2\sum_{i=1}^Mn_i\big]$ in the various kets that appear on the {\em rhs} of Eq. (\ref{Eq:fullPsi(N)})
acts like an implicit and rather nontrivial coupling between all the single-mode Hamiltonians $\{\hat{H}_{\bf k}\}$, and goes a step beyond wavefunctions studied in previous literature.

We now will follow the same steps as in Ref. \onlinecite{Ettouhami2012}, and use the following ansatz for the coefficients $C_{n_i}$:
\begin{equation}
C_{n_i} = (-c_{{\bf k}_i})^{n_i}\sqrt{1- c_{{\bf k}_i}^2}.
\end{equation}
It can then be shown that the expectation value
of the Hamiltonian $\hat{H}_{{\bf k}_j}$ in the state $|\Psi(N)\rangle$ of Eq. (\ref{Eq:fullPsi(N)})
is given by the following expression (see Appendix \ref{App:A}):
\begin{align}
&\langle\Psi(N)|\hat{H}|\Psi(N)\rangle  = \sum_{j=1}^M\Bigg\{
\Big[ 
\varepsilon_{{\bf k}_j} + n_B \bar{v}({\bf k}_j)
\Big]\frac{c_{{\bf k}_j}^2}{1 - c_{{\bf k}_j}^2}
\nonumber\\
&- n_B \bar{v}({\bf k}_j)
\frac{c_{{\bf k}_j}}{1 - c_{{\bf k}_j}^2}
+ n_B \tilde{v}({{\bf k}_j}) \frac{c_{{\bf k}_j}^2}{1 - c_{{\bf k}_j}^2}
\Bigg\},
 \label{Eq:avgHtot}
\end{align}
where $\bar{v}(k_j)$ and $\tilde{v}(k_j)$ are given by:
\begin{subequations}
\begin{align}
\bar{v}(k_j) &= v(k_j)\Big( 1 - \frac{2}{N}\sum_{i=1(\neq j)}^M
\frac{c_{{\bf k}_i}^2}{1-c_{{\bf k}_i}^2}\Big),
\label{Eq:defbarv}
\\
\tilde{v}(k_j) &=  v({{\bf k}_j})\Big[
\frac{1}{N}\sum_{{\bf k}_l \neq 0,\pm {\bf k}_j} \frac{v({\bf k}_j - {\bf k}_l)}{v({\bf k}_j)} \frac{c_{{\bf k}_l}^2}{1 - c_{{\bf k}_l}^2}
\Big].
\label{Eq:deftildev}
\end{align}
\end{subequations}
Minimization of the expectation value in Eq. (\ref{Eq:avgHtot}) with respect to the variational constants $\{c_{{\bf k}_j}\}$ leads to the following equation:
\begin{align}
c_{{\bf k}_j}^2 - 2\Big(\frac{\tilde{\cal E}_{{\bf k}_j}}{n_B\bar{v}({\bf k}_j)}\Big)c_{{\bf k}_j} + 1 =  0.
\label{Eq:Eqckfull}
\end{align}
In the above equation, $\tilde{\cal E}_{\bf k_j}$ denotes the quantity:
\begin{subequations}
\begin{align}
& \tilde{\cal E}_{\bf k_j}   = \varepsilon_{{\bf k}_j} + n_B \bar{v}({\bf k}_j) + \sigma_{\bf k},
\\
& \mbox{with}\quad\sigma_{{\bf k}_j}  = \frac{2}{N}\sum_{i=1(\neq j)}^M n_Bv({\bf k}_i)\Big[
\frac{c_{{\bf k}_i}}{1+c_{{\bf k}_i}} 
\nonumber\\
&+\frac{1}{2}\frac{v({\bf k}_i-{\bf k}_j)}{v({\bf k}_i)}\frac{c^2_{{\bf k}_i}}{1-c^2_{{\bf k}_i}}\Big]
\label{Eq:defsigmak}
\end{align}
\label{Eq:defsigmak:both}
\end{subequations}
Transforming the sum over momentum modes in Eq. (\ref{Eq:defsigmak}) into an integral,\cite{Ettouhami2012} we can rewrite $\sigma_{{\bf k}_j}$ in the form:
\begin{align}
\sigma_{{\bf k}_j}  \simeq	\int_{{\bf k}'} v({\bf k}')\Big[
\frac{c_{\bf k}'}{1+c_{\bf k}'} 
+\frac{1}{2}\frac{v({\bf k}-{\bf k}')}{v({\bf k}')}\frac{c^2_{{\bf k}'}}{1-c^2_{{\bf k}'}}\Big].
\end{align}
Solving Eq. (\ref{Eq:Eqckfull}) for $c_{\bf k}$, we obtain:
\begin{align}
c_{\bf k} = \Big(\frac{\tilde{\cal E}_{\bf k}}{n_B\bar{v}({\bf k})}\Big) - 
\sqrt{\Big(\frac{\tilde{\cal E}_{\bf k}}{n_B\bar{v}({\bf k})}\Big)^2-1},
\label{Eq:newc_k}
\end{align}
where the sign of the second term has been chosen so that $0<c_{\bf k}<1$.
In the particular case where $v({\bf k}) = g$, corresponding to $v({\bf r})=g\delta({\bf r})$ in real space, 
the ratio $[v({\bf k}-{\bf k}')/v({\bf k})]=1$, and the expression of $\sigma_{\bf k}$ can be written in the form:
\begin{subequations}
\begin{align}
\sigma_{{\bf k}_j}  & = \int_{{\bf k}'} \Big[ 
\frac{1}{2}\frac{c^2_{{\bf k}'}}{1-c^2_{{\bf k}'}}
+ v({\bf k}') \frac{c_{\bf k}'}{1+c_{\bf k}'} 
\Big],
\label{Eq:sigmak:a}
\\
&\simeq \frac{1}{2} gn_d + \int_{{\bf k}'} v({\bf k}')
\frac{c_{\bf k}'}{1+c_{\bf k}'},
\label{Eq:sigmak:b} 
\end{align}
\label{Eq:sigmak}
\end{subequations}
where $n_d$ is the density of depleted bosons:
\begin{equation}
n_d = \int_{\bf k} \frac{c^2_{{\bf k}}}{1-c^2_{{\bf k}}}.
\end{equation}

As can be seen, the quantity on the {\em rhs} of Eq. (\ref{Eq:sigmak:b})
does not depend on ${\bf k}$, and we shall henceforth drop the subscript ${\bf k}$
from $\sigma_{\bf k}$, and rewrite Eqs. (\ref{Eq:defsigmak:both}) in the form:
\begin{subequations}
\begin{align}
& \tilde{\cal E}_{\bf k_j}   = \varepsilon_{{\bf k}_j} + n_B \bar{v}({\bf k}_j) + \sigma,
\\
& \mbox{with}\quad\sigma  = \frac{1}{2}gn_d + \int_{\bf k} v(k)\frac{c_{\bf k}}{1+c_{\bf k}}.
\label{Eq:defsigmak2b}
\end{align}
\label{Eq:defsigmak2}
\end{subequations}
In Ref. \onlinecite{Ettouhami2012}, we used a Gaussian form for the interaction potential between bosons of the form:
\begin{equation}
v({\bf r}) = \frac{ge^{-r^2/(2\lambda^2)}}{(2\pi\lambda^2)^{3/2}},
\label{Eq:pot_lambda_RS}
\end{equation}
where $\lambda$ is a positive quantity having the dimensions of length, so as to avoid an ultraviolet divergence in the integral on the {\em rhs} 
of Eq. (\ref{Eq:defsigmak2b}) in three dimensions, which occurs when $v(k)=g$ is a constant because in that case $c_{\bf k}$ from Eq. (\ref{Eq:newc_k})
behaves like $1/k^2$ as $k\to \infty$. In Fourier space, the interaction potential 
of Eq. (\ref{Eq:pot_lambda_RS}) is given by:
\begin{equation}
v({\bf k}) = ge^{-\frac{1}{2}k^2\lambda^2}.
\label{Eq:pot_lambda_FS}
\end{equation}
Let us now introduce dimensionless units, where energies are measured in units of $n_Bv({\bf 0}) = gn_B$,
and wavevectors are measured in units of $k_0 = \sqrt{2m g n_B}/\hbar$. 
Then we can write for $v({\bf k})$ the following expression:
\begin{equation}
v({\bf k}) = g e^{-\frac{1}{2}(8\pi n_B a \lambda^2)\tilde{k}^2},
\label{Eq:pot_lambda_FS2}
\end{equation}
where in the notation of Eq. (\ref{Eq:def:tildek}) $\tilde{\bf k}$ is the dimensionless wavevector $\tilde{\bf k}  = {\bf k}/{k_0}$.
As we mentioned above, the reason behind using an interaction potential with Gaussian form in Ref. \onlinecite{Ettouhami2012} was to avoid the ultraviolet divergence that results from summing over momentum modes
in the integral on the {\em rhs} of Eq. (\ref{Eq:defsigmak2b}). It is not difficult to see from Eq. (\ref{Eq:pot_lambda_FS2}) that such a Gaussian interaction potential $v({\bf k})$ introduces a momentum cut-off around values of $\tilde{k}$ such that $8\pi n_B a \lambda^2 \tilde{k}^2 \sim 1$. Hence, for reasons of numerical efficiency we'll find it expedient to replace the above interaction potential with the following approximation:
\begin{subequations}
\begin{align}
v({\bf k}) & = g \quad k \le \Lambda,
\\
v({\bf k}) & = 0 \quad k > \Lambda,
\end{align}
\end{subequations}
where the ultraviolet momentum cut-off $\Lambda$ is given by:
\begin{equation}
\Lambda = \frac{1}{\sqrt{8\pi n_Ba\lambda^2}}.
\end{equation}
For simplicity, we shall take the characteristic length scale $\lambda$ which governs the range
of the interaction potential $v({\bf r})$ to be the scattering length $a$. The expression of $\Lambda$ becomes:
\begin{equation}
\Lambda = \frac{1}{\sqrt{8\pi n_Ba^3}}.
\end{equation}
We can then write for $c_{\bf k}$ the following expression:
\begin{align}
c_{\bf k}  = 1 + \frac{1}{C_d}(\tilde{k}^2 + \tilde\sigma)
- \sqrt{\Big[ 1 + \frac{1}{C_d}(\tilde{k}^2 + \tilde\sigma) \Big]^2 - 1},
\label{Eq:res_ck}
\end{align}
where we denote by $\tilde\sigma$ the dimensionless quantity:
\begin{align}
\tilde\sigma & = \frac{\sigma}{gn_B},
\label{Eq:def:k:sigma}
\end{align}
and where $C_d$ is given by:\cite{Ettouhami2012}
\begin{equation}
C_d = 1 - \frac{N_d}{N}.
\label{Eq:def:C_d}
\end{equation}
Here, the fraction of depleted bosons $N_d/N$ is given by:\cite{Ettouhami2012}
\begin{align}
\frac{N_d}{N} & = \frac{4\sqrt{2}}{\sqrt\pi}(n_Ba^3)^{\frac{1}{2}}\int_0^\infty d\tilde{k}\;\tilde{k}^2
\left(\frac{1 + Q^2}{\sqrt{Q^2(Q^2 +2)}} 
- 1\right),
\label{Eq:ratioNdN}
\end{align}
where we used the shorthand notation:
\begin{equation}
Q^2 = \frac{\tilde{k}^2+\tilde\sigma}{C_d}.
\label{Eq:Q^2}
\end{equation}

\begin{center}
\begin{table}[tb]
  \begin{tabular}{ | c | c | c | }
    \hline\hline
    {   } $n_Ba^3$ {   } & {   } $\tilde\sigma$ {   } &{   } $C_d$ {   } \\ \hline
    {   } $10^{-5}$ {   } & {   } 0.614 {   } &{  } 0.9967 {  } \\  \hline
    $10^{-4}$ & 0.568 & 0.9897 \\  \hline 
    $10^{-3}$ & 0.452 & 0.9672 \\  \hline
    $10^{-2}$ & 0.263 & 0.8990 \\  \hline
    $10^{-1}$ & 0.172 & 0.7483 \\
    \hline\hline
  \end{tabular}
\caption{Values of $\tilde\sigma$ and $C_d$ for a few representative values of the parameter $n_Ba^3$.}
\label{Tbl:sigma_Cd_vs_nBa3}
\end{table}
\end{center}

If we use the expression (\ref{Eq:res_ck}) of $c_{\bf k}$ in Eq. (\ref{Eq:defsigmak2b}),
and change the variable of integration from ${\bf k}$ to the dimensionless wavevector $\tilde{\bf k} = {\bf k}/k_0$ (see Eq. (\ref{Eq:def:k0})),
we can write for the dimensionless quantity $\tilde{\sigma}$ the following self-consistent equation:
\begin{align}
\tilde{\sigma} & = \frac{1}{2}\Big(\frac{n_d}{n_B}\Big)+\frac{8\sqrt{2}}{\sqrt{\pi}}
(n_Ba^3)^{\frac{1}{2}} 
\nonumber\\
&\times\int_{0}^\Lambda d\tilde{k} \;\tilde{k}^2
\frac{1+Q^2 -\sqrt{Q^2\big(Q^2+2\big)}}
{2+Q^2 -\sqrt{Q^2\big(Q^2+2\big)}}.
\label{Eq:self-consistent-sigma}
\end{align}
A numerical solution to the above equation for $\tilde{\sigma}$ can be found by iteration in the following way.
First, one starts with an initial guess for $\tilde\sigma$. Using this initial guess, one computes the ratio
$N_d/N$ using Eq. (\ref{Eq:ratioNdN}) above, which allows us to find the ``depletion constant" $C_d= 1 - N_d/N$. This 
value of $C_d$ is then used to solve equation (\ref{Eq:self-consistent-sigma}) for $\tilde\sigma$.
This computed value of $\tilde\sigma$ is then used again as an input to find a better estimate of the ratio
$N_d/N$ using Eq. (\ref{Eq:ratioNdN}), and the process is repeated until convergence and a stable solution for 
$C_d$ and $\tilde\sigma$ is found.

\begin{figure}[b]
\includegraphics[width=8.09cm, height=5.5cm]{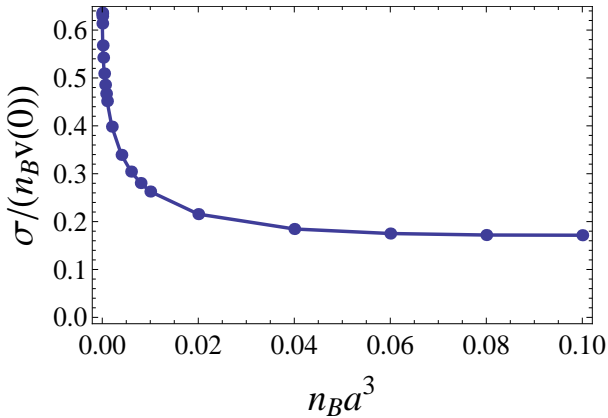}
\caption[]{ 
Plot of $\tilde\sigma$ vs. $n_Ba^3$. 
}\label{Fig:PlotSigmavsnBa3}
\end{figure}

In the following, we shall be mostly interested in dilute Bose gases, for which $n_Ba^3\ll 1$.
Under these circumstances, a numerical solution of the self-consistency equation (\ref{Eq:self-consistent-sigma})
for a few representative values of $n_Ba^3$ between $10^{-5}$ and $10^{-1}$ yields the values for the quantities $\tilde\sigma$ and $C_d$ shown in Table \ref{Tbl:sigma_Cd_vs_nBa3}.
The variation of $\tilde\sigma$ vs. $n_Ba^3$ over the whole range $10^{-6}\le n_Ba^3 \le 0.1$ is shown in Fig. \ref{Fig:PlotSigmavsnBa3}.
As can be seen, $\tilde\sigma$ takes values that are close to $0.6$ for $n_Ba^3 < 10^{-4}$, and decreases to about $0.2$ for $n_Ba^3$ between $0.02$ and $0.1$.

\begin{figure}[htb]
\includegraphics[width=8.09cm, height=5.5cm]{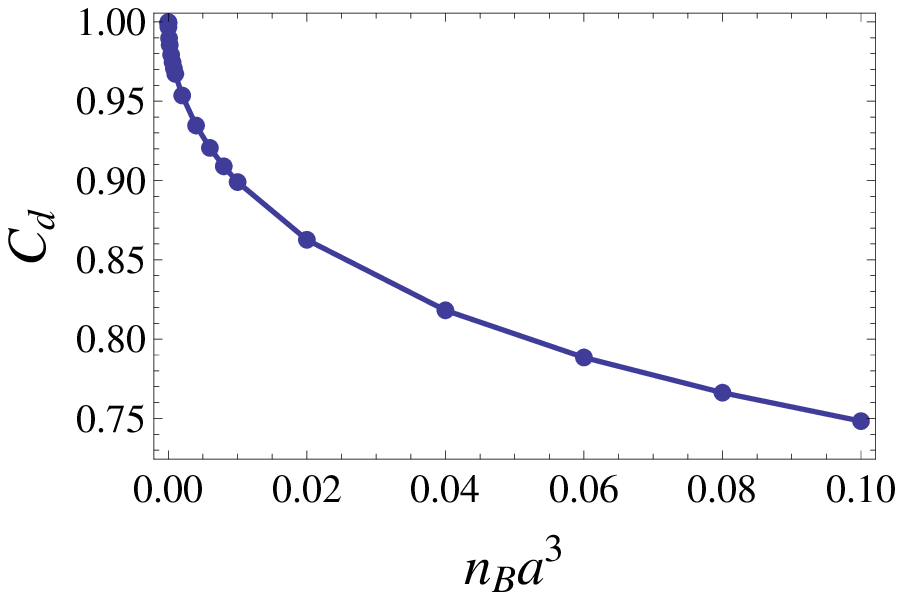}
\includegraphics[width=8.09cm, height=5.5cm]{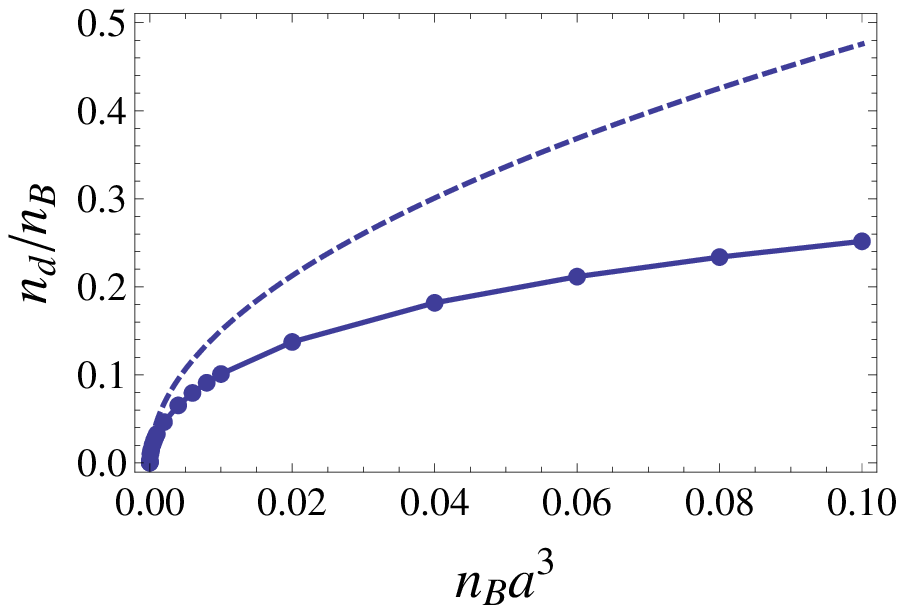}
\caption[]{Upper panel: Plot of $C_d = 1- n_d/n_B$ vs. $n_Ba^3$ in the variational number-conserving approach at $T=0$. Lower panel: Plot of the depleted fraction $n_d/n_B$ vs. $n_Ba^3$ at $T=0$.  The solid line corresponds to the  variational number-conserving approach of this paper, and the dashed line represents the result of the naive Bogoliubov approximation, $n_d/n_B = 8(n_Ba^3)^{\frac{1}{2}}/3\sqrt\pi.$
}\label{Fig:}
\end{figure}

We are now in a position to find the ground state energy $E_0 = \langle\Psi(N)|\hat{H}|\Psi(N)\rangle$ 
of the system, which is given by Eq. (\ref{Eq:avgHtot}). For simplicity, in Eq. (\ref{Eq:deftildev}) we 
shall approximate the interaction potential as $v({\bf r}) = g\delta({\bf r})$, which gives us:
\begin{equation}
\tilde{v}({\bf k}) \simeq \frac{1}{2}g\Big(\frac{N_d}{N}\Big).
\end{equation}
It then follows that Eq. (\ref{Eq:avgHtot}) can be written in the form:
\begin{equation}
E_0 = \sum_{\bf k\neq 0} \frac{1}{1-c_{\bf k}^2}
\Big\{
\big[\varepsilon_{\bf k} + n_B\bar{v}(k) + \frac{1}{2}gn_d \big] c_{\bf k}^2 
- n_B \bar{v}(k)c_{\bf k}
\Big\}.
\end{equation}
Using the expression (\ref{Eq:res_ck}) of the coefficients $c_{\bf k}$ into this last equation and transforming the sum into an integral, we
obtain (in three dimensions):
\begin{align}
\frac{E_0}{V} & = -\frac{1}{2}gn_B^2\Bigg\{
\frac{8\sqrt{2}}{\sqrt{\pi}}(n_Ba^3)^{\frac{1}{2}}
\nonumber\\
&\times\int_0^\Lambda d\tilde{k}\;\tilde{k}^2\Bigg[
C_d \Big( 1 + Q^2
-\sqrt{Q^2(Q^2 +2)}\Big)
\nonumber\\
& + \tilde\sigma\Big(
\frac{1 + Q^2}{\sqrt{Q^2(Q^2 + 2)}} - 1
\Big)
\Bigg] - \Big(\frac{n_d}{n_B}\Big)^2
\Bigg\}.
\label{Eq:En+HF}
\end{align} 
The physical interpretation of the above equation is quite simple: it is just the ground state energy derived in Ref. \onlinecite{Ettouhami2012}, 
plus an extra term 
\begin{equation}
-\frac{1}{2} g n_B^2 \left[-\Big(\frac{n_d}{n_B}\Big)^2\right] = +\frac{1}{2}gn_d^2
\end{equation}
which represents the Fock interaction between depleted bosons.

Numerical evaluation of the integral on the {\em rhs} of Eq. (\ref{Eq:En+HF}) for $n_B a^3 = 10^{-4}$ yields, for $\tilde\sigma = 0.568$ and $C_d= 0.9897$:
\begin{subequations}
\begin{align}
\frac{E_0}{V} & \simeq \frac{1}{2}gn_B^2\cdot(-57.62)(n_Ba^3)^{\frac{1}{2}},
\\
& \simeq \frac{1}{2} gn_B^2 \cdot (-0.5762). 
\end{align}
\label{Eq:CE:10-4}
\end{subequations}
On the other hand, for $n_B a^3 = 10^{-3}$, using $\tilde\sigma = 0.452$ and $C_d= 0.9672$, we have:
\begin{subequations}
\begin{align}
\frac{E_0}{V} & \simeq \frac{1}{2}gn_B^2\cdot(-14.69)(n_Ba^3)^{\frac{1}{2}},
\\
& \simeq \frac{1}{2} gn_B^2 \cdot (-0.4645). 
\end{align}
\label{Eq:CE:10-3}
\end{subequations}
Finally, for  $n_Ba^3 = 10^{-2}$ we let $\tilde\sigma = 0.263$ and $C_d = 0.8990$, and we obtain:
\begin{subequations}
\begin{align}
\frac{E_0}{V} & \simeq \frac{1}{2}gn_B^2\cdot(-2.373)(n_Ba^3)^{\frac{1}{2}},
\\
& \simeq \frac{1}{2} gn_B^2 \cdot (-0.2373).
\end{align}
\label{Eq:CE:10-2}
\end{subequations}

\begin{figure}[tb]
\includegraphics[width=8.09cm, height=5.5cm]{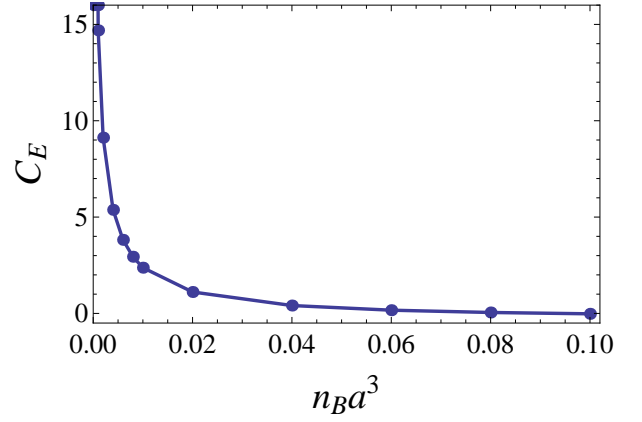}
\caption[]{ 
Plot of the quantity $C_E$ vs. $n_Ba^3$ showing that $C_E$ is not a constant, and varies rather strongly with the expansion parameter $n_Ba^3$.
}\label{Fig:C_EvsnBa3}
\end{figure}

At this point we will introduce some notation, and write the ground state energy $E_0$ at $T=0$ in the form:
\begin{equation}
\frac{E_0}{V}  \simeq -\frac{1}{2}gn_B^2 \, C_E \,(n_Ba^3)^{\frac{1}{2}} < 0,
\label{Eq:def:C_E}
\end{equation}
where $C_E$ is a positive dimensionless quantity whose precise value depends on the parameter $n_Ba^3$.
Just as we discussed in Ref. \onlinecite{Ettouhami2012}, we will not introduce any ``renormalization'' of the ``bare" interaction strength $g$ to convert the negative ground state energy $E_0$ into a positive quantity.  
In this author's view, such ``renormalizations'' where divergent integrals are substracted from $g$ 
and where the ground state energy is converted from a negative value to a positive one while leaving the coefficients of the wavefunction intact,\cite{FetterWalecka,LeggettNJP}
are improper and mathematically unjustified and hence will be avoided in the rest of this paper. 
The negative $E_0$ in Eq. (\ref{Eq:def:C_E}) with $C_E>0$ is what we will use to derive the thermodynamics of the dilute Bose gas in the canonical ensemble in Sec. \ref{Sec:Canonical} below.
For the reader's convenience, the three representative values of $C_E$ listed in Eq. (\ref{Eq:CE:10-4})-(\ref{Eq:CE:10-2}) are summarized in Table \ref{Tbl:CE_vs_nBa3}. The variation of $C_E$ over a wider range of the parameter $n_Ba^3$ is shown in Fig. \ref{Fig:C_EvsnBa3}.

\begin{center}
\begin{table}
  \begin{tabular}{ | c | c | }
    \hline\hline
    {   } $n_Ba^3$ {   } & {   } $C_E$ {   } \\ \hline
    $10^{-4}$ & 57.62  \\  \hline
    $10^{-3}$ & 14.64  \\  \hline
    $10^{-2}$ & 2.37  \\  
    \hline\hline
  \end{tabular}
\caption{Numerical values of the coefficient $C_E$ in the expression of the $T=0$ ground-state energy for a few representative values of the parameter $n_Ba^3$ as obtained from the canonical ensemble description of our variational number-conserving approach.}
\label{Tbl:CE_vs_nBa3}
\end{table}
\end{center}

\subsection{Excitation energies}
\label{Sub:ExcEn}

Having found the ground state energy of the system of interacting bosons in presence of Fock interactions between depleted bosons, we now want to find the excitation spectrum of the system within our variational approach. But, before we do so, we shall digress a little bit and uncover yet another flaw in the standard Bogoliubov theory by finding the excitation energies in presence of Fock interactions between depleted bosons in that theory and showing that these excitation energies diverge in the limit of small momenta. This will be done in the following paragraph.

\subsubsection{Digression: excitation energies in presence of Fock interactions between depleted bosons in the standard Bogoliubov formulation}
\label{SubSub:ExcEnBog}

Let us find the amount of energy $E_{\bf k}$ required to create a single excitation of wavevector ${\bf k}$ above the ground state in the standard Bogoliubov theory when Fock interactions between depleted bosons
is taken into account. This is the quantity:
\begin{align}
E_{\bf k} & = \langle\Psi_B | \alpha_{\bf k} \hat{H}\alpha_{\bf k}^\dagger|\Psi_B\rangle - \langle\Psi_B | \hat{H}|\Psi_B\rangle.
\end{align}
Using the fact that $\hat{H} = \hat{H}_B + \hat{H}_F'$, and the fact that the quantity $\langle\Psi_B | \alpha_{\bf k} \hat{H}_B\alpha_{\bf k}^\dagger|\Psi_B\rangle - \langle\Psi_B | \hat{H}_B|\Psi_B\rangle$
is nothing but the Bogoliubov excitation energy $E_{\bf k}^B=\sqrt{\varepsilon_{\bf k}(\varepsilon_{\bf k} + 2gn_0)}$, we obtain:
\begin{align}
E_{\bf k} & = E_{\bf k}^B +  \langle\Psi_B | \alpha_{\bf k} \hat{H}_F' \alpha_{\bf k}^\dagger|\Psi_B\rangle - \langle\Psi_B | \hat{H}_F'|\Psi_B\rangle.
\end{align}
Now, noticing that:
\begin{equation}
\alpha_{\bf k}\hat{H}_F' \alpha_{\bf k}^\dagger = \alpha_{\bf k}[\hat{H}_F', \alpha_{\bf k}^\dagger] + \alpha_{\bf k}\alpha_{\bf k}^\dagger \hat{H}_F',
\end{equation}
and using the commutation relation $\alpha_{\bf k}\alpha_{\bf k}^\dagger = 1 + \alpha_{\bf k}^\dagger\alpha_{\bf k}$, we can write:
\begin{equation}
E_{\bf k} = E_{\bf k}^B + \langle\Psi_B| \alpha_{\bf k}[\hat{H}_F',\alpha_{\bf k}^\dagger]|\Psi_B\rangle.
\end{equation}
For the purpose of calculating the commutator $[\hat{H}_F',\alpha_{\bf k}^\dagger]$, we shall for simplicity use the linearized approximation to $\alpha_{\bf k}$ and $\alpha_{\bf k}^\dagger$
given in Eq. (\ref{Eq:coh-factors}). Tedious but straightforward algebra leads to the result:
\begin{align}
 \langle\Psi_B| \alpha_{\bf k}[\hat{H}_F,\alpha_{\bf k}^\dagger]|\Psi_B\rangle & = g n_d (u_{\bf k} + v_{\bf k}) 
\notag\\
& - \frac{g}{V}(u_{\bf k}^2 + v_{\bf k}^2)(N_{\bf k} + N_{-\bf k}),
\end{align}
where $n_d = \sum_{\bf k\neq 0}\langle\Psi_B|a_{\bf k}^\dagger a_{\bf k}|\Psi_B\rangle/V$ is the density of depleted bosons.
In this last equation, the second term becomes negligibly small compared to the first term in the thermodynamic limit $V\to\infty$, and so we obtain for the excitation energy:
\begin{subequations}
\begin{align}
E_{\bf k} & = E_{\bf k}^B + g n_d (u_{\bf k} + v_{\bf k}),
\\
& = E_{\bf k}^B + \frac{gn_d(\varepsilon_{\bf k} + gn_0)}{\sqrt{\varepsilon_{\bf k} (\varepsilon_{\bf k} + 2 gn_0)}}.
\end{align}
\end{subequations}
As can be seen, apart from the familiar term $E_{\bf k}^B$
the excitation energy acquires an addtional contribution due to the addition of the Hamiltonian $\hat{H}_F'$. This additional contribution does not vanish in the limit $k\to 0$, but instead diverges like $1/k$, which is an erroneous feature of the standard, number non-conserving Bogoliubov theory, to be added to the list of other erroneous features we already discussed in Ref. \onlinecite{Ettouhami2012}. To remedy this divergence, we shall evaluate the excitation energies directly within the variational approach we developed in this last reference. This will be done in the next paragraph.

\subsubsection{Excitation energies in presence of Fock interactions between depleted bosons: variational approach}

\begin{figure}[tb]
\includegraphics[width=8.09cm, height=5.5cm]{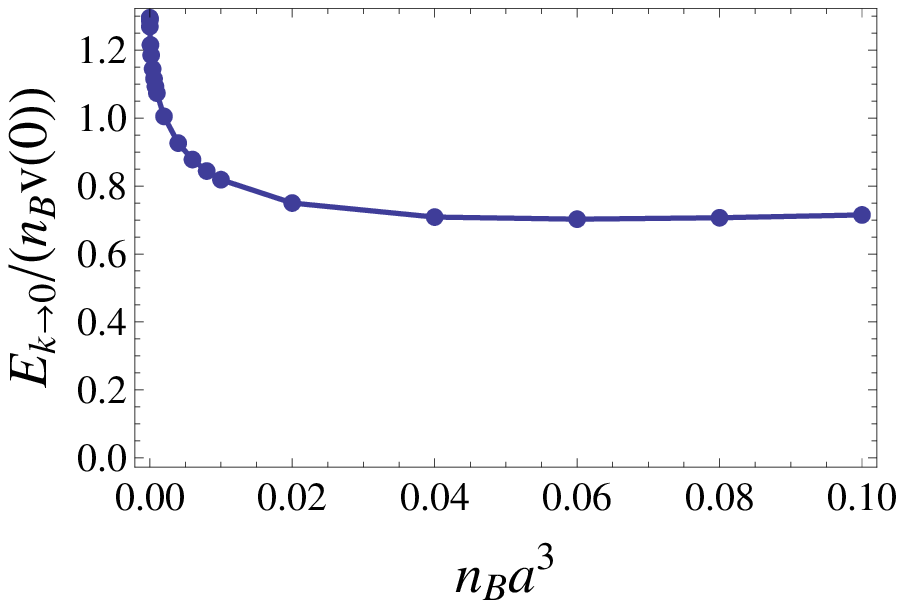}
\caption[]{(Color online)
Plot of the gap $E_{k\to 0}/(n_Bv(0))$ vs. $n_Ba^3$ from Eq. (\ref{Eq:gapEk}).
}\label{Fig:gap}
\end{figure}

We will not give here all the steps regarding the derivation of the excitation energies, and refer the interested reader to the calculation presented in Ref. \onlinecite{Ettouhami2012}, which remains unchanged, 
except for the additonal $gn_d/2$ term on the {\em rhs} of Eq. (\ref{Eq:defsigmak2b}) or equivalently the additonal $n_d/2n_B$ on the {\em rhs} of Eq. (\ref{Eq:self-consistent-sigma}),
with the result for the excitation energy $E_{\bf k}$ of wavevector ${\bf k}$ given by:
\begin{subequations}
\begin{align}
E_{\bf k} & = n_B v({\bf k}) \sqrt{Q^2(Q^2 +2 )},
\label{Eq:EkVar}
\\
Q^2 & = \frac{\tilde{k}^2 + \tilde\sigma}{C_d}.
\end{align}
\end{subequations}
Due to the fact that $Q^2\to \tilde{\sigma}/C_d$ as $k\to 0$, we see that the excitation spectrum $E_{\bf k}$ has a finite gap at $k = 0$: 
\begin{equation}
E_{k\to 0} = n_Bv({\bf 0})\sqrt{\frac{\tilde\sigma}{C_d} \left( \frac{\tilde\sigma}{C_d} + 2 \right)}.
\label{Eq:gapEk}
\end{equation}
Fig. \ref{Fig:gap} shows how the gap varies with the parameter $n_Ba^3$. The gap is largest for very small values $n_Ba^3<10^{-3}$, where $E_{k\to 0} \simeq n_Bv({\bf 0})$, while for $0.01<n_Ba^3<0.1$ the gap is a little smaller, of order $0.8 n_Bv({\bf 0})$. 

In anticipation of the customary criticism that the finite gap in Eq. (\ref{Eq:gapEk}) violates Goldstone's theorem, we here shall repeat a counter-example from Ref. \onlinecite{Ettouhami2012} proving that this theorem does not even apply to this kind of microscopic Hamiltonian. Indeed, if Goldstone's theorem applied to Bogoliubov's Hamiltonian, it would equally apply to the Hartree-Fock Hamiltonian given by the following expression:
\begin{align}
\hat{H}_{HF} & = \frac{v(0)}{2V}\hat{N}(\hat{N}-1)
+\sum_{\bf k\neq 0} \varepsilon_{\bf k}a_{\bf k}^\dagger a_{\bf k}
\nonumber\\
&+ \frac{1}{2V}\sum_{\bf k}\sum_{\bf k'(\neq k)} v({\bf k}-{\bf k}') a_{\bf k}^\dagger a_{\bf k} a_{\bf k'}^\dagger a_{\bf k'},
\end{align}
in which case the excitation spectrum of $\hat{H}_{HF}$ should not have a gap. Unfortunately, this turns out to be not true, as the excitation spectrum of $\hat{H}_{HF}$, which is known exactly,\cite{Ettouhami2012} {\em does} have a gap. We therefore conclude that Goldstone's theorem does not apply to the Hartree-Fock Hamiltonian $\hat{H}_{HF}$, and in a similar fashion does not apply to the Hamiltonian $\hat{H}$ studied in this Section.
We refer the reader to section 8.4 of Ref. \onlinecite{Ettouhami2012} where we discuss questions related to the apparent violation of the Goldstone and Hugenholtz-Pine theorems in quite some detail, and show explicitly that the gap in Eq. (\ref{Eq:gapEk}) does not violate either theorem.


Before we close this Section, we want to attract the reader's attention to the fact that neither the ground state energy in Eq. (\ref{Eq:En+HF}) nor the excitation spectrum in Eq. (\ref{Eq:EkVar}) depend on the density of condensed bosons $n_0$. Hence the canonical partition function $Z_C$ in Eq. (\ref{Eq:ZNTrace2}) will only depend on $(N,V,T)$ as it should, and not also on $N_0$. When in Sec. \ref{Sec:GrandCanonicalFormulation} we will want to derive the grand partition function $Z_G$, we will just have to perform an additional trace over $N$ as in Eq. (\ref{Eq:ZgTrace2b}), and the resulting expression of $Z_G$ will not depend on $N_0$.


With the knowledge of the ground state and excitation energies, we are now equipped to perform traces over eigenfunctions corresponding to these energies and obtain partition functions to study the thermodynamics of interacting bosons. In the following Section, we will study the statistical mechanics of interacting bosons in the canonical formalism using the ground state and excitation energies derived in the current Section. 
A discussion of the grand-canonical formalism will be presented in Sec. \ref{Sec:GrandCanonicalFormulation}.

\section{Statistical mechanics of interacting bosons: canonical number-conserving formulation}
\label{Sec:StatMechCanonicalFormulation}

We now want to discuss the thermodynamics of an interacting Bose gas in the canonical ensemble using the number conserving formalism developed in Ref. \onlinecite{Ettouhami2012} which avoid the use of Bogoliubov's prescription.
We will start by investigating the condensate fraction and the shift in transition temperature $T_c$ in Subsection \ref{Sub:n0:DeltaTc:Canonical} before discussing the thermodynamic properties in Subsection \ref{Sub:thermo:Canonical}.

\subsection{Condensate fraction and shift in transition temperature}
\label{Sub:n0:DeltaTc:Canonical}

In Sec. \ref{Sec:Canonical}, we already derived the expectation value $\langle \hat{N}_{\bf k}\rangle$ of the number of bosons of wavevector ${\bf k}$ as a canonical trace at fixed number of bosons $N$, 
and the main steps of that derivation remain valid in the variational approach of Sec. \ref{Sec:GSEnergyFock}
provided that we replace the coherence factors $u_{\bf k}$ and $v_{\bf k}$ we used in that section by the appropriate expressions for the number-conserving theory:
\begin{subequations}
\begin{align}
u_{\bf k}^2 &= \frac{1}{2}\Big(\frac{\varepsilon_{\bf k} + n_B\bar{v}({\bf k}) + \sigma}{E_{\bf k}}+1\Big), 
\label{Eq:def:u:var}
\\
v_{\bf k}^2 &= \frac{1}{2}\Big(\frac{\varepsilon_{\bf k} + n_B\bar{v}({\bf k}) + \sigma}{E_{\bf k}} - 1\Big), 
\label{Eq:def:v:var}
\end{align}
\label{Eq:coh-factors2}
\end{subequations}
where $\bar{v}({\bf k})$ was defined in Eq. (\ref{Eq:defbarv}). Using Eqs. (\ref{Eq:coh-factors2}) into Eq. (\ref{Eq:avgNk}), we obtain after a few manipulations:
\begin{subequations}
\begin{align}
N_{\bf k} & = \frac{\varepsilon_{\bf k} + n_B\bar{v}({\bf k}) + \sigma}{2E_{\bf k}} \coth\Big(\frac{\beta E_{\bf k}}{2}\Big) - \frac{1}{2},
\label{Eq:NkCanonicalVar}
\\
& =  \frac{Q^2 + 1}{2\sqrt{Q^2(Q^2+2)}} \coth\Big(\frac{\beta E_{\bf k}}{2}\Big) - \frac{1}{2},
\label{Eq:Nk:var}
\end{align}
\end{subequations}
where in going from the first to the second line we introduced the dimensionless quantity $Q^2$ defined in Eq. (\ref{Eq:Q^2}). On the other hand, it is not difficult to show that the quantity $\beta E_{\bf k}$ inside the hyperbolic cotangent can be written in dimensionless form, where temperature $T$ is measured in units of the critical temperature of a non-interacting gas $T_{c0}$, see Eq. (\ref{Eq:def:Tc0}),
as follows:
\begin{align}
 \beta E_{\bf k} = \dfrac{2[\zeta(3/2)]^\frac{2}{3}(n_Ba^3)^\frac{1}{3}}{T/T_{c0}}\sqrt{Q^2(Q^2+2)} .
\label{Eq:def:betaEk}
\end{align}
Before going any further, we pause a moment to note the differences between the expression of $N_{\bf k}$ given in Eq. (\ref{Eq:NkCanonicalVar}) and the similar expression derived within the number non-conserving approach, Eq. (\ref{Eq:N_k:b}). Apart from the appearance of the quantity $\sigma$ on the {\em rhs} of Eq. (\ref{Eq:NkCanonicalVar}), we also note that $n_0$ does not appear on the {\em rhs} of that equation, and instead $n_0$ is replaced with the total density of bosons $n_B$, which we remind the reader is fixed and does not depend on temperature. We thus see that in the variational number-conserving approach the confusion about whether $n_0$ on the {\em rhs} of Eq. (\ref{Eq:N_k:b}) depends or not on temperature does not arise because $n_0$ does not appear in the expression of $N_{\bf k}$ in the first place, hence confirming our earlier claim in Sec. \ref{Sec:Canonical} that $n_0$ on the {\em rhs} of Eq. (\ref{Eq:N_k:b})  should in fact be replaced by $n_B$.

Now, using Eq. (\ref{Eq:def:betaEk}) in Eq. (\ref{Eq:Nk:var}), we obtain after a few manipulations that the condensate fraction in the canonical ensemble using the variational number-conserving approach of Sec. \ref{Sec:GSEnergyFock} is given by:
\begin{widetext}
\begin{align}
\frac{n_0(T)}{n_B} = 1 - \frac{2^\frac{5}{2}}{\sqrt{\pi}}
(n_Ba^3)^\frac{1}{2}
\int_0^\infty d\tilde{k}\,\tilde{k}^2\left[
\frac{Q^2 +1}{\sqrt{Q^2(Q^2+2)}}
\coth\Big(\frac{[\zeta(3/2)]^{\frac{2}{3}}(n_Ba^3)^{\frac{1}{3}}}{T/T_{c0}}\sqrt{Q^2(Q^2+2)}\Big) - 1
\right].
\label{Eq:condfracBogVar}
\end{align}
\end{widetext}

\begin{figure}[tb]
\includegraphics[width=8.09cm, height=5.5cm]{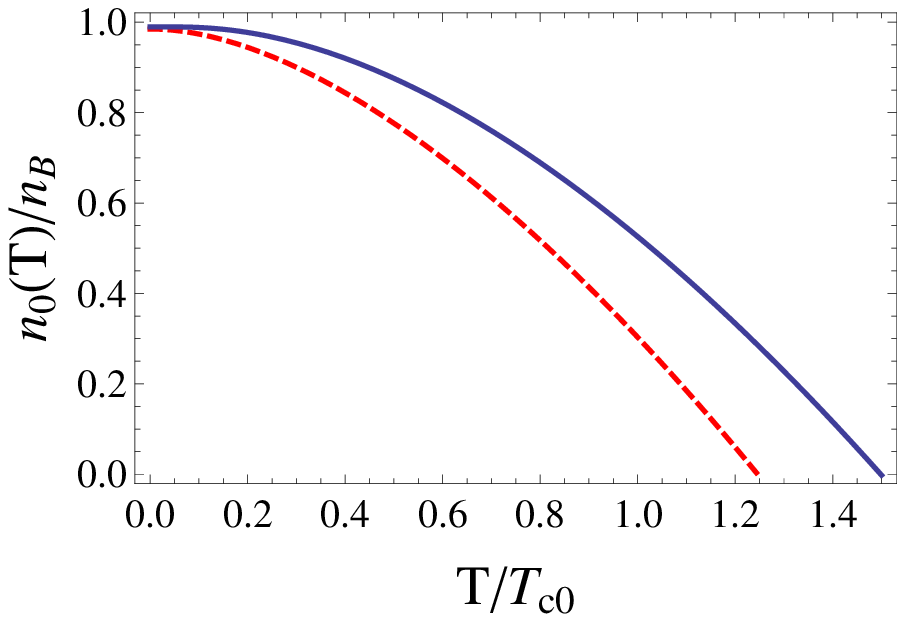}
\includegraphics[width=8.09cm, height=5.5cm]{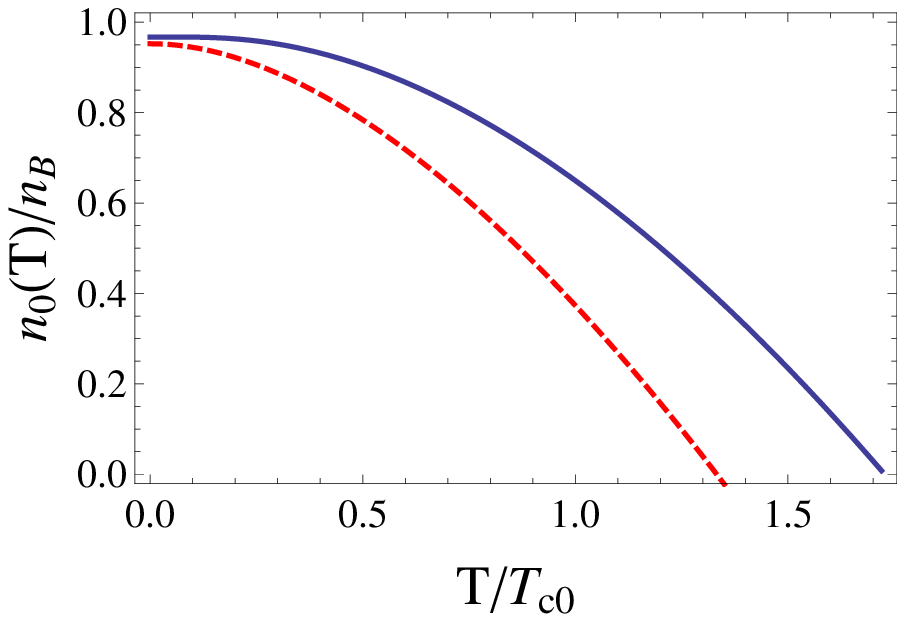}
\caption[]{(Color online)
Plot of the condensate fraction $n_0(T)/n_B$ vs. temperature $T$ for $n_Ba^3 = 10^{-4}$ (upper panel) and $10^{-3}$ (lower panel). The solid curve is the result of the variational number-conserving approach, and the dashed curve is the result obtained from the standard, number non-conserving approximation.
}\label{Fig:Plotn0vsTVar}
\end{figure}

In Fig. \ref{Fig:Plotn0vsTVar} we plot the condensate fraction $n_0(T)/n_B$ vs. temperature $T$ for $n_Ba^3 = 10^{-4}$ (upper panel) and $10^{-3}$ (lower panel). The solid curve is the result of the variational number-conserving approach, and the dashed curve is the result obtained from the standard, number non-conserving approximation. It is seen that the curves obtained from the number-conserving approach lie higher than the curves predicted by the number non-conserving approximation. One can see that it takes higher temperatures to deplete the condensate completely in the number-conserving approach, hence this approach leads to higher critical temperatures than what is predicted by the standard Bogoliubov formulation.

\begin{figure}[htb]
\includegraphics[width=8.09cm, height=5.5cm]{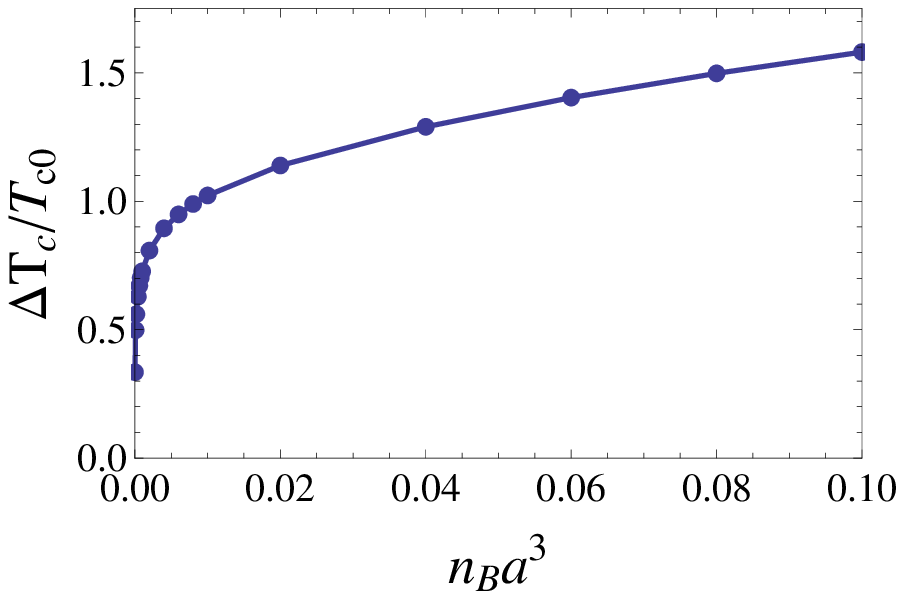}
\includegraphics[width=8.09cm, height=5.5cm]{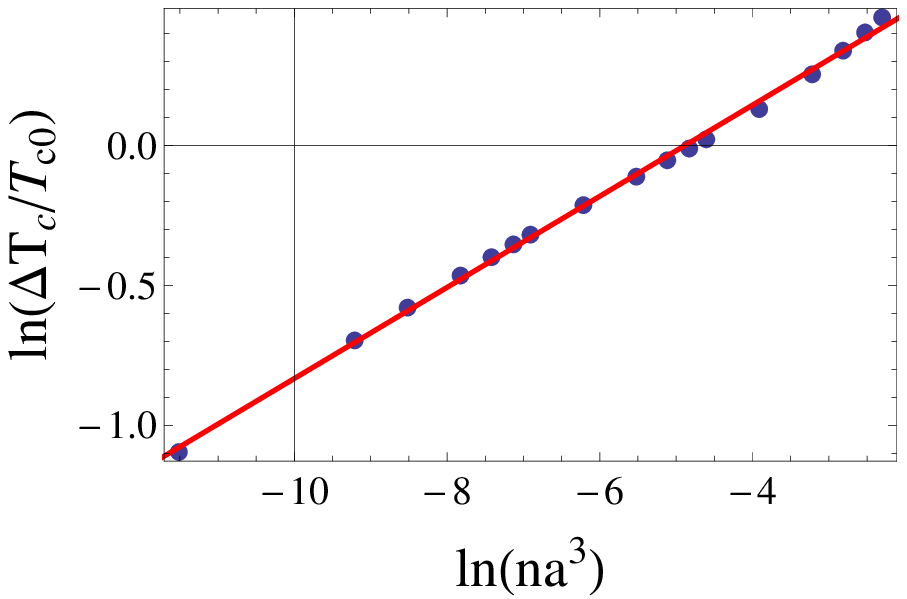}
\caption[]{(Color online)
Upper panel: Plot of $\Delta T_c/T_{c0}$ for various values of the parameter $n_Ba^3$ in the variational number-conserving approach. Lower panel: Plot of $\ln(\Delta T_c/T_{c0})$ for various values of the parameter $\ln(n_Ba^3)$. The data points fall on a straight line of slope 0.163 and intercept 0.796.
}\label{Fig:PlotDeltaTcVar}
\end{figure}

Unlike the case of the number non-conserving approximation where we were able to derive an analytical expression for $\Delta T_c/T_{c0}$ in terms of $n_Ba^3$ for small values of $n_Ba^3$, in the variational number-conserving approach such a derivation is made difficult by the fact that $\tilde\sigma$ depends on $n_Ba^3$, and the fact that this dependence is unknown analytically. Here we shall investigate this dependence numerically, and to this end in the upper panel of Fig. \ref{Fig:PlotDeltaTcVar} we plot $\Delta T_c/T_{c0}$ vs. $n_Ba^3$ in the variational number-conserving approximation (see also Table \ref{Tbl:Tc_vs_nBa3}). 
This plot shows the infinite slope behaviour near the origin which is charateristic of a dependence $\Delta T_c/T_{c0}\propto (n_Ba^3)^\eta$ with $\eta<1$. In the lower panel we plot $\ln(\Delta T_c/T_{c0})$ vs. $\ln(n_Ba^3)$, and it turns out that this plot can be well approximated by a straight line of slope $0.16357$, which seems to point to a dependence of the form:
\begin{equation}
\frac{\Delta T_c}{T_{c0}}\propto(n_Ba^3)^{\frac{1}{6}}
\label{Eq:deltaTc}
\end{equation}
which is reminiscent of the number non-conserving result of Eq. (\ref{Eq:interm3}). The only difference between the variational and standard approach is that while in the latter the curve of $\Delta T_c/T_{c0}$ vs. $n_Ba^3$ has a maximum around $n_Ba^3\approx 0.01$ and decreases monotonically for values $n_Ba^3 \ge 0.01$, no such maximum appears in the number-conserving case, and $\Delta T_c/T_{c0}$ seems to be following the variation 
in Eq. (\ref{Eq:deltaTc}) all the way across the region $0\le n_Ba^3\le 0.1$.

\begin{center}
\begin{table}
  \begin{tabular}{ | c | c | }
    \hline\hline
    {   } $n_Ba^3$ {   } & {   } $T_c/T_{c0}$ {   } \\ \hline
    $10^{-4}$ & 1.498  \\  \hline
    $10^{-3}$ & 1.727  \\  \hline
    $10^{-2}$ & 2.022  \\  
    \hline\hline
  \end{tabular}
\caption{Critical temperature $T_c/T_{c0}$ for a few representative values of the parameter $n_Ba^3$ as obtained from the canonical ensemble description of our variational number-conserving approach.}
\label{Tbl:Tc_vs_nBa3}
\end{table}
\end{center}

\subsection{Thermodynamics in the canonical ensemble}
\label{Sub:thermo:Canonical}

We now turn our attention to the thermodynamic functions of the system. The result derived in Sec. \ref{Sec:Canonical} for the canonical partition function $Z_C(N, T, V)$, Eq. (\ref{Eq:resZN}), remains valid in the number-conserving variational formalism, provided that we use for the excitation energies $E_{\bf k}$ the variational expression from Eq. (\ref{Eq:EkVar}). 
It then follows that the Helmholtz free energy
\begin{equation}
F(N,V,T)=-k_BT\ln {Z_C(N,V,T)}
\end{equation}
is given by
(we here assume that $N\gg 1$ so that $(N-1)\simeq N$):
\begin{align}
F = \frac{1}{2}Vv(0)n_B^2 + E_0 + k_BT\sum_{{\bf k}\neq 0} \ln(1-e^{-\beta E_{\bf k}}).
\end{align}
The internal energy of the gas $U = \langle \hat{H}_B\rangle$ can be obtained using the thermodynamic identity $U = F + \beta (\partial F/\partial\beta)$, valid in the canonical ensemble, with the result:
\begin{align}
U(N,V,T) = \frac{1}{2}Vv(0)n_B^2 + E_0  + \sum_{{\bf k}\neq 0} \frac{E_{\bf k}}{e^{\beta E_{\bf k}} - 1}.
\label{Eq:U(N,V,T)}
\end{align}

If we call excess internal energy $U_{exc}(N,V,T)$ the temperature-dependent term on the {\em rhs} of Eq. (\ref{Eq:U(N,V,T)}), {\em i.e.}
\begin{subequations}
\begin{align}
U_{exc}(N,V,T) & = U(N,V,T) - \frac{1}{2}Vv(0)n_B^2 - E_0 ,
\\
& = \sum_{{\bf k}\neq 0} \frac{E_{\bf k}}{e^{\beta E_{\bf k}} - 1},
\end{align}
\end{subequations}
then in dimensionless units this quantity is given by:
\begin{align}
&\frac{U_{exc}}{(n_BV)\big(n_Bv({\bf 0})\big)} = \dfrac{16}{\sqrt{2\pi}}(n_B a^3)^\frac{1}{2}
\nonumber\\
&\times\int_0^\infty d\tilde{k} 
\frac{\tilde{k}^2\sqrt{Q^2(Q^2 + 2)}}{\exp\Big(\dfrac{2[\zeta(3/2)]^\frac{2}{3}(n_Ba^3)^\frac{1}{3}}{T/T_{c0}}\sqrt{Q^2(Q^2+2)} \Big) - 1}.
\label{Eq:U_variationalApproach}
\end{align}

\begin{figure}[tb]
\includegraphics[width=8.09cm, height=5.5cm]{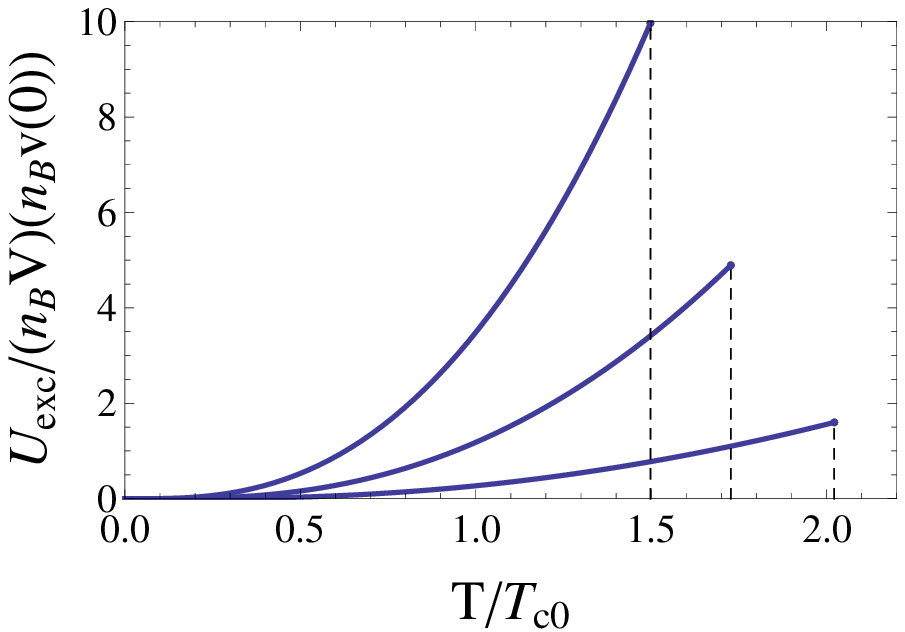}
\includegraphics[width=8.09cm, height=5.5cm]{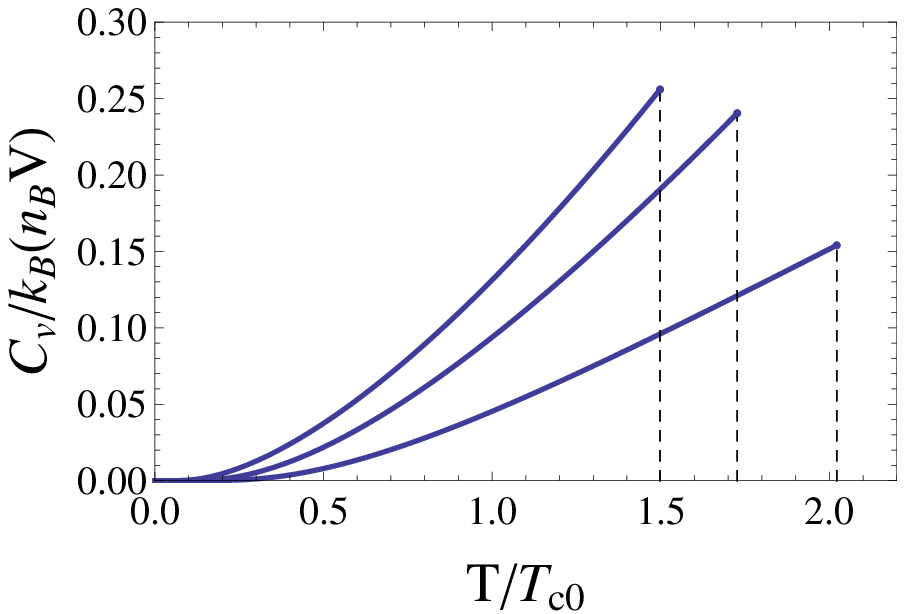}
\caption[]{ 
Upper panel: plot of the excess internal energy $U_{exc}$ of Eq. (\ref{Eq:U_variationalApproach}) as a function of the reduced temperature $T/T_{c0}$ in the canonical formulation of our variational number-conserving theory for $n_Ba^3 = 10^{-4}$, $10^{-3}$ and $10^{-2}$ from top to bottom. Lower panel: plot of the specific heat $C_v$ vs. $T/T_{c0}$ from Eq. (\ref{Eq:Cv_variationalApproach}) in the canonical formulation of our variational number-conserving theory for $n_Ba^3 = 10^{-4}$, $10^{-3}$ and $10^{-2}$ from top to bottom. In both panels, the dashed vertical line at the tip of the curve indicates the location of the critical temperature $T_c$ for the corresponding value of $n_Ba^3$.
}\label{Fig:U_vs_T_variationalApproach}
\end{figure}

The heat capacity at constant volume $C_v = (\partial U/\partial T)_{N,V}$ is given by:
\begin{align}
C_v = \frac{1}{k_BT^2}\sum_{\bf k\neq 0} \frac{E_{\bf k}^2 e^{\beta E_{\bf k}}}{\big( e^{\beta E_{\bf k}} - 1 \big)^2},
\label{Eq:C_v}
\end{align}
and in dimensionless units can be written in the form:
\begin{align}
& \frac{C_v}{k_B(Vn_B)} = \frac{2^{\frac{3}{2}}}{\pi^\frac{1}{2}}(n_Ba^3)^\frac{1}{2}\int_0^\infty d\tilde{k}\,\dfrac{ \tilde{k}^2\,\big(\beta E_{\bf k}\big)^2}{\sinh^2\Big(\frac{\beta E_{\bf k}}{2}\Big)},
\label{Eq:Cv_variationalApproach}
\end{align}
where we remind the reader that the quantity $\beta E_{\bf k}$ is given in dimensionless units by the {\em rhs} of Eq. (\ref{Eq:def:betaEk}).
In Fig.  \ref{Fig:U_vs_T_variationalApproach}, we show plots of the excess internal energy $U_{exc}(T)$ and specific heat $C_v(T)$ as obtained in the canonical formalism based on our variational number-conserving theory, 
Eqs. (\ref{Eq:U_variationalApproach}) and (\ref{Eq:Cv_variationalApproach}). In both plots, the dashed vertical line at the tip of the curve indicates the location of the critical temperature $T_c$ for the corresponding value of $n_Ba^3$.
As can be seen, at any given value of temperature, the excess internal energy $U_{exc}$ and heat capacity $C_v$ decrease as $n_Ba^3$ increases, reflecting the fact that higher repulsive interactions tend to reduce the number of depleted bosons, hence reducing the values of $U_{exc}$ and $C_v$.

\begin{figure}[h]
\includegraphics[width=8.09cm, height=5.5cm]{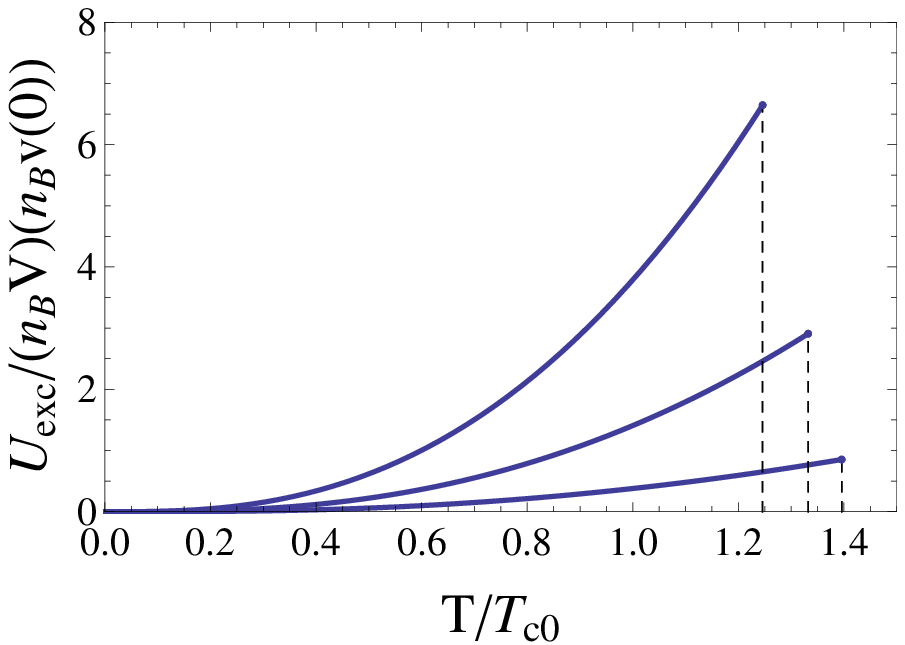}
\includegraphics[width=8.09cm, height=5.5cm]{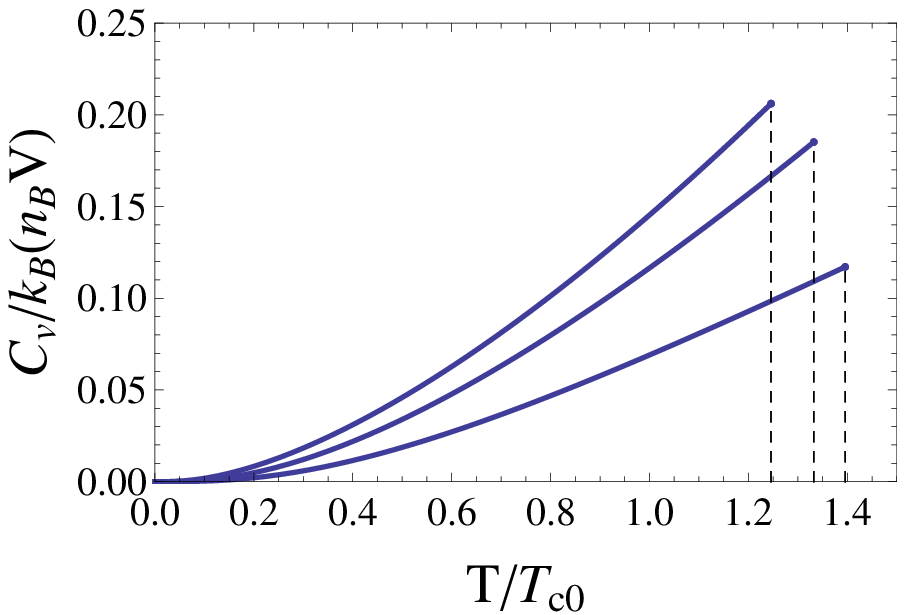}
\caption[]{  Upper panel: plot of the excess internal energy $U_{exc}$ of Eq. (\ref{Eq:UexcNaiveBog}) as a function of the reduced temperature $T/T_{c0}$ in the naive Bogoliubov theory for $n_Ba^3 = 10^{-4}$, $10^{-3}$ and $10^{-2}$ from top to bottom. Lower panel: plot of the specific heat $C_v$ of Eq. (\ref{Eq:CvNaiveBog}) vs. $T/T_{c0}$ in the naive Bogoliubov theory for $n_Ba^3 = 10^{-4}$, $10^{-3}$ and $10^{-2}$ from top to bottom. In both panels, the dashed vertical line at the tip of the curve indicates the location of the critical temperature $T_c$ for the corresponding value of $n_Ba^3$.
}\label{Fig:UCvBog}
\end{figure}

For completeness, let us examine what the above expressions of $U_{exc}$ and $C_v$ become in the naive Bogoliubov theory. In that case, $\tilde\sigma=0$, $C_d=1$ and $Q^2 = \tilde{k}^2$, and we can write:
\begin{subequations}
\begin{align}
&\frac{U_{exc}}{(n_BV)\big(n_Bv({\bf 0})\big)} = \frac{16}{\sqrt{2\pi}}(n_B a^3)^\frac{1}{2}
\nonumber\\
&\times\int_0^\infty d\tilde{k} 
\frac{\tilde{k}^2\sqrt{\tilde{k}^2(\tilde{k}^2 + 2)}}{\exp\Big(2[\zeta(3/2)]^\frac{2}{3}(n_Ba^3)^\frac{1}{3}\Big(\dfrac{T_{c0}}{T}\Big)\sqrt{\tilde{k}^2(\tilde{k}^2+2)} \Big) - 1} ,
\label{Eq:UexcNaiveBog}
\\
{}\nonumber
\\
 &\frac{C_v}{k_B(Vn_B)} = \frac{2^{\frac{3}{2}}}{\pi^\frac{1}{2}}(n_Ba^3)^\frac{1}{2}
\nonumber\\
&\times\int_0^\infty d\tilde{k}\,\frac{ \tilde{k}^2\,
\left[ 2[\zeta(3/2)]^\frac{2}{3}(n_Ba^3)^\frac{1}{3}\Big(\dfrac{T_{c0}}{T}\Big)\sqrt{\tilde{k}^2(\tilde{k}^2+2)} \right]^2}
{\sinh^2\left(
 [\zeta(3/2)]^\frac{2}{3}(n_Ba^3)^\frac{1}{3}\Big(\dfrac{T_{c0}}{T}\Big)\sqrt{\tilde{k}^2(\tilde{k}^2+2)}
\right)}.
\label{Eq:CvNaiveBog}
\end{align}
\end{subequations}
In Fig. \ref{Fig:UCvBog} we plot the excess internal energy (upper panel) and specific heat (lower panel) in the naive Bogoliubov theory for $n_Ba^3= 10^{-4}, 10^{-3}$, and $10^{-2}$. In both plots, the dashed vertical line at the tip of each curve indicates the location of the critical temperature $T_c$ for the corresponding value of $n_Ba^3$, with $T_c=1.246T_{c0}$ for $n_Ba^3 = 10^{-4}$,  $T_c=1.332T_{c0}$ for $n_Ba^3 = 10^{-3}$ and $T_c=1.396T_{c0}$ for $n_Ba^3 = 10^{-2}$. These plots are qualitatively similar to the plots obtained from the variational approach shown in Fig. \ref{Fig:U_vs_T_variationalApproach}. There is a quantitative difference coming from the fact that the critical temperature $T_c$ is a little bit higher in the variational approach than in the naive Bogoliubov approach for any given value of the parameter $n_Ba^3$ owing to the fact that when the conservation of the number of bosons is enforced, the ground state depletion becomes extremely small, and hence higher temperatures are needed in order to completely deplete the state ${\bf k}=0$ of all its bosons.

\section{Statistical mechanics of interacting bosons: grand-canonical formulation}
\label{Sec:GrandCanonicalFormulation}

Let us recapitulate what we have done so far. After having realized that the field-theoretic calculation of the grand partition function based on Bogoliubov's prescription involved an incomplete trace where the number of bosons in the condensate $N_0$ is not traced over, we generalized the variational approach of Ref. \onlinecite{Ettouhami2012} to include Fock interactions between depleted bosons, and derived the thermodynamics of a dilute Bose gas in the canonical ensemble within this method. The advantage of using the variational method lies in the fact that this approach allows us to avoid Bogoliubov's prescription, which on one hand eliminates the spurious temperature dependence introduced by this prescription in the BBP approach, and on the other hand allows us to evaluate traces in the canonical ensemble cleanly, with all the relevant variables being traced over. The result is a canonical partition function $Z_C(N, V, T)$ that does not depend on $N_0$ at all, but depends on the total number of bosons $N$ instead, as it should. In this section, we want to explore how one can perform the extra trace over $N$ to derive the grand canonical partition function of the system. This will be done in the following Subsection.

\subsection{Using saddle-point techniques to evaluate the grand partition function}

In this Subsection, we want to examine the question of how to formulate the grand-canonical description of an interacting Bose gas using the canonical formulation of Bogoliubov's theory from Sec. \ref{Sec:GSEnergyFock} as a starting point. In this Section, we shall use the expression of the ground state energy $E_0$ given by the standard Bogoliubov theory, in spite of the grave reservations \cite{Ettouhami2012}
this author has about the way this expression is usually derived\cite{FetterWalecka} and in particular about the sign of the constant $C_E$, which in this approach is given by:
\begin{equation}
C_E = -\frac{128}{15\sqrt\pi},
\end{equation}
which results in an overall {\em positive} correction to the Gross-Pitaevskii result $\frac{1}{2}gn_B^2$.
Our goal behind using the standard Bogoliubov theory in this section is to show for the record how to derive a grand-canonical formulation for the naive Bogoliubov approach. On a more practical level, using Bogoliubov's theory with constant $C_E$ allows us to ignore the complication arising from the variation of $C_E$ with the expansion parameter $n_Ba^3$, which in turn allows us to avoid taking derivatives of $C_E$ with respect to $n_B$ at various steps of the calculation.

\begin{figure}[tb]
\includegraphics[width=8.09cm, height=5.5cm]{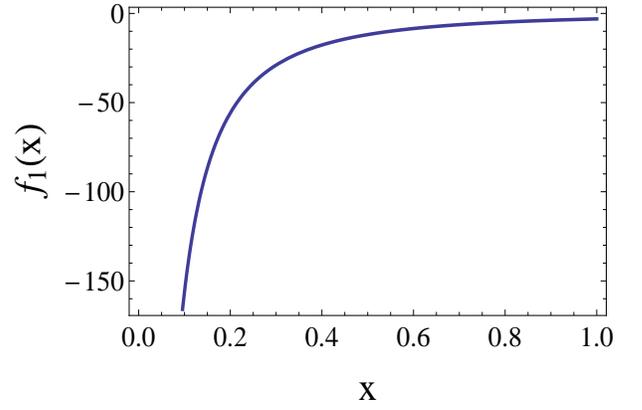}
\caption[]{Plot of the function $f_1(x)$ of Eq. (\ref{Eq:f1}) of the text. 
}\label{Fig:plotf1}
\end{figure}

We shall start by rewriting the expression of the grand-canonical partition function $Z_G(\mu, V, T)$, which is given by:
\begin{subequations}
\begin{align}
Z_G(\mu,V,T) & = \sum_{N = 0}^\infty {Z_C}(N, V, T) e^{\beta \mu N},
\label{Eq:ZgZN}
\\
& =  \sum_{N = 0}^\infty e^{-\beta ( F_N - \mu N)},
\label{Eq:ZgFN}
\end{align}
\end{subequations}
where ${Z_C}$ is the canonical partition function and $F_N$ is the canonical free energy for a system of $N$ bosons, and where in going from the first to the second line use has been made of the fact that 
${Z_C} = \exp(-\beta F_N)$. Now, in Sec. \ref{Sec:GSEnergyFock} we saw that the canonical free energy $F_N$ is given by:
\begin{align}
F_N = \frac{v({\bf 0})}{2V}N(N-1) + E_0 + \sum_{\bf k\neq 0} k_BT\ln\Big(1 - e^{-\beta E_{\bf k}}\Big).
\label{Eq:F_N1}
\end{align}
Dividing both sides by $V$ and transforming the sum to an integral, we obtain:
\begin{align}
\frac{F_N}{V} = \frac{1}{2}v({\bf 0}) n_B^2 + \frac{E_0}{V} +k_BT \int_{\bf k} \ln\Big(1 - e^{-\beta E_{\bf k}}\Big),
\label{Eq:F_N2}
\end{align}
where we approximated $(N-1)\simeq N$ and used the shorthand notation $\int_{\bf k} = \int d{\bf k}/(2\pi)^3$. Introducing the dimensionless wavevector $\tilde{k}=k/k_0$, we obtain:
\begin{align}
\frac{F_N}{V} & = \frac{1}{2} v({\bf 0}) n_B^2 + \frac{E_0}{V} + \frac{16\sqrt{2}}{\sqrt\pi}(k_BTn_B) (n_Ba^3)^\frac{1}{2} \times
\nonumber\\
&\times \int_0^\infty d{\tilde{k}}\,\tilde{k}^2  \ln\Big(1 - e^{-\beta E_{\tilde k}}\Big),
\label{Eq:F_N3}
\end{align}
where the excitation spectrum $E_{\tilde k}$ is given by:
\begin{equation}
E_{\tilde k} = n_B v({\bf k})\sqrt{\tilde{k}^2(\tilde{k}^2 + 2)}.
\end{equation}
Now, using Eq. (\ref{Eq:def:C_E}) in Eq. (\ref{Eq:F_N3}), we finally obtain:
\begin{align}
\frac{F_N}{V} & = \frac{1}{2} v({\bf 0}) n_B^2\big[ 1 - C_E(n_B a^3)^\frac{1}{2}\big]
\nonumber\\
& + (k_BTn_B) (n_Ba^3)^\frac{1}{2} f_1(\beta n_B v({\bf k})),
\label{Eq:F_N4}
\end{align}
where we denote by $f_1(x)$ the function:
\begin{align}
f_1(x) =  \frac{16\sqrt{2}}{\sqrt\pi} \int_0^\infty d{\tilde{k}}\,\tilde{k}^2  \ln\Big(1 - e^{- x \sqrt{k^2(k^2 + 2)}}\Big).
\label{Eq:f1}
\end{align}
A plot of this function is shown in Fig. \ref{Fig:plotf1}. It is seen that $f_1(x)$ takes large negative values for $0<x<1$, and goes to zero negatively for $x>1$.

\begin{figure}[tb]
\includegraphics[width=8.09cm, height=5.5cm]{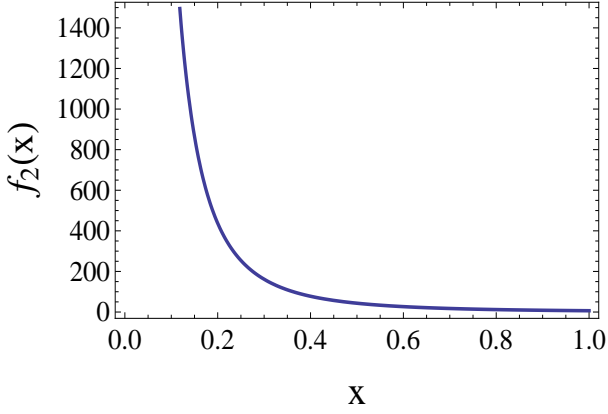}
\caption[]{ 
Plot of the function $f_2(x)$ of Eq. (\ref{Eq:f2}).
}\label{Fig:plotf2}
\end{figure}

Replacing the expression of $F_N$ from Eq. (\ref{Eq:F_N4}) into Eq. (\ref{Eq:ZgFN}), 
we find that the grand partition function 
\begin{equation}
Z_G =\sum_{N=0}^\infty e^{V(\beta\mu n_B - \beta F_N/V)}
\end{equation}
can be written in the form (note that $n_B=N/V$ here is not an actual density of bosons but merely a dummy variable which is being summed over):
\begin{subequations}
\begin{align}
\frac{Z_G }{V}& =  \frac{1}{V}\sum_{N=0}^\infty \exp \Bigg\{
V\Big[
\beta\mu n_B 
\nonumber\\
& -  \frac{1}{2} v({\bf 0}) n_B^2\Big( 1 - C_E(n_B a^3)^\frac{1}{2}\Big) 
\nonumber\\
&- (k_BTn_B) (n_Ba^3)^\frac{1}{2} f_1(\beta n_B v({\bf k}))
\Big]
\Bigg\}.
\label{Eq:Zg1}
\end{align}
\end{subequations}
Now, the quantity on the {\em rhs} can be interpreted as a Riemann sum, which in the limit $V\to\infty$ converges {\em exactly} to an integral
over the variable $n_B = N/V$, with the result:
\begin{align}
Z_G  & = V \int_0^\infty dn_B \exp\left(V g(n_B)\right),
\label{Eq:def:ZG:g}
\end{align}
where $g(n_B)$ is the following function: 
\begin{align}
g(n_B) & = 
\beta\mu n_B -  \frac{1}{2} v({\bf 0}) n_B^2\Big[ 1 - C_E(n_B a^3)^\frac{1}{2}\Big]
\nonumber\\
& - (k_BTn_B) (n_Ba^3)^\frac{1}{2} f_1(\beta n_B v({\bf k})).
\label{Eq:Zg2}
\end{align}
The integral in Eq. (\ref{Eq:def:ZG:g}) can be calculated to a very good approximation when $V\to\infty$ using saddle-point methods. To this end, we start by finding the value of $n_B$ for which the argument of the exponential is maximal. Setting the derivative $g'(n_B)$ to zero, we find that the value of $n_B$ maximizing the exponential satisfies the following equation:
\begin{align}
\mu & = n_B v({\bf 0})\left\{ 1 + (n_B a^3)^\frac{1}{2}\Big[f_2\big(\beta n_B v({\bf k})\big) - C_E\Big] \right\}
\nonumber\\
&+\frac{3}{2}k_BT(n_Ba^3)^\frac{1}{2} f_1\big(\beta n_B v({\bf k})\big)
\label{Eq:mu-nB}
\end{align}
where we defined:
\begin{align}
f_2(x) & = \frac{16\sqrt{2}}{\sqrt\pi}\int_0^\infty d\tilde{k}\, \tilde{k}^2 \frac{v({\bf k})}{v({\bf 0})}
\frac{\sqrt{k^2(k^2+2)}e^{-x \sqrt{k^2(k^2+2)}}}{1 - e^{-x \sqrt{k^2(k^2+2)}}}.
\label{Eq:def:f2}
\end{align}
For the rest of this section we shall place ourselves in the
particular case where the interaction potential is a delta function in real space, $v({\bf r}) = g\delta({\bf r})$. Then $v({\bf k}) = v({\bf 0}) = g$, and the above expression
of $f_2$ reduces to (note that $f_2$ in this case is simply the derivative of $f_1$, $f_2(x)=f_1'(x)$):
\begin{align}
f_2(x) & = \frac{16\sqrt{2}}{\sqrt\pi}\int_0^\infty d\tilde{k} \,\tilde{k}^2 
\frac{\sqrt{k^2(k^2+2)}e^{-x \sqrt{k^2(k^2+2)}}}{1 - e^{-x \sqrt{k^2(k^2+2)}}}.
\label{Eq:f2}
\end{align}
A plot of this function is shown in Fig. \ref{Fig:plotf2}. It is seen that $f_2(x)$ takes large positive values for $0<x<1$ and tends to zero positively for $x>1$.

In order to solve Eq. (\ref{Eq:mu-nB}) for $n_B$ anlytically, it is necessary to make an assumption about the value of the quantity $n_B a^3$. If we denote by $n_B^*$ the value of $n_B$ that solves Eq. (\ref{Eq:mu-nB}), we shall assume that $\sqrt{n_B^* a^3} \ll 1$. This will allow us organize the solution as an expansion in powers of the small parameter $\sqrt{n_B^* a^3}$.
Then, to zero-th order in this small parameter, we simply have:
\begin{equation}
n_B^* \simeq \frac{\mu}{v({\bf 0})}.
\label{Eq:nB*0}
\end{equation}
The next order approximation to the value of $n_B^*$ is obtained by letting $n_B = \mu/v({\bf 0})$ in the $(n_Ba^3)^\frac{1}{2}$ terms in Eq. (\ref{Eq:mu-nB}), with the result:
\begin{align}
n_B^* & = \frac{\mu}{v({\bf 0})}\Big\{
1 + \Big(\frac{\mu a^3}{v({\bf 0})}\Big)^\frac{1}{2}\Big[
\frac{5}{4} C_E - \frac{3}{2}\frac{k_BT}{\mu} f_1(\beta\mu)
\nonumber\\
& -f_2(\beta\mu) 
\Big]
\Big\}.
\label{Eq:nB*1}
\end{align}
As mentioned above, the functions $f_1(x)$ and $f_2(x)$ assume very large values at small arguments, and become small for values of $x$ that are greater than unity. Therefore, the expansion in powers of $\sqrt{n_B^*a^3}$ in Eq. (\ref{Eq:nB*1}) is only valid for values of $\beta\mu$ such that $\beta\mu > 1$, {\em i.e.} $k_BT < \mu$; otherwise, the expansion (\ref{Eq:nB*1}) is no longer valid, and a numerical method would be needed to solve Eq. (\ref{Eq:mu-nB}). Assuming the condition $\beta\mu > 1$ to be true, we now want to perform a Taylor expansion of $g(n_B)$ around $n_B^*$, writing:
\begin{subequations}
\begin{align}
g(n_B) & = g(n_B^*) + g'(n_B^*)(n_B - n_B^*) 
\nonumber\\
&+ \frac{1}{2} g''(n_B*)(n_B - n_B^*)^2,
\label{Taylor1}
\\
& = g(n_B^*)  - \frac{1}{2} |g''(n_B*)|(n_B - n_B^*)^2,
\label{Eq:Taylor2}
\end{align}
\end{subequations}
where in going from the first equality to the second we used the fact that $g'(n_B^*) = 0$ and we assumed that $g''(n_B^*) < 0$, {\em i.e.} that $n_B^*$ is a maximum of $g(n_B)$. 
The grand-canonical partition function now becomes:
\begin{align}
Z_G & = V e^{Vg(n_B^*)}\int_0^\infty dn_B e^{-\frac{1}{2}V |g''(n_B^*)|(n_B - n_B^*)^2},
\\
& = V e^{Vg(n_B^*)}\int_{-\infty}^\infty dn_B e^{-\frac{1}{2}V |g''(n_B^*)|(n_B - n_B^*)^2},
\end{align}
where in going from the first to the second line we extended the lower limit of integration to $-\infty$ because the exponential decays extremely rapidly away from $n_B^*$ when $V\to \infty$.
At this point the integral can be easily calculated, and we obtain the following result for the grand-partition function $Z_G$:
\begin{equation}
Z_G = \sqrt{\frac{2\pi V}{ |g''(n_B*)|}}\exp\left(Vg(n_B^*)\right).
\label{Eq:result1Zg}
\end{equation}
To lowest order in $(n_B^* a^3)^\frac{1}{2}$, the second derivative $g''(n_B^*)$ is given by:
\begin{equation}
g''(n_B^*) = -\beta v({\bf 0}),
\end{equation}
and is indeed negative if $v(\bf 0) > 0$. On the other hand, upon using the value of $n_B^*$ from Eq. (\ref{Eq:nB*1}) into the expression of $g(n_B^*)$ we obtain:
\begin{align}
Z_G & = \sqrt{\frac{2\pi V}{ \beta v(\bf 0)}} \exp\Bigg\{
V\frac{\beta\mu^2}{2v(\bf 0)}\Big[
1 + \Big(\frac{\mu a^3}{v(\bf 0)}\Big)^\frac{1}{2}\times 
\nonumber\\
&\times \Big(C_E - \frac{2k_BT}{\mu}f_1(\beta\mu)\Big)
\Big]
\Bigg\}.
\end{align}
Given this result for the grand-canonical partition function, one can immediately write the expression of the grand-potential $\Omega = -k_BT \ln {Z_G}$, which is given by:
\begin{align}
\Omega &= \frac{1}{2}k_BT\ln\left(\frac{ \beta v(\bf 0)}{2\pi V} \right) - \frac{V\mu^2}{2v(\bf 0)}\Big[
1 + \Big(\frac{\mu a^3}{v(\bf 0)}\Big)^\frac{1}{2}\times 
\nonumber\\
&\times \Big(C_E - \frac{2k_BT}{\mu}f_1(\beta\mu)\Big)
\Big]
\end{align}
The important thing to note in this result is that the {\em rhs} only depends on the thermodynamic variables $(\mu, T, V)$, as should be the case 
in a truly grand-canonical formulation:
\begin{align}
\Omega = \Omega(\mu, T, V).
\end{align}
In particular, previous formulations almost invariably express $\Omega$ as a function of $\mu$, $T$, $V$, but also of $n_0(T)$ 
which, in an incomplete tracing procedure is never traced over
(sometimes, both $n_0$ and the density of bosons $n_B$, which is considered to be temperature-independent, appear in the expression of $\Omega$). Here on the contrary, both $n_0$ and $n_B$ 
have been traced over and do not appear on the {\em rhs} of the above equation for $\Omega(\mu, T, V)$. Moreover, we are going to show that $n_B$ itself depends on temperature, and cannot be taken to be a temperature-independent quantity as in previous formulations.

\begin{figure}[tb]
\includegraphics[width=8.09cm, height=5.5cm]{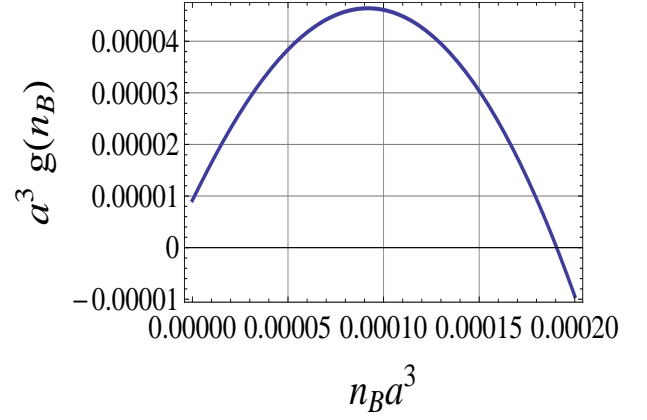}
\caption[]{(Color online) Plot of the dimensionless quantity $a^3 g(n_B)$ vs. $n_Ba^3$ for $(\mu a^3/g) = 10^{-4}$ and $T/T_0 = 0.2$. One can see that $g(n_B)$ has a pronounced maximun around $n_B^* a^3 = 10^{-4}$, in agreement with the zeroth-order approximation of Eq. (\ref{Eq:nB*0}). 
}\label{Fig:plotg(n_B)}
\end{figure}

\subsection{Finding the average number of bosons $\llangle\hat{N}\rrangle$ in the system}

Let us now find the value of the average number of bosons $\llangle \hat{N}\rrangle$ for given values of $\mu$, $T$ and $V$. In the rest of this section, we will use double angular brackets to denote the grand-canonical average. We do so in order to emphasize that this average consists of a double trace procedure, by contrast with the canonical average which involves a single trace. 
We have:
\begin{align}
\llangle \hat{N}\rrangle & = \frac{1}{Z_G}\sum_{N= 0}^\infty N e^{\beta \mu N} {Z_C},
\nonumber\\
& = - \frac{\partial \Omega}{\partial \mu}.
\end{align}
Performing the calculation, we find after a few steps:
\begin{align}
\llangle \hat{N}\rrangle &= \frac{V\mu}{v(\bf 0)}\Bigg\{
1 + \Big(\frac{\mu a^3}{v(\bf 0)}\Big)^\frac{1}{2}\Big[
\frac{5}{4}C_E - \frac{3}{2}\frac{1}{\beta \mu} f_1(\beta\mu)
\nonumber\\
& - f_2(\beta\mu)
\Big]
\Bigg\}.
\label{Eq:avgNGC}
\end{align}
Note that this value of $\llangle \hat{N}\rrangle$ agrees with the value $n_B^*$ of $n_B$ which maximizes the argument $g(n_B)$ of the exponential, see Eq. (\ref{Eq:nB*1}).
More importantly, the expression on the {\em rhs} of Eq. (\ref{Eq:avgNGC}) shows that $\llangle\hat{N}\rrangle$ varies with the thermodynamic variables $\mu$, $T$ and $V$:
\begin{equation} 
\llangle\hat{N}\rrangle =  \llangle\hat{N}\rrangle(\mu,T,V),
\end{equation}
as it should in a grand-canonical description where the system is in contact with a particle reservoir  and
the average number of particles in the system varies, in particular, when temperature is varied. 
As $T\to 0$, using the fact that $\lim_{x\to \infty} f_1(x) = \lim_{x\to\infty} f_2(x)=0$, we obtain that the average number of bosons $\llangle \hat{N}\rrangle$
approaches the limiting value:
\begin{equation}
\llangle \hat{N}\rrangle|_{T=0} = \frac{V\mu}{v({\bf 0})}\left[
1 + \frac{5}{4}C_E\Big(\frac{\mu a^3}{v({\bf 0})}\Big)^{\frac{1}{2}}
\right]. 
\label{Eq:avgNT=0}
\end{equation}

\begin{figure}[tb]
\includegraphics[width=8.09cm, height=5.5cm]{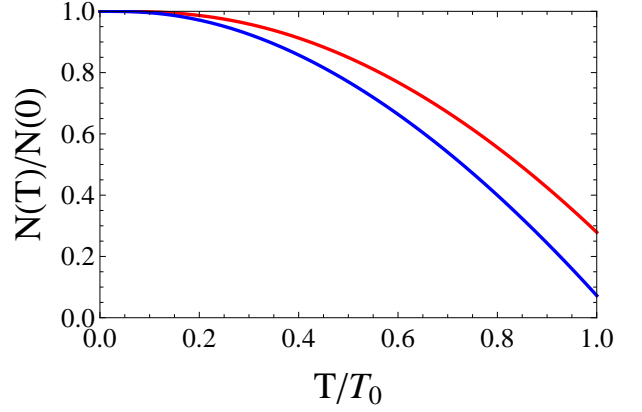}
\caption[]{(Color online) Plot of the ratio $\llangle \hat{N}\rrangle(T)/\llangle \hat{N}\rrangle(0)$ as a function of temperature in the grand-canonical ensemble. The upper curve is for $(\mu a^3/v({\bf 0}))=10^{-3}$,
and the lower curve is for $(\mu a^3/v({\bf 0}))=10^{-4}$.
}\label{Fig:AvgN}
\end{figure}

In order to be able to draw plots, we now need to introduce dimensionless units. By analogy with the canonical case where temperatures were measured in units of the critical temperature of an ideal Bose gas $T_{c0}$, 
we here shall introduce the characteristic temperature $T_0$ such that:
\begin{equation}
k_BT_0 = \frac{2\pi\hbar^2}{m}\left[
\frac{\mu/v({\bf 0})}{\zeta(3/2)}
\right]^\frac{2}{3}.
\label{Eq:def:T0}
\end{equation}
We shall start by writing:
\begin{align}
\beta\mu = \frac{\mu a^3}{v({\bf 0})}\left(\frac{v({\bf 0})}{k_BT a^3}\right).
\label{Eq:beta-mu}
\end{align}
Then, using the definition of $T_0$, Eq. (\ref{Eq:def:T0}), and the fact that $v({\bf 0})= g = 4\pi a\hbar^2/m$, we obtain after a few manipulations:
\begin{align}
\frac{v({\bf 0})}{k_BT a^3} = \frac{1}{T/T_0}\left[
\frac{2\sqrt{2}\zeta(3/2)}{\mu a^3/v({\bf 0})}
\right]^\frac{2}{3}
\label{Eq:g_over_kTa3}
\end{align}
On the other hand, going back to Eq. (\ref{Eq:beta-mu}), we find that we can write $\beta\mu$ in dimensionless units in the form:
\begin{align}
\beta\mu = \big[2\sqrt{2}\zeta(3/2)\big]^\frac{2}{3}\frac{\big[\mu a^3/v({\bf 0})\big]^\frac{1}{3}}{T/T_0}.
\label{Eq:betamu:dim}
\end{align}
We now can write the dimensionless quantity $a^3 g(n_B)$ in the form:
\begin{align}
a^3 g(n_B) & = \frac{v({\bf 0})}{k_BTa^3}\Big\{
\Big(\frac{\mu a^3}{v({\bf 0})}\Big)(n_Ba^3) 
\nonumber\\
& - \frac{1}{2}(n_Ba^3)^2\big[ 1 - C_E (n_Ba^3)^\frac{1}{2}) \big]
\nonumber\\
&- \Big(\frac{k_BTa^3}{v({\bf 0})}\Big)(n_Ba^3)^\frac{3}{2} f_1\Big(n_Ba^3 \frac{v({\bf 0})}{k_BT a^3}\Big)
\Big\}.
\label{Eq:dim:g}
\end{align}
In Fig. \ref{Fig:plotg(n_B)}, we plot the dimensionless function $a^3g(n_B)$ for $(\mu a^3/v({\bf 0}))=10^{-4}$ and $T/T_0 = 0.2$, where $v({\bf 0})/(k_BTa^3)$ on the {\em rhs} of Eq. (\ref{Eq:dim:g}) is expressed in dimensionless units using Eq. (\ref{Eq:g_over_kTa3}). It is seen that $g(n_B)$ has a pronounced maximum around $n_Ba^3 = 10^{-4}$, in agreement with the zeroth-order approximation of Eq. (\ref{Eq:nB*0}). 
On the other hand, in Fig. \ref{Fig:AvgN} we plot the ratio $\llangle \hat{N}\rrangle/\big[\llangle \hat{N}\rrangle|_{T=0}\big]$ as obtained by taking the ratio of Eqs. (\ref{Eq:avgNGC}) and (\ref{Eq:avgNT=0})
for a couple of values of the dimensionless parameter $ (\mu a^3/v({\bf 0}))$, where on the {\em rhs} of Eq. (\ref{Eq:avgNGC}) the dimensionless expression of $\beta\mu$ given in Eq. (\ref{Eq:betamu:dim}) has been used.  
The two curves in the figure correspond to $ (\mu a^3/v({\bf 0})) = 10^{-3}$ and $10^{-4}$ from top to bottom. The average density of bosons decreases quite substantially with temperature for 
both values of the parameter $(\mu a^3/v({\bf 0}))$, and appear to go to zero for temperatures in the range of $T_0$. This shows that the variation of $\llangle\hat{N}\rrangle$ with temperature is not a small perturbative effect that can be neglected, and has to be taken into account in any grand canonical formulation of the statistical mechanics of an interacting Bose gas.

\subsection{Condensate fraction in the grand-canonical ensemble}

Having found the average number of bosons $\llangle\hat{N}\rrangle(\mu, V,T)$, we now want to study the condensate fraction 
\begin{equation}
\frac{\llangle \hat{N}_0\rrangle(\mu,V,T)}{\llangle\hat{N}\rrangle(\mu, V, T=0)} = \frac{\llangle a_0^\dagger a_0\rrangle(\mu,V,T)}{\llangle\hat{N}\rrangle(\mu, V, T=0)},
\end{equation}
and in particular how this quantity varies with temperature $T$. In order to do so, we shall start by
deriving the average number of depleted bosons in the single-particle state of momentum ${\bf k}$, $\llangle \hat{N}_{\bf k}\rrangle =\llangle a_{\bf k}^\dagger a_{\bf k}\rrangle$. We have:
\begin{subequations}
\begin{align}
\llangle \hat{N}_{\bf k}\rrangle  &= \frac{1}{Z_G} \mbox{Tr}_G\Big(e^{-\beta(\hat{H}-\mu\hat{N})}a_{\bf k}^\dagger a_{\bf k}\Big),
\label{Eq:Nk-GCTrace}
\\
& = \frac{1}{Z_G}\sum_{N=0}^\infty \mbox{Tr}_N\Big(e^{-\beta(\hat{H}-\mu\hat{N})}a_{\bf k}^\dagger a_{\bf k}\Big).
\label{Eq:Nk-CanTrace}
\end{align}
\end{subequations}
In Eq. (\ref{Eq:Nk-GCTrace}), we denote by $\mbox{Tr}_G$ the grand-canonical trace: it consists of a trace over all eigenvectors of the Hamiltonian $\hat{H}$ at a given number of bosons $N$, plus a trace over all values of $N$ from $0$ to $+\infty$. The fact that the grand-canonical trace consists of two separate traces is explicitly spelled out in Eq. (\ref{Eq:Nk-CanTrace}), where we separated out the canonical trace, denoted by $\mbox{Tr}_N$ (the index $N$ emphasizing that the trace is performed at a fixed number of bosons), and the summation over all values of the total number of bosons $N$. Now, since $\hat{H}$ commutes with $\hat{N}$, 
$\exp[-\beta(\hat{H}-\mu\hat{N})] = \exp(\beta\mu\hat{N})\exp(-\beta\hat{H})$, and we can write:
\begin{align}
\llangle \hat{N}_{\bf k}\rrangle  &= \frac{1}{Z_G}\sum_{N=0}^\infty e^{\beta\mu N} \mbox{Tr}_N\Big(e^{-\beta\hat{H}}a_{\bf k}^\dagger a_{\bf k}\Big),
\nonumber\\
&= \frac{1}{Z_G}\sum_{N=0}^\infty e^{\beta\mu N}{Z_C}\times\frac{1}{{Z_C}} \mbox{Tr}_N\Big(e^{-\beta\hat{H}}a_{\bf k}^\dagger a_{\bf k}\Big),
\nonumber\\
&= \frac{1}{Z_G}\sum_{N=0}^\infty e^{\beta\mu N}{Z_C} \langle a_{\bf k}^\dagger a_{\bf k}\rangle_N,
\label{Eq:NkGC}
\end{align}
where in the last equation we used the definition of the average number of bosons in the single-particle state of momentum ${\bf k}$ in the canonical ensemble (we again use the subscript $N$ to emphasize that the average is taken at fixed number of bosons $N$):
\begin{equation}
\langle a_{\bf k}^\dagger a_{\bf k}\rangle_N = \frac{1}{{Z_C}} \mbox{Tr}_N\Big(e^{-\beta\hat{H}}a_{\bf k}^\dagger a_{\bf k}\Big).
\end{equation}
Now, in Eq. (\ref{Eq:NkGC}) we use the fact ${Z_C} = \exp(-\beta F_N)$ to write:
\begin{align}
\llangle \hat{N}_{\bf k}\rrangle & = \frac{1}{Z_G}\sum_{N=0}^\infty e^{\beta(\mu N - F_N)} \langle a_{\bf k}^\dagger a_{\bf k}\rangle_N,
\nonumber\\
& = \frac{V}{Z_G}\cdot\frac{1}{V}\sum_{N=0}^\infty e^{V(\beta \mu N/V - F_N/V)}\langle a_{\bf k}^\dagger a_{\bf k}\rangle_N.
\end{align}
Transforming the sum into an integral over the dummy variable $n_B=N/V$, we can write:
\begin{align}
\llangle \hat{N}_{\bf k}\rrangle & = \frac{V}{Z_G}\int_0^\infty dn_B\; e^{V(\beta\mu n_B - F_N/V)} \langle a_{\bf k}^\dagger a_{\bf k}\rangle_N,
\nonumber\\
& =  \frac{V}{Z_G}\int_0^\infty dn_B \; e^{Vg(n_B)} \langle a_{\bf k}^\dagger a_{\bf k}\rangle_N.
\end{align}
Using again the Taylor expansion of $g(n_B)$ around $n_B^*$, Eq. (\ref{Eq:Taylor2}), we obtain (note that we are extending the lower limit of integration to $-\infty$):
\begin{align}
\llangle \hat{N}_{\bf k}\rrangle & =  \frac{Ve^{Vg(n_B^*)}}{Z_G}\int_{-\infty}^\infty dn_B e^{-\frac{1}{2}V|g''(n_B^*)|(n_B-n_B^*)^2} \langle a_{\bf k}^\dagger a_{\bf k}\rangle_N.
\end{align}
Now, in the limit $V\to\infty$, the exponential is so sharply peaked around $n_B^*$ that a very good approximation to the integral can be obtained by using the value $n_B^*$ that maximizes $g(n_B)$ in the expression of
$\langle a_{\bf k}^\dagger a_{\bf k}\rangle_N$. Doing that, and taking this quantity outside the integral, we find:
\begin{align}
\llangle \hat{N}_{\bf k}\rrangle = \frac{1}{Z_G}\sqrt{\frac{2\pi V}{|g''(n_B^*)|}} \, e^{Vg(n_B^*)} \langle a_{\bf k}^\dagger a_{\bf k}\rangle\big|_{n_B=n_B^*}.
\end{align}
Given the expression of $Z_G$ found in Eq. (\ref{Eq:result1Zg}), we finally can write (note that the single angular brackets average on the {\em rhs} of Eq. (\ref{Eq:AvgNGCa}) is a canonical average taken at $n_B=n_B^*$):
\begin{subequations}
\begin{align}
\llangle \hat{N}_{\bf k}\rrangle & \simeq \langle \hat{N}_{\bf k}\rangle\big|_{n_B=n_B^*},
\label{Eq:AvgNGCa}
\\
& = \frac{\varepsilon_{\bf k} + gn_B^*(T)}{2E_{\bf k}\big(n_B^*(T)\big)}\coth\left(\frac{\beta E_{\bf k}\big(n_B^*(T)\big)}{2}\right) - \frac{1}{2}.
\end{align}
\end{subequations}
We therefore see that in the grand-canonical ensemble the average number of bosons in the single-particle state of momentum $\bf k$ is given by the same expression as in the canonical ensemble, Eq. (\ref{Eq:avgNkn_B}),
except that now the total density of bosons $n_B$ is replaced by the average density $n_B^*(\mu, V, T)$.

\begin{figure}[tb]
\includegraphics[width=8.09cm, height=5.5cm]{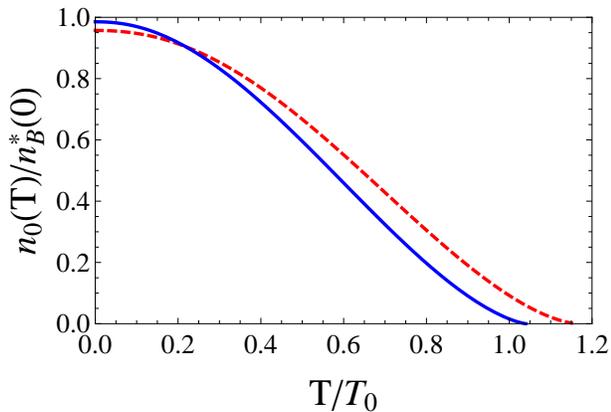}
\caption[]{(Color online) Plot of the ratio $\llangle \hat{n}_0\rrangle(T)/n_B^*(0)$ vs. temperature $T$ in the grand canonical ensemble. The solid line is for $(\mu a^3/v({\bf 0})) = 10^{-4}$, the dashed line is for 
$(\mu a^3/v({\bf 0})) = 10^{-3}$.
}\label{Fig:plotCondFracGC}
\end{figure}

Taking the sum over all vectors ${\bf k}$, we find that the condensate fraction in the grand-canonical ensemble is given by (note that this result is identical to the one in Eq. (\ref{Eq:condfracBog}), except for the fact that $n_B$ in this last equation is now replaced with $n_B^*(T)$):
\begin{widetext}
\begin{align}
\frac{\llangle n_0(T)\rrangle}{n_B^*(T)} =
1 - \frac{2^\frac{5}{2}}{\sqrt{\pi}}
(n_B^*(T)a^3)^\frac{1}{2}
 \int_0^\infty d\tilde{k}\,\tilde{k}^2
\Bigg[
\frac{\tilde{k}^2 +1}{\sqrt{\tilde{k}^2(\tilde{k}^2+2)}}
\coth\Bigg([\zeta(3/2)]^{2/3}
\big(n_B^*(T)a^3\big)^\frac{1}{3}\frac{\sqrt{\tilde{k}^2(\tilde{k}^2+2)}}{T/T_{c0}}\Bigg) - 1
\Bigg].
\end{align}
Rearraging terms, the above expression can be written in the form:
\begin{align}
\frac{\llangle n_0(T)\rrangle}{n_B^*(0)}& = \frac{n_B^*(T)}{n_B^*(0)}
\Bigg\{1 - \frac{2^\frac{5}{2}}{\sqrt{\pi}}\Big(\frac{n_B^*(T)}{n_B^*(0)}\Big)^\frac{1}{2}
(n_B^*(0)a^3)^\frac{1}{2}
\nonumber\\
&\times \int_0^\infty d\tilde{k}\,\tilde{k}^2
\Bigg[
\frac{\tilde{k}^2 +1}{\sqrt{\tilde{k}^2(\tilde{k}^2+2)}}
\coth\Bigg([\zeta(3/2)]^{2/3}\Big(\frac{n_B^*(T)}{n_B^*(0)}\Big)^\frac{1}{3}\big(n_B^*(0)a^3\big)^\frac{1}{3}\frac{\sqrt{\tilde{k}^2(\tilde{k}^2+2)}}{T/T_{c0}}\Bigg) - 1
\Bigg]
\Bigg\}.
\end{align}
\end{widetext}
A plot of the condensed fraction $\llangle n_0(T)\rrangle/\llangle n_B(0)\rrangle$ vs. temperature $T$ in the grand canonical ensemble is shown in Fig. \ref{Fig:plotCondFracGC}, with the solid curve corresponding to 
$(\mu a^3/v({\bf 0})) = 10^{-4}$ and the dashed curve to $ (\mu a^3/v({\bf 0})) = 10^{-3}$. We again note that higher values of $(\mu a^3/v({\bf 0}))$ lead to higher values of the critical temperature $T_c$, which is qualitatively the same behaviour of $T_c(a)$ we found previously in the canonical ensemble.

This concludes our discussion of the grand-canonical formulation of the statistical mechanics of interacting bosons. Our take-away from this section is that any such formulation should necessarily take into account the fact that the total number of bosons in the system is not fixed, but is a function of the grand-canonical variables $(\mu, V, T)$.

\section{Discussion and Conclusions} 
\label{Sec:Conclusions}

In this paper, we have discussed many aspects of the statistical mechanics of a dilute gas of interacting bosons. 
In doing so, we have covered a lot of ground and brought forth a number of new ideas, and so we will offer the following thoughts by way of conclusion:

\begin{enumerate}
\item
The standard BBP formulation of the statistical mechanics of interacting bosons performs an incomplete trace over the occupation numbers $N_i$ of single particle states of momentum ${\bf k}_i$, where the occupation number $N_0$ of the condensate is not traced over. The result of this partial tracing procedure is equivalent to a canonical trace at fixed boson number $N$. 
The realization that previous statistical treatments of interacting Bose gases are canonical in nature and not grand-canonical as commonly thought is {\em the} main result of this paper, a result that sets the record straight on several decades of research on the statistical mechanics of interacting bosons using BBP type of methods.

\item
Another important result of this paper is the realization that the discontinuous jump in the condensate density $n_0(T)$ at the critical transition temperature $T_c$ that is commonly obtained by these BBP type of theories originates from the inappropriate generalization of the Bogoliubov prescription $\Psi({\bf r},\tau)\simeq \sqrt{n_0} + \psi({\bf r},\tau)$ to finite temperatures. Indeed, not only is the ${\bf k}=0$ Fourier component of $\Psi({\bf r},\tau)$, which is a flucuating field in $\tau$-space, replaced with a scalar $\sqrt{n_0}$, but furthermore this scalar is erroneously assumed to have a temperature dependence even though no thermal averaging process is involved. This state of 
great confusion that is directly caused by Bogoliubov's prescription undescores the importance of using methods that circumvent this type of ad-hoc approximations.
Within the number-conserving variational approach used in this paper, where Bogoliubov's prescription is avoided altogether, no such issues arise, as the condensate fraction $n_0(T)/n_B$ vanishes continuously at $T=T_c$ and the jump discontinuity discussed above does not occur.

\item
A hallmark of grand-canonical formulations is that the total number of particles in the system is not constant, but depends on the thermodynamic variables $(\mu, V, T)$. To the best of the author's knowledge, none of the previous formulations of the statistical mechanics of interacting bosons made any attempt (let alone succeeded) at reproducing this important property, as in all these theories the density of bosons in the system $n_B$ is commonly taken to be a constant. A prominent example of this is the common use of the relation $n_B = n_0(T) + n_1(T)$ with a temperature-independent $n_B$ (most notably to find the condensate fraction $n_0(T)/n_B$), which is only correct to do in the canonical ensemble. In Sec. \ref{Sec:GrandCanonicalFormulation}, we have shown how one can devise a formulation which is truly grand-canonical by tracing over the number of bosons in the system using saddle-point techniques.
When this is done, the average number of bosons $\llangle \hat{N}\rrangle$ that is obtained does vary with temperature, and we have shown that this variation is not small, and may therefore not be neglected. 

\item
Since previous field-theoretic formulations of the BBP approach were essentially canonical in nature, and not grand-canonical in the least, the validity of many higher order perturbation theory results \cite{Shi1998}
for the spectral density function, damping effects and the lifetime of quasiparticles becomes highly questionable. On one hand, the perturbative methods used to derive these results assume the underlying statistical ensemble to be grand-canonical, while these studies are essentially canonical. On the other hand, the Bogoliubov prescription $\Psi({\bf r},\tau)\simeq \sqrt{n_0(T)} + \psi({\bf r},\tau)$ introduces a spurious temperature dependence in the expression of the normal and anomalous Green's functions $G_{11}({\bf k},\omega)$ and $G_{12}({\bf k},\omega)$ making any results derived within these theories, one is sorry to say, of highly dubious character. To be clear, this is not a statement that this author is taking lightly, nor is there any derogatory or polemical intent in it. The fact is, if we can all agree that Bogoliubov's prescription introduces an erroneous temperature dependence in the expression of $n_0$ on the {\em rhs} of this prescription (which at the most should be regarded as $n_0(T=0)$ and not $n_0(T)$), and that this erroneous temperature dependence trickles down to the expression of the Green's functions, then one is automatically led to the conclusion that any results based on such erroneous Green's functions cannot be taken at face value and need to be thoroughly and carefully re-examined. 

\item
In addition to the four items above, which constitute the most important results of this paper, we have derived other, auxiliary rsults along the way which, while not as important, should not be completely overlooked. 
An example of this would be our prediction that the shift in critical temperature $\Delta T_c$ due to the repulsive interactions between bosons varies with the dimensionless parameter $n_Ba^3$ like $\Delta T_c \propto(n_B a^3)^{\eta}$, with the exponent $\eta=1/6$ appearing to be quite robust, as it is obtained both in the naive Bogoliubov theory and in our number-conserving variational approximation.

\item
Finally, the variation of the ``gap paramter" $\tilde\sigma$ and the prefactor $C_E$ in the expression of the ground state energy of the dilute Bose gas with the interaction parameter $(n_Ba^3)$
are also interesting results in their own right. Having been derived somewhat hastily here, these results deserve a more thorough examination, perhaps in a future contribution.  The author understands that the variational theory studied in Ref. \onlinecite{Ettouhami2012} and in the present paper, having two variational parameters to be determined self-consistently, is definitely more cumbersome that the sleek-looking and elegant Bogoliubov method. 
However, as we mentioned in the introduction, this should not be an excuse for us to continue using Bogoliubov's theory totally unabated, 
after having witnessed all the shortcomings of this theory whether it be at $T=0$ or at finite temperatures, just because this theory predicts a gapless spectrum. 
If anything, this study should give us a fresh impetus to look for a more satisfying theory of Bose systems: one that rigorously conserves the number of bosons, and at the same time produces a gapless excitation spectrum.
But then again, such a theory may well turn out to be cumbersome and not be as sleek and elegant as the number non-conserving Bogoliubov theory we currently have.
 
\end{enumerate}
 
An important lesson of this paper is that there is indeniable value in non field-theoretic, good old first principles calculations. As shown in Sec. \ref{Sec:Canonical}, it is only when we ``looked under the hood," in a figurative sense, and analyzed the thermal averaging process using a back to basics, no-nonsense approach based on tracing over actual eigenvalues of the Bogoliubov Hamiltonian that we discovered the very important fact that field-theoretic formulations of the statistical mechanics of Bose systems correspond to an incomplete tracing procedure and are canonical in nature, not grand-canonical as they purport to be. Hence, first principles calculations based on actual eigenvalues and eigenfunctions of the Hamiltonian should be used, whenever it is possible, as a sanity check to convoluted field-theoretic calculations where the heavy formalism, sophisticated though it is, can conceal seemingly insignificant details which have great physical importance. It is to be hoped that the ideas presented in this study will help correct common misconceptions about the theories of interacting bosons, hence helping us formulate improved theories for these systems in the future.

\appendix

\section{Expectation value of the Hamiltonian in the number-conserving approach}
\label{App:A}

In this Appendix, we briefly describe how we calculate the expectation value of the total Hamiltonian $\hat{H} = \sum_{\bf k_j\neq 0}\hat{H}_{{\bf k}_j}$ where
\begin{subequations}
\begin{align}
\hat{H}_{{\bf k}_j} & = \frac{1}{2} \varepsilon_{{\bf k}_j}\big(a_{{\bf k}_j}^\dagger a_{{\bf k}_j}
+ a_{-{\bf k}_j}^\dagger a_{-{\bf k}_j}\big) 
+ \frac{v({\bf k}_j)}{2V}\big(a_0^\dagger a_0 a_{{\bf k}_j}^\dagger a_{{\bf k}_j}
\notag\\
& + a_0^\dagger a_0 a_{{\bf k}_j}^\dagger a_{{\bf k}_j}  
+ a_0 a_0 a_{{\bf k}_j}^\dagger a_{-{\bf k}_j}^\dagger
+ a_0^\dagger a_0^\dagger a_{{\bf k}_j} a_{-{\bf k}_j}\big)  
\nonumber\\
& +  \frac{1}{2V}\sum_{\bf k_l \neq 0,\pm k} v({\bf k}_j - {\bf k}_l) a^\dagger_{{\bf k}_j}a_{{\bf k}_j}a^\dagger_{{\bf k}_l}a_{{\bf k}_l},
\label{Eq:App:Hk}
\end{align}
\end{subequations}
in the variational ground state $|\Psi(N)\rangle$ of Eq. (\ref{Eq:fullPsi(N)}), namely (note that $Z$ here is a normalization constant, chosen so that $\langle \Psi(N)|\Psi(N)\rangle = 1$, not a partition function):
\begin{align}
|\Psi(N)\rangle & = Z\sum_{n_1=0}^\infty\cdots\sum_{n_M=0}^\infty C_{n_1}\cdots C_{n_M} 
\notag\\
&\times|N-2\sum_{i=1}^M n_i;n_1,n_1;\ldots;n_M,n_M\rangle.
\end{align}
If we denote by $\hat{H}_{\bf k_j}^{(0)}$ the part consisting of the first six terms of $\hat{H}_{\bf k_j}$ and  $\hat{H}_{\bf k_j}^{(1)}$ the part consisting of the last term on the {\em rhs} of Eq. (\ref{Eq:App:Hk}),
then the expectation value $\langle\Psi(N) | \hat{H}_{{\bf k}_j}^{(0)} |\Psi(N)\rangle$ was already calculated in Appendix  B of Ref. \onlinecite{Ettouhami2012}, with the result:
\begin{align}
&\langle\Psi(N)|\hat{H}_{{\bf k}_j}^{(0)}|\Psi(N)\rangle  = 
\varepsilon_{{\bf k}_j}\frac{c_{{\bf k}_j}^2}{1 - c_{{\bf k}_j}^2}
\nonumber\\
&+ n_B v({\bf k}_j)\frac{c_{{\bf k}_j}^2}{1 - c_{{\bf k}_j}^2}
\Big[
1 - \frac{2}{N}\sum_{i=1(\neq j)}^M\frac{c_{{\bf k}_i^2}}{1 - c_{{\bf k}_i}^2}
\Big]
\nonumber\\
& - n_B v({\bf k}_j)\frac{c_{{\bf k}_j}}{1 - c_{{\bf k}_j}^2}
\Big[
1 - \frac{2}{N}\sum_{i=1(\neq j)}^M\frac{c_{{\bf k}_i^2}}{1 - c_{{\bf k}_i}^2}
\Big].
\end{align}
Let us now calculate the expectation value $\langle \Psi(N)| \hat{H}_{\bf k_j}^{(1)}|\Psi(N)\rangle$. We have:
\begin{align}
&a^\dagger_{{\bf k}_j}a_{{\bf k}_j}a^\dagger_{{\bf k}_l}a_{{\bf k}_l} |\Psi(N)\rangle = Z \sum_{n_1=0}^\infty \cdots \sum_{n_M=0}^\infty C_{n_1}\cdots C_{n_M} n_j n_l
\notag\\
&\times |N - 2\sum_{i = 1}^M n_i; n_1;\ldots; n_M\rangle.
\end{align}
Hence:
\begin{subequations}
\begin{align}
\langle\Psi(N)|a^\dagger_{{\bf k}_j}a_{{\bf k}_j}a^\dagger_{{\bf k}_l}a_{{\bf k}_l} |\Psi(N)\rangle & = \frac{\sum_{n_j=0}^\infty n_j C_{n_j}^2}{\sum_{n_j=0}^\infty C_{n_j}^2}
\notag\\
&\times \frac{\sum_{n_j=0}^\infty n_l C_{n_l}^2}{\sum_{n_l=0}^\infty C_{n_l}^2}
\\
& = \frac{c_{{\bf k}_j}^2}{1 - c_{{\bf k}_j}^2}\cdot\frac{c_{{\bf k}_l}^2}{1 - c_{{\bf k}_l}^2}.
\end{align}
\end{subequations}
Using the above result, we obtain that the expectation value of the Hamiltonian $\hat{H}_{\bf k_j}^{(1)}$ is given by:
\begin{align}
\langle\Psi(N)| \hat{H}_{\bf k_j}^{(1)} |\Psi(N)\rangle & = \frac{1}{2V}\sum_{{\bf k}_l \neq 0,\pm {\bf k}_j} v({\bf k}_j - {\bf k}_l)  \frac{c_{{\bf k}_j}^2}{1 - c_{{\bf k}_j}^2}
\notag\\
&\times\frac{c_{{\bf k}_l}^2}{1 - c_{{\bf k}_l}^2}.
\end{align}
Rearranging terms, we obtain for the total Hamiltonian $\hat{H} = \sum_{j=1}^M\hat{H}_{{\bf k}_j}$ the following expression:
\begin{widetext}
\begin{align}
\langle\Psi(N)|\hat{H}|\Psi(N)\rangle & = \sum_{j=1}^M\Bigg\{
\Big[ 
\varepsilon_{{\bf k}_j} + n_B v({\bf k}_j)
\Big(
1 - \frac{2}{N}\sum_{i=1(\neq j)}^M\frac{c_{{\bf k}_i}^2}{1 - c_{{\bf k}_i}^2}
\Big) \Big]\frac{c_{{\bf k}_j}^2}{1 - c_{{\bf k}_j}^2}
 - n_B v({\bf k}_j)
\Big(
1 - \frac{2}{N}\sum_{i=1(\neq j)}^M\frac{c_{{\bf k}_i}^2}{1 - c_{{\bf k}_i}^2}
\Big)\frac{c_{{\bf k}_j}}{1 - c_{{\bf k}_j}^2}
\notag\\
& + n_B v({{\bf k}_j}) \frac{c_{{\bf k}_j}^2}{1 - c_{{\bf k}_j}^2}\Big[
\frac{1}{N}\sum_{{\bf k}_l \neq 0,\pm {\bf k}_j} \frac{v({\bf k}_j - {\bf k}_l)}{v({\bf k}_j)} \frac{c_{{\bf k}_l}^2}{1 - c_{{\bf k}_l}^2}
\Big]
\Bigg\}.
\end{align}
\end{widetext}
This is Eq. (\ref{Eq:avgHtot}) from the text, with $\bar{v}({\bf k})$ and $\tilde{v}({\bf k})$ defined as in Eqs. (\ref{Eq:defbarv}) and (\ref{Eq:deftildev}), respectively.

Taking the partial derivative of $\langle \hat{H} \rangle = \langle\Psi(N)|\hat{H}|\Psi(N)\rangle$ with respect to $c_{{\bf k}_j}$, using steps that are similar to those in Appendix C of Ref. \onlinecite{Ettouhami2012}, we obtain:
\begin{align}
\frac{\partial\langle \hat{H} \rangle}{\partial c_{{\bf k}_j} } = \frac{1}{(1 - c_{{\bf k}_j}^2)^2}
\Big[
2\tilde{\cal E}_{{\bf k}_j} c_{{\bf k}_j} - n_B \bar{v}({\bf k}_j) (1 + c_{{\bf k}_j}^2)
\Big],
\label{Eq:App:minH1}
\end{align}
where $\tilde{\cal E}_{{\bf k}_j}$ is given by:
\begin{subequations}
\begin{align}
\tilde{\cal E}_{{\bf k}_j} & \simeq \varepsilon_{{\bf k}_j} + n_B\bar{v}({\bf k}_j) + \frac{2}{N} \sum_{l = 1}^M n_B v({\bf k}_l)\frac{c_{{\bf k}_l}}{1 + c_{{\bf k}_l} }
\notag\\
& + \frac{1}{N} \sum_{l = 1}^M v({\bf k}_l - {\bf k}_j)\frac{c_{{\bf k}_l}^2}{1 - c_{{\bf k}_l}^2 }. 
\\
&= \varepsilon_{{\bf k}_j} + n_B\bar{v}({\bf k}_j) + \sigma_{\bf k_j},
\end{align}
\end{subequations}
where $\sigma_{\bf k_j}$ is defined in Eq. (\ref{Eq:defsigmak}).
Now, if we set the {\em rhs} of Eq. (\ref{Eq:App:minH1}) to zero, we obtain:
\begin{equation}
c_{\bf k_j}^2 - 2\Big(\frac{\tilde{\cal E}_{\bf k_j}}{n_B\bar{v}({\bf k_j})} \Big) c_{\bf k_j} + 1 = 0,
\end{equation}
which is nothing but Eq. (\ref{Eq:Eqckfull}) of the text.


\begin{thebibliography}{99}
\bibitem{Remark} One might argue that $\Psi_0$ acquires a temperature dependence in the step leading to Eq. (\ref{Eq:avgPsiPsi}). 
This is simply not true, because no thermal averaging is involved in that particular step, which merely states that the variance and covariance of the Gaussian weight $e^{-S_0[\psi,\psi^*]/\hbar}$ are given by the matrix elements of the Green's function matrix. Under these circumstances, Bogoliubov's prescription should be seen as a convenient ``trick" or a shortcut to derive the expressions of $G_{11}$ and $G_{12}$, for if we keep the $\tau$ dependence of $\Psi_0$ in the quadratic action, we will not be able to derive these expressions. Note that the same thing can be said about the Dyson's equations method, where $\Psi_0$ has to be a $\tau$-independent scalar for us to be able to find $G_{11}$ and $G_{12}$ using these equations.
\bibitem{Remark-ideal} Incidentally, we note that the trace in Eq. (\ref{Eq:ZgTrace2}) is the same one used to derive the grand partition function of an ideal Bose gas.
\end{thebibliography}
\end{document}